\documentclass[fleqn,usenatbib]{mnras}

\usepackage[T1]{fontenc}

\DeclareRobustCommand{\VAN}[3]{#2}
\let\VANthebibliography\thebibliography
\def\thebibliography{\DeclareRobustCommand{\VAN}[3]{##3}\VANthebibliography}
\def\eagle{{\sc eagle}}

\usepackage{graphicx}	
\usepackage{amsmath}	
\usepackage{amssymb}	

\title[Formation pathways of slow rotators in EAGLE]{The diverse nature and formation paths of slow rotator galaxies in the \eagle\ simulations}

\author[Claudia del P. Lagos et al.]{
Claudia del P. Lagos$^{1,2}$\thanks{E-mail: claudia.lagos@icrar.org}, Eric Emsellem$^{3,4}$, Jesse van de Sande$^{5,2}$, Katherine E. Harborne$^{1,2}$,
\newauthor
Luca Cortese$^{1,2}$, Thomas Davison$^{6,3}$, Caroline Foster$^{5,2}$, Ruby J. Wright$^{1,2}$
\\
$^{1}$International Centre for Radio Astronomy Research (ICRAR), M468, University of Western Australia, 35 Stirling Hwy, Crawley, WA 6009, Australia.\\
$^{2}$ARC  of Excellence for All Sky Astrophysics in 3 Dimensions (ASTRO 3D).\\
$^3$European Southern Observatory, Karl-Schwarzschild-Str. 2, 85748 Garching, Germany.\\
$^{4}$Univ. Lyon, Univ. Lyon1, ENS de Lyon, CNRS, Centre de Recherche Astrophysique de Lyon, UMR5574,69230 Saint-Genis-Laval, France.\\
$^5$Sydney Institute for Astronomy, School of Physics, A28, The University of Sydney, NSW 2006, Australia.\\
$^6$Jeremiah Horrocks Institute, University of Central Lancashire, Preston PR1 2HE, UK.
}

\date{Accepted XXX. Received YYY; in original form ZZZ}

\pubyear{2020}

\begin{document}
\label{firstpage}
\pagerange{\pageref{firstpage}--\pageref{lastpage}}
\maketitle

\begin{abstract}
We use a sample of $z=0$ galaxies visually classified as slow rotators (SRs) in the \eagle\ hydrodynamical simulations to explore the effect of galaxy mergers on their formation, characterise their intrinsic galaxy properties, and study the connection between quenching and kinematic transformation. SRs that have had major or minor mergers (mass ratios $\ge 0.3$ and $0.1-0.3$, respectively) tend to have a higher triaxiality parameter and ex-situ stellar fractions than those that had exclusively very minor mergers or formed in the absence of mergers (``no-merger'' SRs). No-merger SRs are more compact, have lower black hole-to-stellar mass ratios and quenched later than other SRs, leaving imprints on their $z=0$ chemical composition. For the vast majority of SRs we find that quenching, driven by active galactic nuclei feedback, precedes kinematic transformation, except for satellite SRs, in which these processes happen in tandem. However, in $\approx 50$\% of these satellites, satellite-satellite mergers are responsible for their SR fate, while environment (i.e. tidal field and interactions with the central) can account for the transformation in the rest. By splitting SRs into kinematic sub-classes, we find that flat SRs prefer major mergers; round SRs prefer minor or very minor mergers; prolate SRs prefer gas-poor mergers. Flat and prolate SRs are more common among satellites hosted by massive halos ($>10^{13.6}\,\rm M_{\odot}$) and centrals of high masses ($M_{\star} > 10^{10.5}\, \rm M_{\odot}$). Although \eagle\ galaxies display kinematic properties that broadly agree with observations, there are areas of disagreement, such as inverted stellar age and velocity dispersion profiles. We discuss these and how upcoming simulations can solve them.  
\end{abstract}




\section{Introduction}

The advent of integral field spectroscopy (IFS) and large IFS surveys, such as ATLAS$^{\rm 3D}$ \citep{Cappellari11},
the Sydney-AAO Multi-Object Integral-Field Spectrograph (SAMI) Galaxy Survey \citep{Croom12,Bryant15}, the Calar Alto Legacy Integral Field Area Survey (CALIFA; \citealt{Sanchez12}),
MASSIVE \citep{Ma14} and the Mapping Nearby Galaxies at Apache Point Observatory (MaNGA) survey \citep{Bundy15}, have contributed to significantly expand our understanding of galaxy kinematics and their connection to intrinsic galaxy properties and their environment (e.g. see \citealt{Cappellari16} for a review on kinematics of early-type galaxies). Among the kinematic parameters that have been most studied in the literature is the stellar spin parameter, $\lambda_{\rm r}$, first introduced by \citet{Emsellem07}.  $\lambda_{\rm r}$ provides a measurement of how rotationally supported a galaxy is, and strongly correlates with the stellar rotation-to-velocity dispersion ratio \citep{Emsellem11,VandeSande17,Harborne20b}. The study of galaxies in the $\lambda_{\rm r}$-ellipticity ($\epsilon)$ plane led \citet{Emsellem07,Emsellem11} to coin the terms {\it slow} and {\it fast} rotators. 

IFS surveys have unveiled various correlations between $\lambda_{\rm r}$ and galaxy properties. \citet{Emsellem11, VandeSande17b, Veale17, Brough17,Wang20} show that the fraction of low $\lambda_{\rm r}$ galaxies, or slow rotators (SRs), increases with stellar mass, and by  $10^{11.3}-10^{11.5}\,\rm M_{\odot}$, about half of the galaxies are classified as SR. In addition, \citet{Emsellem11,Cappellari16,Brough17} show that most SRs live in high density environments, typical of massive groups or galaxy clusters. However, when galaxies are studied at fixed stellar mass, it is yet unclear whether this environmental trend holds \citep{Brough17,Greene17,Graham19,Wang20}. Despite this uncertainty, it is well known from optical surveys that visually classified early-type galaxies, red and low star formation rate (SFR) galaxies become more common as we move to high density environments (e.g. \citealt{Dressler80, Peng10,Deeley17,Davies19}), even after controlling by stellar mass. \citet{Weijmans14,Foster17,Li18c,Krajnovic18} find that SRs tend to have a higher occurrence of triaxial or prolate intrinsic shapes compared to fast rotators, which are mostly oblate, axisymmetric systems (often with bars). The intrinsic stellar populations of SRs indicate flat $\rm \alpha/Fe$ metallicity radial profiles, uniform old stellar ages,  and declining metallicity radial profiles (where the central parts are more metal-rich than the outer parts; \citealt{Kuntschner10,Bernardi19,Krajnovic20}).

An outstanding question is what causes morphological or kinematic transformation in galaxies, and whether the same processes are responsible for quenching their star formation.
Several simulations have suggested that an effective way of transforming the kinematics of galaxies is via galaxy mergers (e.g. \citealt{Jesseit09,DiMatteo09,Bois11,Naab14,Penoyre17,Choi17,Lagos16b,Lagos18b,Lagos18a,Schulze18}). Although the exact remnant of a galaxy merger is dependent on many of the merger parameters involved (e.g. mass ratio, gas mass, orbital parameters, etc; e.g. \citealt{DiMatteo09,Naab14,Lagos18a}), some general trends have been reported in the literature. Among the most interesting ones is the fact that gas poor mergers tend to decrease $\lambda_{\rm r}$ \citep{Naab14,Lagos18b}, a series of minor mergers or a single major merger can have a similar effect \citep{Naab14,Choi17b,Lagos18b}, and that circular orbits preferentially
produce fast rotators \citep{Li18,Lagos18a}. One common conclusion among simulations is that even if a SR remnant is formed after a merger, continuous accretion and star formation can quickly rebuild the galaxy disk and turn the galaxy into a fast rotator \citep{Naab14,Sparre16,Penoyre17,Lagos16b,Walo-Martin20}. The latter suggests that quenching either prior or during the kinematic transformation is required to produce a SR.
Another possible way of transforming galaxies is via environmental effects, such as interactions between galaxies or with the tidal field of the group or cluster (e.g. \citealt{Choi17}). With the aim of isolating the effect of environment, \citet{Cortese19} focused on the relation between the change in SFR and $\lambda_{\rm r}$ of $z=0$ satellite galaxies in \eagle\ since they were accreted, finding no correlation between the two. This suggests that quenching and kinematic transformation are distinct processes (see also \citealt{Correa19,Wright19,Tacchella19} for similar conclusions regarding the connection of quenching with other morphological indicators in simulations). 

Many of the conclusions above have been achieved by separating fast and slow rotators using parametric selections in the $\lambda_{\rm r}-\epsilon$ plane. However, the population of galaxies obtained by these parametric forms is diverse, encompassing galaxies that are likely to have different origins. Those include what would be considered classic ellipticals (round, non-rotating objects), relatively flat SRs (flat, non-rotating objects), prolate galaxies (those that display little rotation and rotate along the minor axis), and $2\sigma$ galaxies (which have counter rotating disks that tend to cancel each other's angular momentum yielding a net low rotational velocity) (e.g. \citealt{Emsellem11,Cappellari16,vandeSande20}). In addition, simulations suggest that studying the kinematic properties of galaxies beyond $\lambda_{\rm r}$ can yield important information regarding the formation histories of galaxies \citep{Bois11,Naab14,Schulze20}. 
\citet{vandeSande20} analysed $\approx 1,800$ SAMI galaxies and compared the visual classification of the kinematic maps of galaxies with how they would be classified if they were to use a parametric selection, finding that no simple parametric cut in the $\lambda_{\rm r}-\epsilon$ plane can truly provide a high completeness, low contamination sample of galaxies visually classified as non rotators. The reason why contamination is a lot higher than in the original work of \citet{Emsellem11} is likely the poorer spatial resolution in SAMI compared to the survey ATLAS$^{\rm 3D}$ used by \citet{Emsellem11}. Because other large IFS surveys, such as MaNGA, generally have similarly limited spatial resolution, a high contamination in the parametric selection of SRs to SAMI is also expected. This lends significant weight to the process of visual classification if we are to understand the formation mechanisms of truly non- or weakly-rotating galaxies and the possible connection between kinematic transformation and quenching.

Very few examples exist of visual kinematic classification of galaxies in simulations. Among these are the work of \citet{Li18}, who used visual classification of galaxies in the Illustris simulation to find prolate galaxies and study their formation mechanisms. They found that the vast majority of prolate galaxies in their simulation have had galaxy mergers of nearly radial orbits. \citet{Ebrova20} used visual classification of Illustris galaxies to identify those with kinematically decoupled cores (KDCs) and found that they were long lived, with the vast majority of them forming after major mergers. \citet{Schulze18} visually classified the kinematic maps of early-type galaxies in the Magneticum simulation, finding a diverse family among SRs, including non-rotators, prolates and $2\sigma$ galaxies. \citet{Schulze18} found that the parametric selection of SRs of \citet{Emsellem11} led to significant contamination, with many galaxies classed as ``rotators'' being misclassified as SR. These works show that visual classification of simulated galaxies can yield new, important information about the formation of galaxies. 

In this paper we aim to understand the formation pathways of SRs and possible connection to quenching using the \eagle\ simulations. \eagle\ is a state-of-the-art cosmological hydrodynamical simulation suite \citep{Schaye14,Crain15}. Its largest cosmological box has a good compromise between volume, $(100\,\rm Mpc)^3$, and spatial resolution, $700$~pc, that allows us to have a statistically significant
sample of galaxies (several thousands with stellar masses, $M_{\star}>10^{10}\,\rm M_{\odot}$) and with enough structural detail to be able to study their stellar kinematic properties. \eagle\ has been compared to several observations of the structural and kinematic properties of galaxies in observations, finding that the simulation can reproduce reasonably well the size-stellar mass relation of active and passive galaxies across cosmic time (\citealt{Furlong15,Lange16,Rosito19}), the stellar angular momentum-stellar mass relation \citep{Lagos16b}, the fraction of SRs vs. stellar mass \citep{Lagos18b}, and the distribution of stellar rotation-to-dispersion velocity ratio \citep{vandeSande19,Walo-Martin20}. This makes \eagle\ well suited for our experiment. Because we are interested in separating truly SR galaxies from the rest of the galaxies, we go through a similar exercise as \citet{vandeSande20}, and visually inspect galaxies in \eagle\ at $z=0$ to (i) select SRs, and (ii) separate different classes of SRs (flat vs. round SRs, prolate and $2\sigma$ galaxies). We then take advantage of the plethora of galaxy properties \eagle\ allows us to measure to investigate whether the different merger histories of SRs in \eagle\ leave imprints on their intrinsic galaxy properties and kinematic class at $z=0$ that could in principle be used to connect to observed SRs and to understand whether quenching and kinematic transformation happen in tandem or not. 

This paper is organised as follows. $\S$~\ref{EagleSec} provides a brief summary of the \eagle\ simulations, how we compute kinematic properties of galaxies and visually classify them, and build the galaxy merger history of galaxies. We also compare the properties of SRs between the visually-selected vs. parametric-selected ones in \eagle.
$\S$~\ref{SRsLocally} analyses the merger history, kinematic transformation and quenching of star formation, and the stellar populations of the galaxies that are visually classified as SRs in \eagle. $\S$~\ref{KinClassesSec} analyses the connection between the different kinematic classes of SRs in \eagle\ with their merger history, and finally in $\S$~\ref{Conclusions} presents a discussion of the main results and our conclusions.

\section{The EAGLE simulation}\label{EagleSec}

The \eagle\ simulation suite
(described in detail in \citealt{Schaye14}, hereafter
S15, and \citealt{Crain15}, hereafter C15) consists of a large number of cosmological
hydrodynamic simulations with different resolutions, cosmological volumes and subgrid models,
adopting a \citet{Planck14} cosmology.
S15 introduced a reference model, within which the parameters of the
sub-grid models governing energy feedback from stars and accreting black holes (BHs) were calibrated to ensure a
good match to the $z=0.1$ galaxy stellar mass function, 
the sizes of present-day disk galaxies and the BH-stellar mass relation (see C15 for details on the tuning of parameters).

\begin{table}
\begin{center}
  \caption{Specifications of the \eagle\ Ref-L100N1504 simulation used in this paper. The rows list:
    (1) initial particle masses of gas and (2) dark
    matter, (3) comoving Plummer-equivalent gravitational
    softening length, and (4) maximum physical
    gravitational softening length. Units are indicated in each row. \eagle\
    adopts (3) as the softening length at $z\ge 2.8$, and (4) at $z<2.8$. This simulation
    has a side length of $L=100$~$\rm cMpc^3$. {Here, pkpc and ckpc refer to proper and comoving kpc, respectively.} }\label{TableSimus}
\begin{tabular}{l l l l}
\\[3pt]
\hline
& Property & Units & Value \\
\hline
(1)& gas particle mass & $[\rm M_{\odot}]$ & $1.81\times 10^6$\\
(2)& DM particle mass & $[\rm M_{\odot}]$ & $9.7\times 10^6$\\
(3)& Softening length & $[\rm ckpc]$ & $2.66$\\
(4)& max. gravitational softening & $[\rm pkpc]$& $0.7$ \\
\end{tabular}
\end{center}
\end{table}

Table~\ref{TableSimus} summarises the numerical parameters
of the simulation used in this work.
Throughout the text we use pkpc to denote proper kiloparsecs and
cMpc to denote comoving megaparsecs.
A key aspect of \eagle\ is the use of
state-of-the-art sub-grid models that capture unresolved physics.
The sub-grid physics modules adopted by \eagle\ include: (i) radiative cooling and
photoheating \citep{Wiersma09b}, (ii) star formation \citep{Schaye08}, (iii) stellar evolution and chemical enrichment \citep{Wiersma09},
(iv) stellar feedback \citep{DallaVecchia12}, and (v) BH growth and active galactic nucleus (AGN) feedback 
\citep{Rosas-Guevara13}. In addition,
the fraction of atomic and molecular gas in a gas particle is calculated in post-processing
following \citet{Rahmati13} and \citet{Lagos15}.
\eagle\ employs {\sc SUBFIND}
 (\citealt{Springel01}; \citealt{Dolag09}) to identify self-bound overdensities of particles within halos (i.e. substructures). These substructures are the galaxies in \eagle.

Throughout the text we will refer to ``central'' and ``satellite'' galaxies, where 
the central corresponds to the galaxy hosted by the main subhalo of a Friends-of-Friends halo, 
while other subhalos within the group host satellite galaxies \citep{Qu17}.
\citet{Lagos18b} computed the stellar spin parameters of galaxies in \eagle\ for the simulation of Table~\ref{TableSimus}, using the definition of \citet{Emsellem07}:

\begin{eqnarray}
\lambda_{\rm r} &=& \frac{\sum_{\rm i} L_{\rm i}\, r_{\rm i} |V_{\rm i}|}{\sum_{\rm i} L_{\rm i}\, r_{\rm i} \sqrt{V^2_{\rm i}+\sigma^2_{\rm i}}},\label{lambdaR}
\end{eqnarray}

\noindent where $V_i$ and $\sigma_i$ are the $r$-band luminosity-weighted
line-of-sight mean and standard deviation velocities in a pixel $i$ of a cubic grid for each galaxy, and 
$r_i$ is the distance from the centre of the galaxy to the $i$th pixel (i.e. the circular radius). Each cubic grid is computed using a cell of side $1.5$~pkpc, which \citet{Lagos18b} showed produce well-converged results. 
As in \citet{Emsellem11},
to measure these quantities within $r$, we only include pixels enclosed by the ellipse
of major axis $r$, ellipticity $\epsilon(r)$ and position angle $\theta_{\rm PA}$(r). 
$\epsilon(r)$ is computed within circular apertures of radii $r$ using the diagonalised inertia tensor of the galaxy's
luminosity surface density (see Eqs~$1$-$3$ in \citealt{Lagos18b} which follow \citealt{Cappellari07}). Here, we adopt $r=r_{\rm 50}$, the half-light radius in the r-band to make our measurements comparable to observations from local Universe IFU surveys. Note that our method of measuring $\epsilon(r)$ can be biased low compared to what is done in observations, where isophotes are commonly used. More details on how this was computed are presented in Section~$2.1$ of \citet{Lagos18b}. We measure $\lambda_{\rm r}$ and $\epsilon(r)$ in two orientations: with galaxies viewed through the z-axis of the simulation (considered to be random) and orienting them edge-on (using the stellar specific angular momentum). As we measure both these quantities within $r_{\rm 50}$, throughout the text we refer to them as $\lambda_{\rm r_{50}}$ and $\epsilon_{\rm r_{50}}$ for random orientations, and $\lambda_{\rm r_{50}, edge-on}$ and $\epsilon_{\rm r_{50},edge-on}$ for the edge-on case.

\citet{Lagos18b} showed that the fraction of SRs (using a variety of definitions based on $\lambda_{\rm r_{\rm 50}}$ and $\epsilon_{\rm r_{50}}$) decreases steeply with decreasing stellar mass, being $\approx 0.1$ at $10^{10}\,\rm M_{\odot}$. Considering this and that the quantities above are well converged at stellar masses above $10^{10}\,\rm M_{\odot}$ (see Appendix~A in \citealt{Lagos18b}), in this study we focus solely on galaxies above this stellar mass threshold, which results in $3,638$ galaxies at $z=0$.

\subsection{Galaxy mergers}\label{galmergerssec}

We use the merger trees available in the \eagle\ database \citep{McAlpine15} to identify galaxy mergers.
These merger trees were created using the $\tt D-Trees$ algorithm of \citet{Jiang14}. \citet{Qu17}
described how this algorithm was adapted to work with \eagle\ outputs.
Galaxies that went through mergers have more than one progenitor, and for our purpose,
we track the most massive progenitors of the merged galaxies, and compare the 
kinematic properties of those with that of the merger remnant.
The trees stored in the public database of \eagle\ connect $29$ epochs.
The time span between snapshots range from $\approx 0.3$~Gyr to $\approx 1$~Gyr.
\citet{Lagos16b} showed that these timescales are appropriate to study the effect of galaxy mergers 
on the specific angular momentum of galaxies, as $\lesssim 1$~Gyr correspond to the merger settling time. Here, we study the merger history from a loockback time of $0$ to $10$~Gyrs of $z=0$ galaxies.
We classify galaxy mergers as major mergers when the stellar mass ratio between the secondary and the primary galaxy, $M_{\rm \star,sec}/M_{\rm \star,prim}$, is $\ge 0.3$. Minor mergers are those in which this ratio is between $0.1-0.3$. We classify mergers with smaller mass ratios as ``very minor mergers''. The distinction between very minor mergers and higher-mass ratio mergers is important, as the remnants of the former can have drastically different properties \citep{Karademir19}. Even with this classification of mergers, $\approx 21$\% of galaxies with $M_{\star}\ge 10^{10}\,\rm M_{\odot}$ do not have mergers identified in the last $10$~Gyr. Table~\ref{mergers} summarises the number of galaxies we find in each of these merger classes.

\begin{table}
    \centering
        \caption{Number of galaxies in \eagle\ at $z=0$ that have $M_{\star}\ge 10^{10}\rm \,M_{\odot}$ and that went through $\ge1$ major mergers in the last $10$~Gyrs; through $\ge 1$ minor mergers and $=0$ major mergers (in the same time period); through $\ge 1$ very minor mergers and $=0$ minor/major mergers (in the same time period); have not had any mergers in the last $10$~Gyrs; have had $\ge 1$ dry mergers; have had no dry mergers, but $\ge 1$ wet mergers.}
    \begin{tabular}{c|c}
    \hline
        Sample & Number\\
        \hline
        all $M_{\star}\ge 10^{10}\,\rm M_{\odot}$ &  3638\\
        major mergers &  1113\\
        minor mergers (and no major mergers)& 1042\\
        very minor mergers (and no minor/major mergers) & 708\\
        w/o mergers & 775\\
        dry mergers ($M_{\rm \star,sec}/M_{\rm \star,prim}>0$) & 650\\
        wet mergers (no dry mergers; $M_{\rm \star,sec}/M_{\rm \star,prim}>0$) & 2213\\
        \hline
    \end{tabular}
    \label{mergers}
\end{table}

In addition, we compute the total star-forming gas ($M_{\rm SFgas}$)-to-stellar mass ratio involved in the galaxy merger: $M_{\rm SFgas,total}/M_{\rm \star,total}=\sum M^i_{\rm SFgas}/\sum M^i_{\star}$, where $i=0,1$ (for two galaxies involved in a merger). This fraction provides a measurement of whether a merger is gas-rich or poor, with a threshold at $M_{\rm SFgas,total}/M_{\rm \star,total}\approx 0.1$ separating gas-poor and gas-intermediate or rich mergers. This threshold comes from the distribution of $M_{\rm SFgas,total}/M_{\rm \star,total}$ in galaxy mergers in \eagle\ presented by \citet{Lagos18a}. We split galaxies between those that went through dry and wet mergers, by selecting those that had $\ge 1$ dry mergers over the last $10$~Gyr, and those that did not but had $\ge 1$ wet mergers over the same period (statistics of those are presented in Table~\ref{mergers}). The logic of this split is that dry mergers on average happen later compared to wet mergers, and hence in the presence of dry mergers, the past history of wet mergers is less relevant.

\subsection{Building mock kinematic maps of \eagle\ galaxies}\label{buildingmocks}

An important aspect of this paper is the visual classification of \eagle\ galaxies in a way that resembles the SAMI survey classification of \citet{vandeSande20}. Hence, we aim to build stellar kinematic maps that mimic SAMI in terms of spatial and velocity sampling, as well as seeing. For this purpose we generate mock kinematic cubes for each \eagle\ galaxy with $M_{\star}\ge 10^{10}\,\rm M_{\odot}$ using the R-package {\sc SimSpin} \citep{Harborne20a}. 

{\sc SimSpin} takes an N-body or hydrodynamical SPH simulation and produces a kinematic data cube in the style of an IFS observation. We have designed these mock observations to reflect the observational parameters of the SAMI survey \citep{Scott18}: kinematic cubes have a spatial pixel size of $0.5$~arcsec and a velocity pixel size of $65\,\rm km\,s^{-1}$ \citep{Green18}.

In each case, the stellar particle properties (initial mass, age and metallicity) are used to assign a flux to each particle. We logarithmically interpolate the GALEXEV synthesis models (\citealt{Bruzual03}; BC03) for simple stellar populations to generate a spectral energy distribution (SED) for each stellar particle using {\sc ProSpect} \citep{Robotham20}. In cases in which the metallicities lie outside the boundaries of the BC03 range, we extrapolate to find a solution as in \citet{Trayford15}.

Each galaxy has been projected to a distance such that the projected half-stellar mass radius is equivalent to a consistent number of pixels within the aperture to reduce the effects of spatial sampling. The velocities of each particle have been convolved with a Gaussian function to mimic the instrumentation effects, using a kernel of $2.65$\AA\ to match the line-spread function of the blue observing arm of the SAMI spectrograph \citep{VandeSande17}. We have further included a realistic level of seeing in these mock-observations by convolving each spatial plane in the data-cube with a Gaussian point-spread function with FWHM of $1$~arcsec. These images are produced at several inclinations, oriented using the inertia tensor. Unless otherwise specified, we use the images produced at an inclination of $60$~degrees.
{This inclination is chosen as we are trying to balance two requirements: (i) to avoid edge-on inclinations as those hamper the visual classification of, specially, velocity dispersion maps when searching for decoupled cores, local peaks of $\sigma$, etc.; (ii) to avoid orientations too close to face-on as those would make all galaxies appear round. The chosen $60$~degrees is a good compromise, corresponds to the average inclination of galaxies in the Universe, and is one in which intrinsically flat galaxies are still easy to identify as such.}

\begin{figure}
        \begin{center}
                \includegraphics[trim=0mm 2mm 0mm 2.1mm, clip,width=0.5\textwidth]{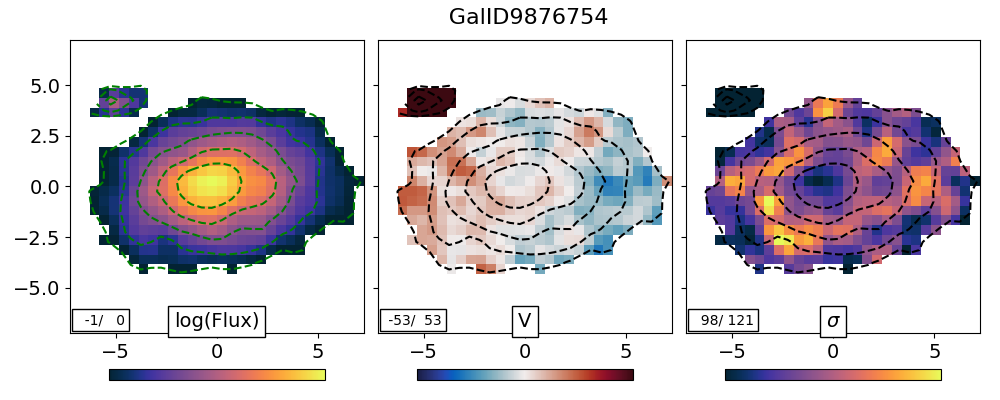}
                \includegraphics[trim=0mm 2mm 0mm 2.1mm,clip,width=0.5\textwidth]{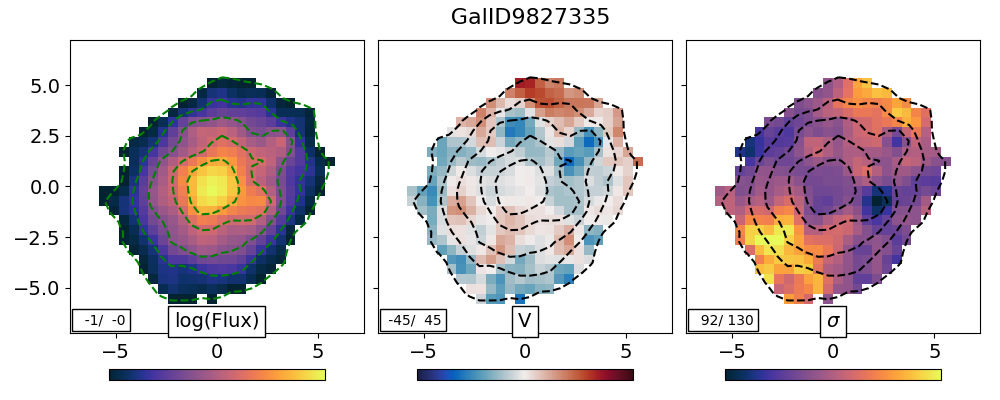}
                \includegraphics[trim=0mm 2mm 0mm 2.1mm,clip,width=0.5\textwidth]{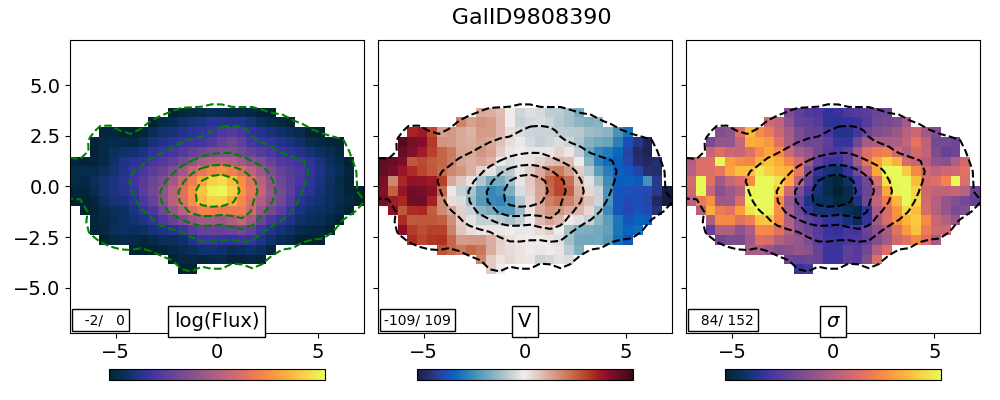}
                 \includegraphics[trim=0mm 2mm 0mm 2.1mm,clip,width=0.5\textwidth]{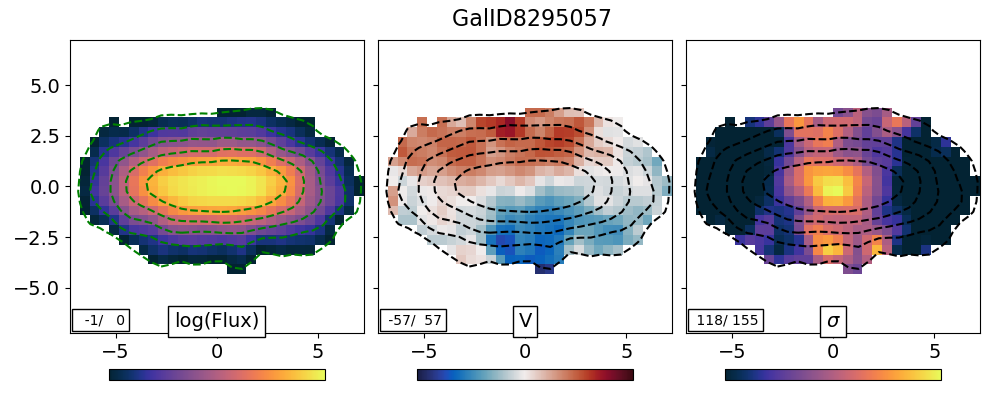}
                  \includegraphics[trim=0mm 2mm 0mm 2.1mm,clip,width=0.5\textwidth]{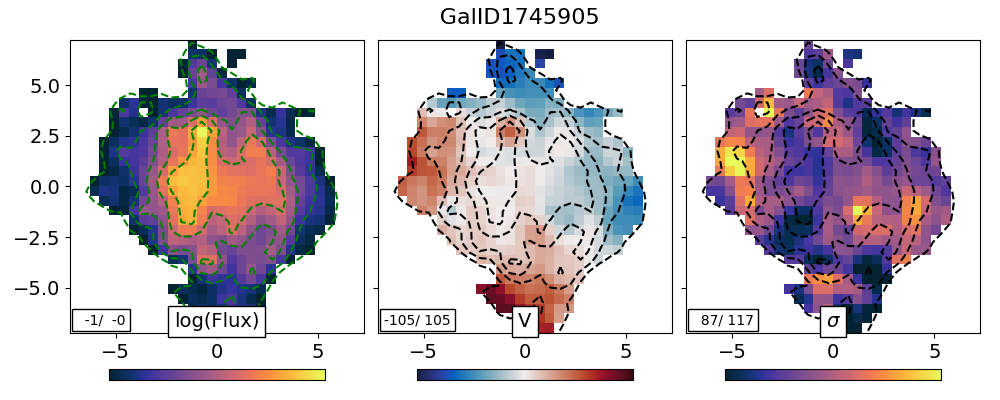}
                  \includegraphics[trim=0mm 2mm 0mm 2.1mm,clip,width=0.5\textwidth]{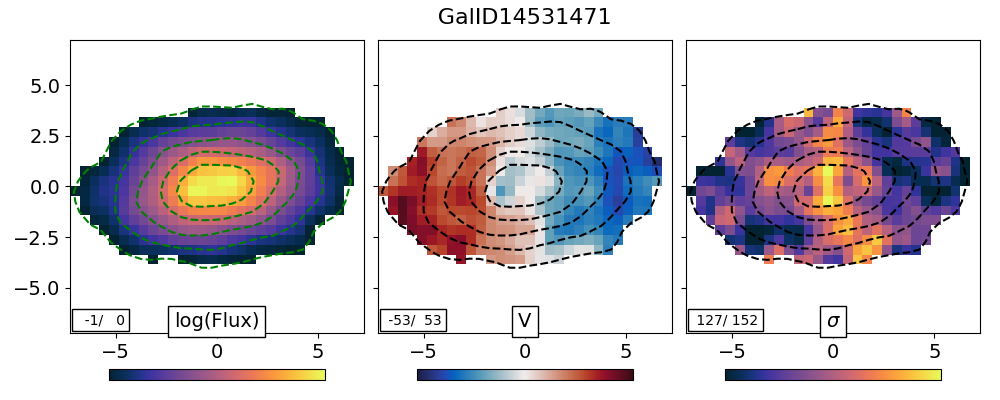}
                \caption{Examples of flux (left), LOS stellar velocity (middle) and velocity dispersion (right) maps for $z=0$ galaxies in \eagle. Units in the x- and y-axes are pkpc. Colour bar's minima and maxima are shown at the bottom left of each panel (with velocities in $\rm km/s$). From top to bottom, we show examples of galaxies with $100$\% agreement among classifiers that belong to the flat SR, round SR, $2\sigma$, prolate, unclear and rotator kinematic classes, respectively (see $\S$~\ref{secclassification} for details). The Galaxy ID is shown at the top of each row, and can be matched to the IDs in the \eagle\ database \citep{McAlpine15}.} 
                \label{ExampleImages}
        \end{center}
\end{figure}

Flux, line-of-sight (LOS) velocity and velocity dispersion maps are constructed from these mock data cubes and visualised using {\sc Pynmap}\footnote{\url{https://github.com/emsellem/pynmap}}. Flux maps are simply the sum of the flux in each pixel throughout the cube; LOS velocity maps are the flux weighted mean of the velocities at each pixel; and LOS velocity dispersion maps are the flux weighted standard deviation of the velocities in each pixel. For more information about the construction of these data products, we direct the reader to \citet{Harborne20a}. Fig.~\ref{ExampleImages} shows examples of the maps generated with {\sc SimSpin} and visualised using {\sc Pynmap}. In some cases the central stellar velocity dispersion is lower than in the outskirts (see third and bottom right hand panels of Fig.~\ref{ExampleImages}). We find this to be a frequent feature in massive galaxies in \eagle. In fact, $\approx 55$\% of galaxies with $M_{\star}>10^{10}\,\rm M_{\odot}$ have $\sigma_{\star}(0.5\,\rm r_{50})<\sigma_{\star}(r_{50})$, where $\sigma_{\star}(0.5\,\rm r_{50})$ and $\sigma_{\star}(r_{50})$ are the stellar velocity dispersions measured using particles within $0.5\,\rm r_{50}$ and $\rm r_{50}$, respectively. This is further discussed in $\S$~\ref{subsect_stellarpops}.

\subsection{Visual classification of simulated kinematic maps}\label{secclassification}

In previous papers we have classified galaxies as slow and fast rotators using parametric criteria based on the distribution of galaxies in the $\lambda_{\rm r_{50}}-\epsilon_{\rm r_{50}}$ plane. Recently, \citet{vandeSande20} have questioned the applicability of these criteria which, for the most part, have been built with higher resolution data, highlighting that a visual classification of kinematic maps yields different classifications to those obtained by parametric criteria. 
{\citet{Harborne20b}, using numerical simulations of galaxies of different disk/bulge ratios, quantified how resolution affects the derived $\lambda_{\rm R}$. They found that lower resolution leads to artificially low $\lambda_{\rm R}$, which can lead to galaxies being misclassified as being below the line of slow rotators in the $\lambda_{\rm R}-\epsilon$ plane. A similar result was presented in \citet{Graham18}.}
In addition, \citet{Naab14} showed that the details of the kinematic maps of galaxies can yield important information about the formation history of SRs, making visual classification desirable to advance our understanding of galaxy evolution.

Here, we take advantage of the maps generated in $\S$~\ref{buildingmocks} to go through a similar classification campaign as presented in \citet{vandeSande20} for SAMI. The aim is to isolate ``unambiguous'' SRs in \eagle\ and understand their relation to assembly history as well as environment. Here, unambiguous refers to galaxies that visually look like SRs. We first select all galaxies with $M_{\star}\ge 10^{10}\,\rm M_{\odot}$, which are expected to have well-converged internal stellar kinematics. From this sample, we take a very conservative selection in $\lambda_{\rm r_{50},edge-on} \le 0.2$. \citet{vandeSande20} decomposed the galaxy population in bins of stellar mass and used mixture models to determine the existence of a distinct population of low $\lambda_{\rm r_{50}}$ in SAMI and several simulations, including \eagle. A cut at $\lambda_{\rm r_{50},edge-on} \le 0.2$ comfortably includes all galaxies that belong to the population of low $\lambda_{\rm r_{50}}$ in \eagle. This selection in stellar mass and $\lambda_{\rm r_{50}, edge-on}$ yields 559 galaxies at $z=0$. 

We ask $5$ members of our team to independently classify those maps into $6$ different kinematic classes: flat SRs (FSR), round SRs (RSR), $2\sigma$ galaxies (that display two clear peaks in the stellar velocity dispersion map), prolate galaxies (Prol; those displaying rotation along the minor axis), unclear (Uncl) and rotators. We purposely avoid giving any instructions to the classifiers and simply let them assess what they expect for these different classes. We believe this provides a truly independent classification and avoid confirmation bias. We then compiled these classifications and analyse the level of agreement. Fig.~\ref{ExampleImages} shows $6$ examples of the kinematic classes above, for which all classifiers agreed. For the Uncl cases, we find that those generally are similar to the example shown in Fig.~\ref{ExampleImages}, in which there is a lot of substructure that is assigned to the same subhalo. This is a well-known shortcoming of $3$-dimensional subhalo finders \citep{Canas19}, which tends to get worse in high density environments.

\begin{figure}
        \begin{center}
                \includegraphics[trim=3mm 8mm 5mm 14mm, clip,width=0.2575\textwidth]{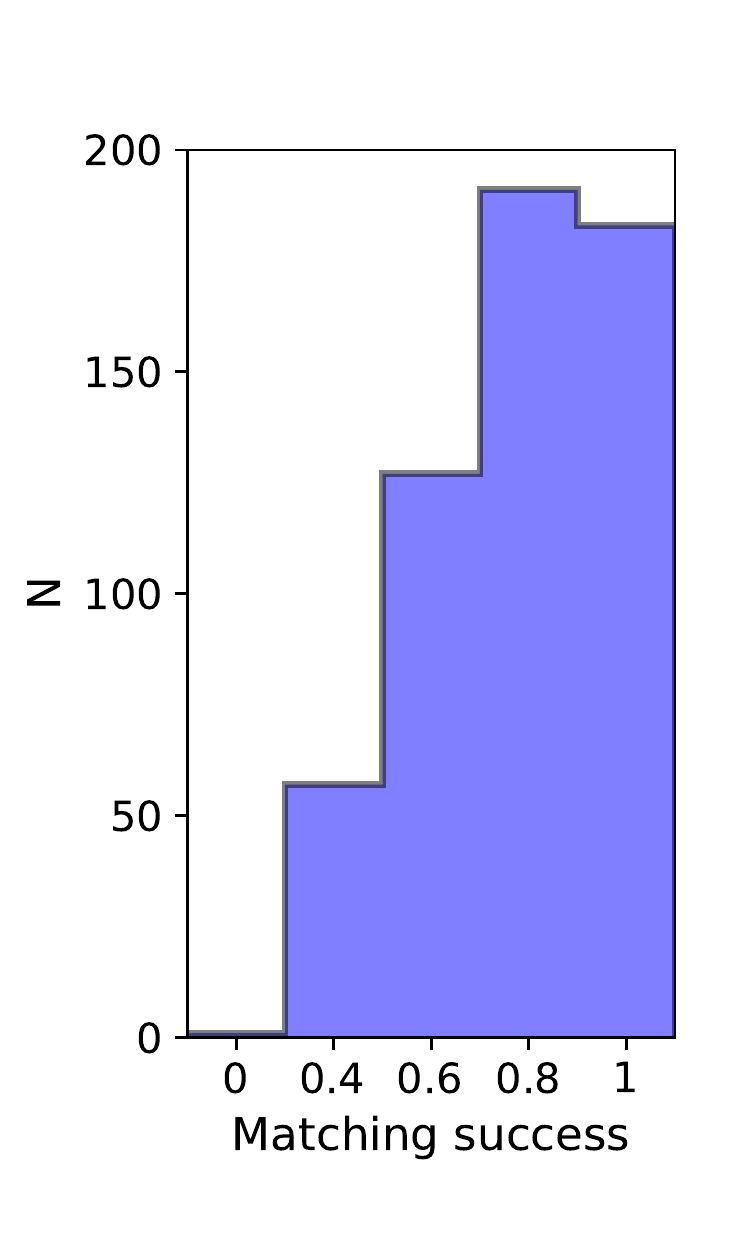}
                \includegraphics[trim=14mm 6mm 5mm 12mm, clip,width=0.2125\textwidth]{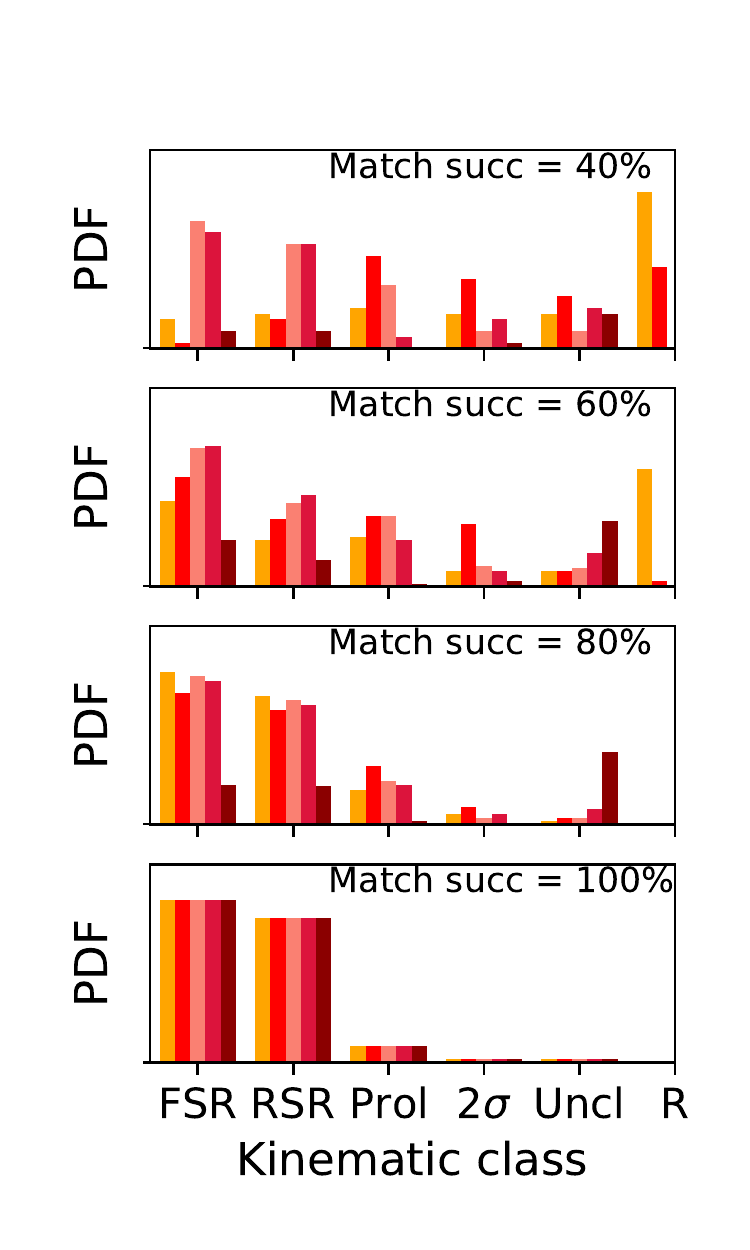}
                \caption{{\it Left panel:} Distribution of the matching success, with 0\% indicating no agreement between the kinematic classes of classifiers. The majority of galaxies have an agreement of $\ge 60$\% among classifiers. {\it Right panels:} Probability density function of kinematic classes in four bands of matching success, as labelled in each panel, with each colour showing a different classifier.}
                \label{KinematicMatching}
        \end{center}
\end{figure}

\begin{figure}
        \begin{center}
                \includegraphics[trim=24.5mm 0mm 7mm 5mm, clip,width=0.5\textwidth]{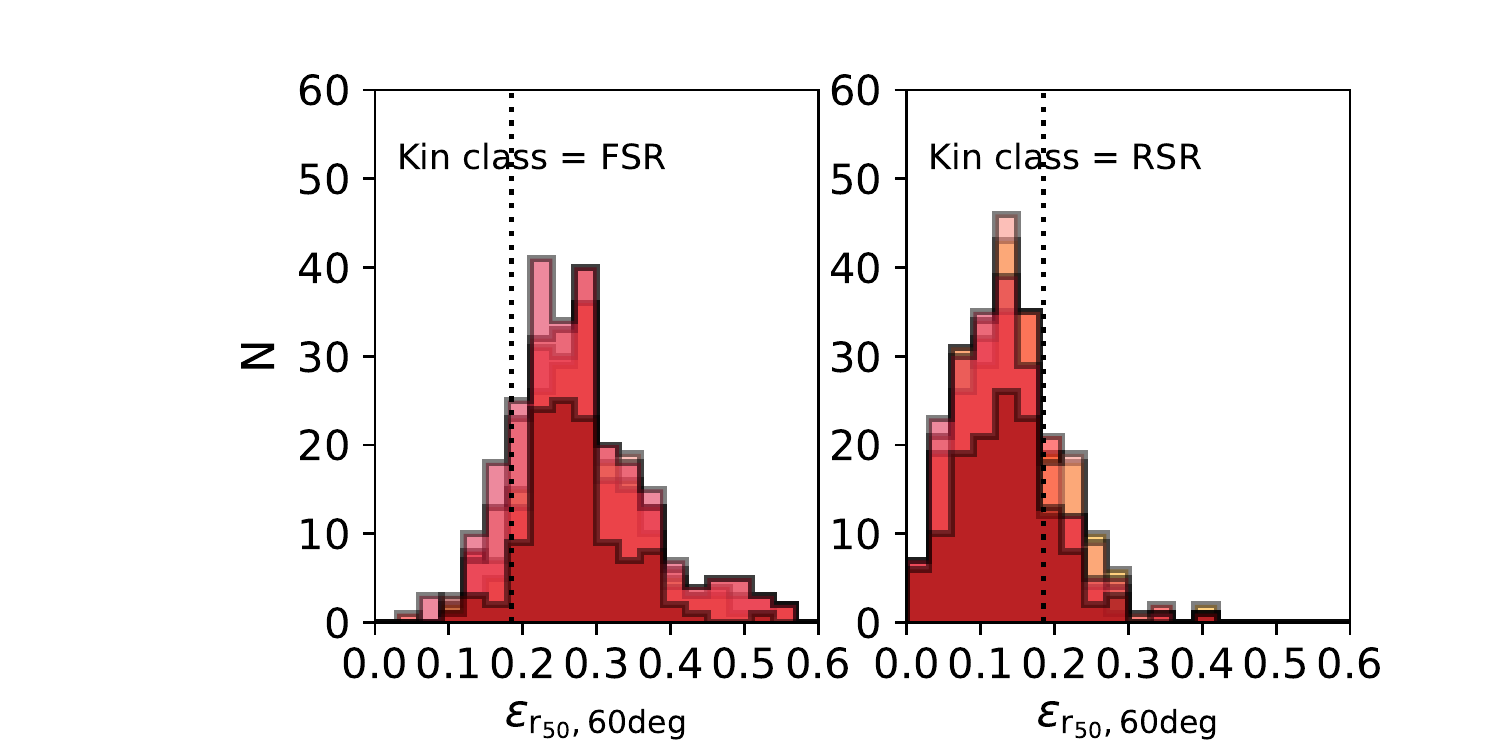}
                \caption{Distribution of ellipticities for flat (left) and round (right) SRs. Each coloured histogram shows a different classifier. Ellipticities here are measured directly from the {\sc SimSpin} maps, which adopted an inclination of $60$~degrees. The vertical line shows $\epsilon_{\rm r_{50},60deg}=0.2$, which we consider a reasonable threshold to separate flat and round SRs.} 
                \label{EllipClass}
        \end{center}
\end{figure}
\begin{figure*}
\begin{center}
\includegraphics[trim=19mm 4.5mm 43mm 11mm, clip,width=0.99\textwidth]{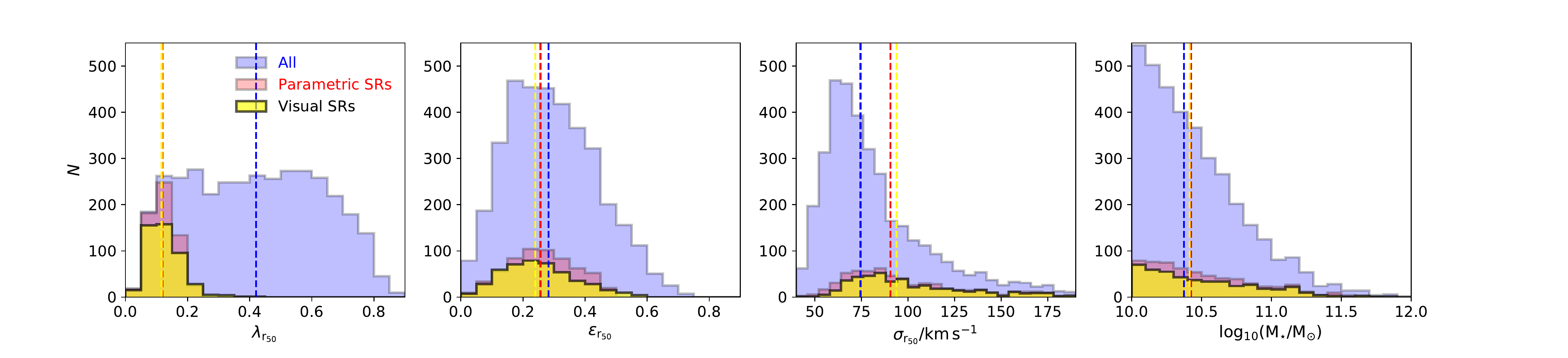}
\caption{Distribution of $\lambda_{\rm r_{50}}$,  $\epsilon_{\rm r_{50}}$, stellar velocity dispersion, and stellar mass of $z=0$ galaxies in \eagle\ with $M_{\star}\ge 10^{10}\,\rm M_{\odot}$.
The distributions are shown for all galaxies, SRs classified following the parametric selection of \citet{vandeSande20}, and the visually classified SRs (with a confidence $\ge 60$\%), as labelled. Vertical lines show the medians of each distribution.}
\label{CompWithSAMI}
\end{center}
\end{figure*}
The left panel of Fig.~\ref{KinematicMatching} shows the distribution of matching success among classifiers. Most galaxies can be kinematically classified with an agreement $\ge 60$\% (3 out of 5 classifiers agree on the class). By adopting this threshold, we are left with $501$ of the initially $559$ classified galaxies (i.e. $\approx 90$\% of the sample). 
The right panel of Fig.~\ref{KinematicMatching} shows the distribution of kinematic classes in the different levels of agreement of each independent classifier. For the cases in which 2/5 agree, we find that the conflict arises in whether galaxies are rotators/unclear or not. $2\sigma$ galaxies are also hard to classify, with most of them being in the matching success panels of $40$\% and $60$\%. We note that \eagle\ produces galaxies of diverse kinematic classes, which are also seen in observations \citep{Emsellem11}. \citet{Schulze18} via visual classification of the kinematic maps of early-type galaxies in the Magneticum simulations also found similarly diverse kinematic classes.
Table~\ref{kinclassesnum} presents the number of galaxies classified in each kinematic class with a confidence level $\ge 60$\%.

\begin{table}
    \centering
        \caption{Number of galaxies at $z=0$ visually classified with a confidence $\ge 60$\% in each kinematic class.}
    \begin{tabular}{c|c}
    \hline
        Sample & N\\
        \hline
        All (confidence $\ge 60$\%) &  501\\
        Flat SR &  238\\
        Round SR & 192\\
        Prolate & 49\\
        $2\sigma$ & 12\\
        Unclear & 9\\
        Rotator & 1\\
        \hline
    \end{tabular}
    \label{kinclassesnum}
\end{table}

The classification between FSR and RSR so far adopted can be subjective. In order to determine whether there is an obvious ellipticity threshold distinguishing between the two sub-classes, we turn to the ellipticity distribution of the visual classes, FSRs and RSRs. This is shown in Fig.~\ref{EllipClass} for each classifier. Overall, a threshold of $\epsilon_{\rm r_{50},60deg}=0.2$ appears appropriate for all classifiers. From hereon, we use this threshold to classify galaxies between FSR and RSR.

From these findings, we will consider as SRs in \eagle\ all galaxies visually classified as FSR, RSR and prolate. Unless otherwise specified, we only consider SRs in which there is $\ge 60$\% agreement among classifiers and refer to this sample as unambiguous SRs. This threshold was chosen to be similar to that adopted in \citet{vandeSande20}. Table~\ref{mergersSRs} presents the breakdown in the incidence of different types of mergers in the unambiguous SRs and the breakdown between centrals and satellites. Because some of the subsamples are rather small, we tend to sub-divide them in ways that we always have $\ge 10$ galaxies to measure medians from. 

\begin{table}
    \centering
        \caption{Number of galaxies in our unambiguous SR sample at $z=0$ that went through $\ge1$ major mergers in the last $10$~Gyrs; through $\ge 1$ minor mergers and $=0$ major mergers (in the same time period); through $\ge 1$ very minor mergers and $=0$ minor/major mergers (in the same time period); have not had any mergers in the last $10$~Gyrs; have had $\ge 1$ dry mergers; have had no dry mergers, but $\ge 1$ wet mergers. We also show the breakdown between centrals and satellites in each group.}
    \begin{tabular}{c|c|c|c}
    \hline
        Sample & All & Cens & Sats\\
        \hline
        Visual SRs (confidence $\ge 60$\%) &  479 & 293 & 186\\
        major mergers &  225 & 149 & 76\\
        minor mergers (and no major mergers)& 145 & 87 & 58\\
        very minor mergers (and no minor/major mergers) & 72 & 41 & 31\\
        w/o mergers & 37 & 16 & 21\\
        dry mergers & 178 & 113 & 65\\
        wet mergers (no dry mergers) & 264 & 164 & 100\\
        \hline
    \end{tabular}
    \label{mergersSRs}
\end{table}

\subsection{Parametric vs. visually classified slow rotators in {\sc EAGLE}}\label{CompWithSAMISec}

As discussed in the introduction, most simulation-based papers have adopted a parametric selection of SRs to analyse their formation history. Thus, it is important to understand how different our visual classification of SRs is from parametric selections.

Fig.~\ref{CompWithSAMI} shows the distribution of $\lambda_{\rm r_{50}}$, $\epsilon_{\rm r_{50}}$, $\sigma_{\rm r_{50}}$ and stellar mass of all galaxies in \eagle\ at $z=0$ with $M_{\star}\ge 10^{10}\,\rm M_{\odot}$, and the subsamples of galaxies selected as SRs based on the parametric classification of \citet{vandeSande20} and the visually identified SRs. The parametric classification of \citet{vandeSande20} is as follows:

\begin{equation}
    \lambda_{\rm r_{50}} < 0.12 + 0.25\, \epsilon_{\rm r_{50}}, \, {\rm for}\, \epsilon_{\rm r_{50}}\le 0.5.
\end{equation}

\noindent Most visually classified SRs fall within the classification of \citet{vandeSande20} with a small fraction ($\approx 12$\%) falling in the region of low $\epsilon_{\rm r_{50}}$ and elevated $\lambda_{\rm r_{50}}$. We visually inspect the galaxies that are in the unambiguous SR sample and have $\lambda_{\rm r_{\rm 50}} > 0.2$ ($\approx 6$\% of the sample). We find these are a mix bag of $2\sigma$ galaxies, galaxies that have some rotation in the outskirts but none in the central parts, and galaxies that have contamination from substructure but not enough as to fall in the ``unclear'' category, so they can still be easily identified as SR. Something in common among these galaxies is that they have $0.1<\lambda_{\rm r_{\rm 50},edge-on}<0.2$, so by the original criterion of \cite{Emsellem07} they would not be considered SRs.
There is another, even smaller fraction ($\approx 2.7$\%) of SRs in the unambiguous SR sample with $\epsilon_{\rm r_{50}}>0.5$. The success rate of the \citet{vandeSande20} classification in \eagle\ is $\approx 85$\%, which is similar to the success rate obtained by the authors using a visually classified sample of SAMI galaxies ($\approx 90$\%). The downside is that this parametric selection has a high contamination rate, selecting 295 galaxies that are not SRs (40 of those have a lower visual classification confidence, $<60$\%, and 205 have $\lambda_{\rm r_{50},edge-on}>0.2$). Even in the best case scenario (in which we drop our confidence threshold down to $40$\%), the purity of the selection (fraction of unambiguous SRs) would be $65$\% in \eagle. 

In general, we find that the visually classified SRs prefer $\epsilon_{\rm r_{50}}\lesssim 0.5$, with most of them having $0.1\lesssim \epsilon_{\rm r_{50}}\lesssim 0.5$. Note that these values of $\epsilon_{\rm r_{50}}$ cover a wider range than the SR selection criterion of \citet{Cappellari16}, who imposes a threshold $\epsilon_{\rm r_{50}}\le 0.4$ for a galaxy to be considered a SR. Thus, the criterion of \citet{vandeSande20} works better in \eagle, albeit with a high contamination.
From the first and third panels of Fig.~\ref{CompWithSAMI}, it is clear that visually classified SRs tend to populate the lower $\lambda_{\rm r_{50}}$ and higher $\sigma_{\rm r_{50}}$ regions of the parametric SRs distributions. There is also a small tendency of the visual SRs to have lower $\epsilon_{\rm r_{50}}$ and higher stellar masses than the parametric SRs. In addition to the properties in Fig.~\ref{CompWithSAMI}, we investigated several other galaxy properties and found
that the specific SFR (sSFR) and $r_{\rm 50}$ were on average $23$\% lower and $10$\% larger, respectively, in the visual SRs compared to the parametric ones. Visual SRs also have a higher incidence of galaxy mergers, with the mean number of mergers in this sample being $\approx 4.3$ compared to $3.7$ in the parametric SRs.
All the evidence above shows the importance of the visual classification of the kinematic maps we present in this paper required to isolate a sample of unambiguous SRs in the simulation, from which we can study their kinematic transformation.

An interesting result in Fig.~\ref{CompWithSAMI} regarding the entire galaxy population in \eagle, is that the $\lambda_{\rm r_{50}}$ distribution shows signs of a bimodality, with peaks at $\approx 0.2$ and $\approx 0.6$. \citet{vandeSande20} present a detailed quantification of the existence of a bimodality in $\lambda_{\rm r_{50}}$ at fixed stellar mass, and conclude that even though this bimodality is clear in SAMI (see also \citealt{Graham18} for a similar analysis in MaNGA), it appears less clear in \eagle. For massive galaxies, \citet{vandeSande20} showed that, although two beta functions were required for a good fit, the one peaking at low $\lambda_{\rm r_{\rm 50}}$ in \eagle\ had a prominent tail towards high values of $\lambda_{\rm r_{\rm 50}}$. One important difference with the analysis of \citet{vandeSande20} is that here we include all \eagle\ galaxies with $M_{\star}\ge 10^{10}\,\rm M_{\odot}$, while \citet{vandeSande20} analysed a subsample of the simulation selected to have the same stellar mass distribution as the SAMI survey, which ends up biased towards high masses {(with a peak at $10^{10.3}\rm\, M_{\odot}$)}. This possibly means that the bimodality in observations may be stronger than reported, in which case a volume complete sample would be needed to confirm that. 
Another important result from Fig.~\ref{CompWithSAMI} is that the sample of visually classified SRs in \eagle\ is only a fraction of the galaxies that would be associated with the low $\lambda_{\rm r_{50}}$ population. This population of SRs is not distinct enough to be cleanly separated by statistical means, lending support to our approach of visually classifying galaxies to study the formation mechanisms of SRs in \eagle.

\begin{figure}
\begin{center}
\includegraphics[trim=0mm 6mm 0mm 0mm, clip,width=0.49\textwidth]{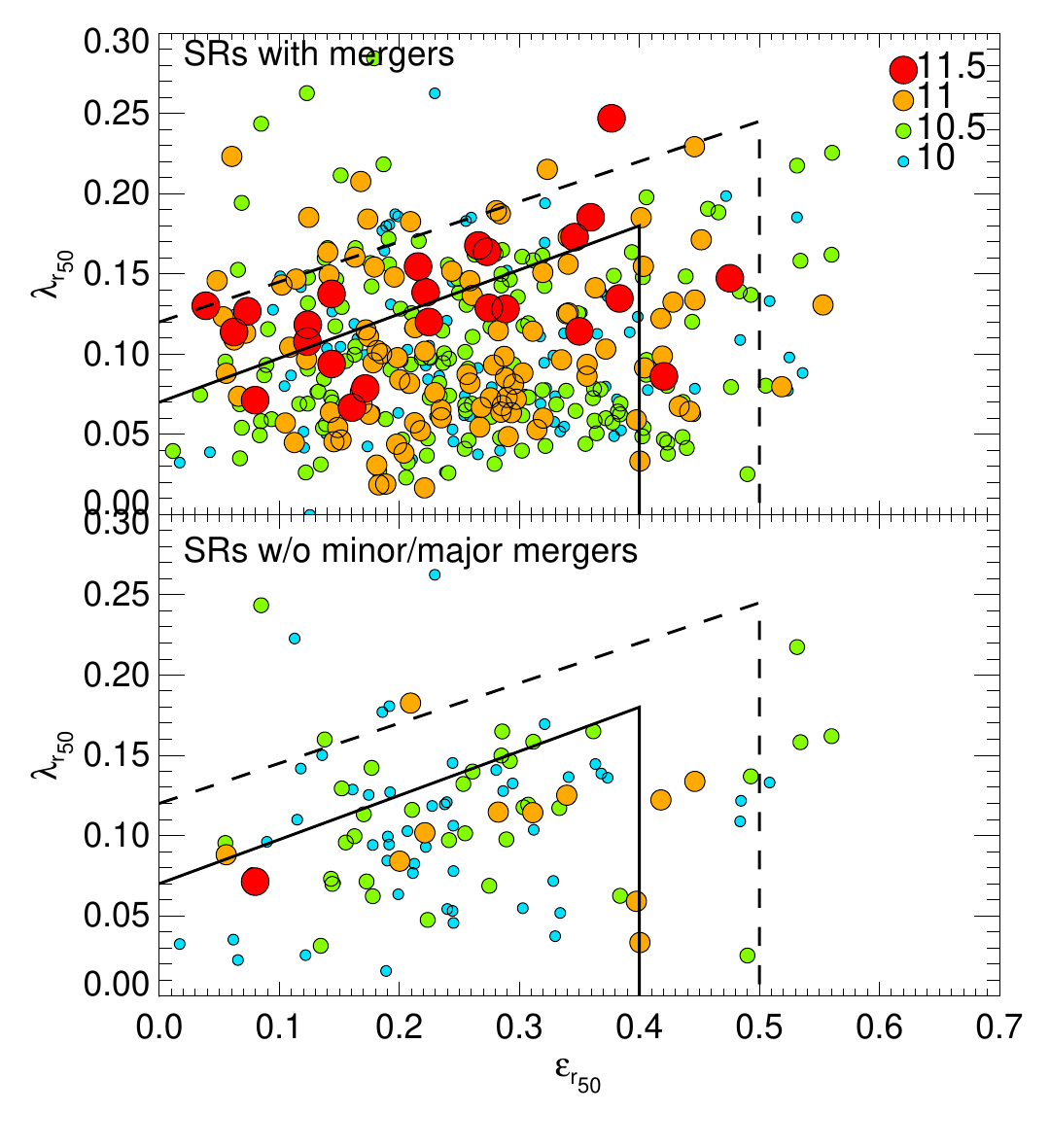}
\caption{$\lambda_{\rm r_{50}}$ as a function of $\epsilon_{\rm r_{50}}$ at $z=0$ for SRs in 
\eagle. The top panel shows 
SRs that have had mergers with mass ratios $\ge 0.1$ in the last $10$~Gyr, while 
the bottom panel shows the complement SRs. Sizes and colours of the symbols correspond to different stellar masses, as labelled in the top panel. For reference we show as solid and dashed lines the parametric classifications of \citet{Cappellari16} and \citet{vandeSande20}, respectively.}
\label{LambdaREllipSRs}
\end{center}
\end{figure}

\begin{figure}
\begin{center}
\includegraphics[trim=0mm 6mm 0mm 0mm, clip,width=0.49\textwidth]{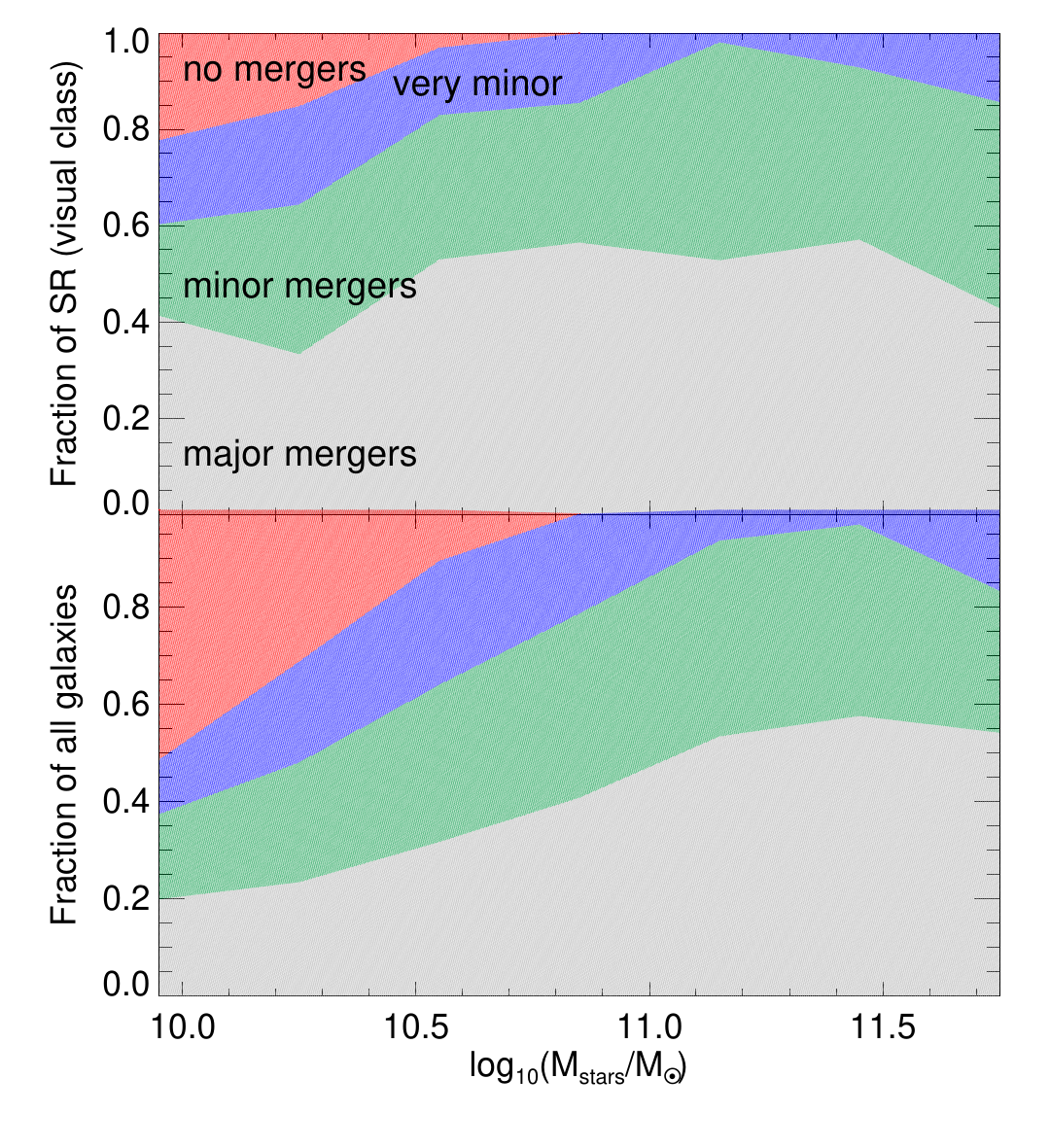}
\caption{The top panel shows the fraction of SRs that had $\ge 1$ major mergers (grey shaded region), no major mergers but $\ge 1$ minor mergers  (green), no minor or major mergers, but $\ge 1$ very minor mergers (blue) and no mergers (red) over the last $10$~Gyr, as a function of stellar mass. 
The bottom panel shows the same but for all galaxies regardless of their kinematic class.}
\label{LambdaREllipSRs2}
\end{center}
\end{figure}

\section{The properties of slow rotators in EAGLE}\label{SRsLocally}

In this section we analyse various properties of \eagle\ galaxies selected as SRs based on the visual classification presented in $\S~\ref{secclassification}$ and that have a classification confidence $\ge 60$\%. 

We study the distribution of SRs in the $\lambda_{\rm r_{50}}$-$\epsilon_{\rm r_{50}}$ in Fig.~\ref{LambdaREllipSRs}. We separate SRs that had $\ge 1$ minor or major mergers in the last $10$~Gyrs, from those that did not. We show for reference the parametric SR classifications of \citet{Cappellari16} and \citet{vandeSande20}. 
Fig.~\ref{LambdaREllipSRs} shows that the most massive SRs have had $\ge 1$ minor or major mergers in the last $10$~Gyrs, while in the subset of SRs without mergers or exclusively very minor mergers, we preferentially find lower-mass galaxies. This is quantified in the top panel of Fig.~\ref{LambdaREllipSRs2}, where we show the contribution of the four subsets of SRs selected based on their merger history as a function of stellar mass. The subset of ``no mergers'' is only present at $10^{10}\,\rm M_{\odot}<M_{\star}<10^{10.5}\,\rm M_{\odot}$, while most galaxies in the ``very minor mergers'' subset are preferentially in the $10^{10}\,\rm M_{\odot}<M_{\star}<10^{11}\,\rm M_{\odot}$ range. The fraction of SRs that have not experienced mergers is much smaller than the ``no mergers'' fraction of the entire galaxy population at fixed stellar mass (see bottom panel of Fig.~\ref{LambdaREllipSRs2}). On the other hand, about $40-50$\% of SRs had $\ge 1$ major mergers in the last $10$~Gyrs, even at relative low stellar masses ($10^{10}\,\rm M_{\odot}<M_{\star}<10^{10.5}\,\rm M_{\odot}$), which is twice the incidence of major mergers seen in the overall galaxy population at the same stellar mass. Minor and very minor mergers are represented in similar fractions in the SRs and all galaxies samples at fixed stellar mass.

The significantly lower fraction of ``no mergers'' and higher fraction of major mergers among SRs shows the importance of the latter in producing SRs in \eagle. In the coming sections we analyse intrinsic properties of SRs selected by their merger history to understand whether there are observable properties that are expected to be systematically different among these SRs.

\subsection{Intrinsic properties of slow rotators}
 
We focus on intrinsic properties of SRs that have attracted interest in the literature, including: intrinsic shape, velocity anisotropy and sizes. In addition, as we are interested in quenching and galaxy mergers in SRs, we also explore their BH masses and stellar ex-situ fraction, $f_{\rm ex-situ}$.

\begin{figure}
	\begin{center}
		\includegraphics[trim=4mm 5mm 4mm 1mm, clip,width=0.315\textwidth]{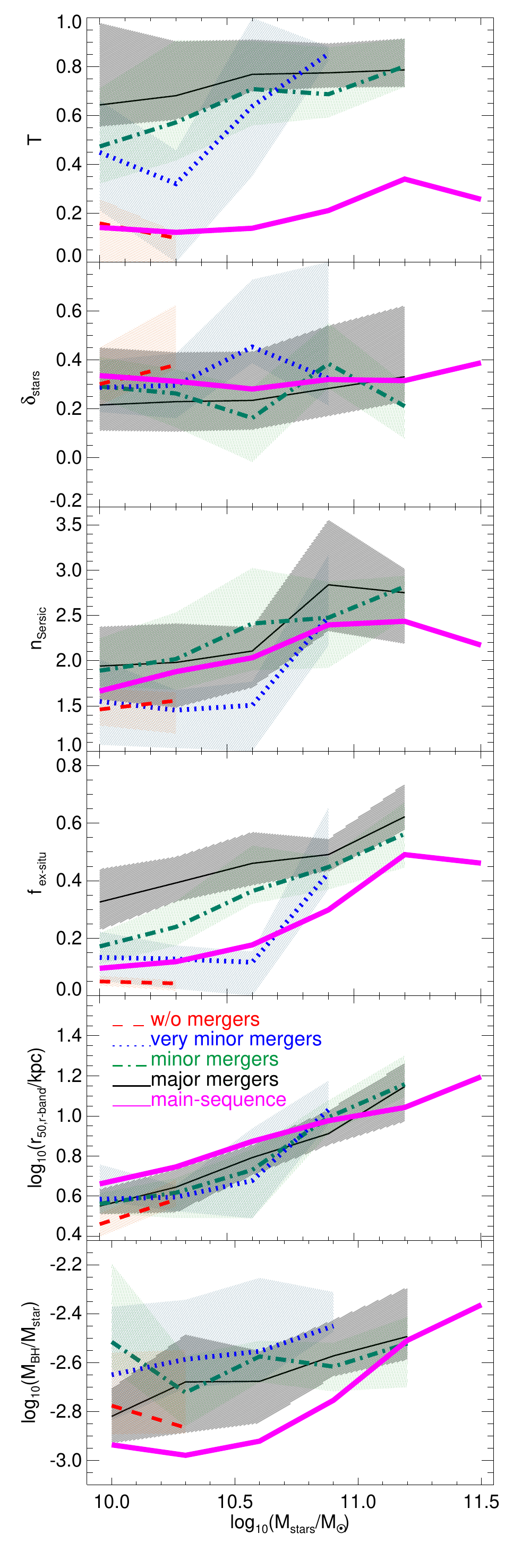}
		\includegraphics[trim=9mm 5mm 0mm 1mm, clip,width=0.1395\textwidth]{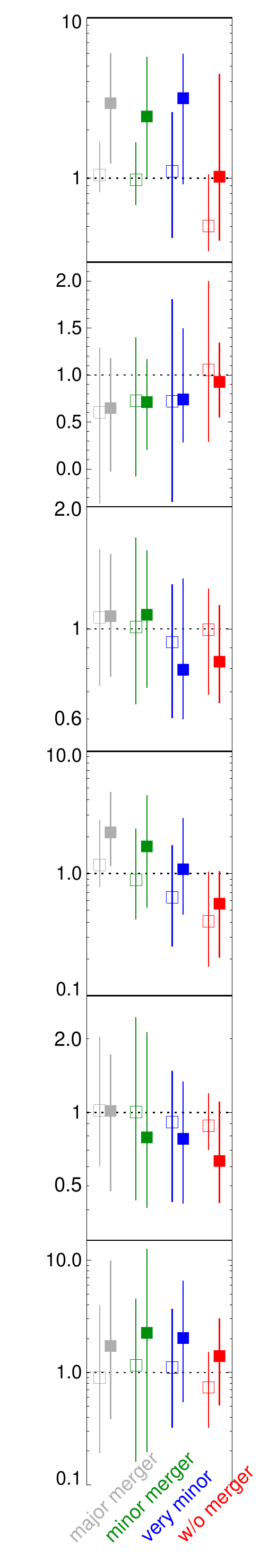}
		\caption{{\it Left panels:} Triaxiality (top), anisotropy stellar velocity dispersion (second), 3D S\'ersic index (third), ex-situ stellar fraction (fourth),  half-light radius (fifth) and BH-to-stellar mass ratio (bottom) as a function of stellar mass 
		for SRs at $z=0$ in \eagle. We show separately SRs that have had $\ge 1$ major mergers (solid lines), $= 0$ major but $\ge 1$ minor mergers (dot-dashed lines), $= 0$ major/minor mergers but $\ge 1$ very minor mergers (dotted lines), and $= 0$ mergers (dashed lines) in the last $10$~Gyrs. Lines with shaded regions show the median and $25^{\rm th}-75^{\rm th}$ 
		percentile range, respectively, and we only show bins with $\ge 10$ galaxies. For reference, the thick, magenta line shows the median relation for main sequence galaxies in \eagle\ (those with a $\rm sSFR>0.01\, Gyr^{-1}$). Parameters are calculated considering all their 
		stellar particles within the inner $30$~pkpc. {\it Right panels:} median and $25^{\rm th}-75^{\rm th}$ percentile range of the ratio between the properties in the left panel for the $4$ different SRs subsamples of the left panels selected based on their merger history, and two stellar-mass matched samples of fast rotators (filled squares) and SRs (empty squares). The horizontal dotted line marks equality.}
		\label{ShapeSRs}
	\end{center}
\end{figure}

The left panels of Fig.~\ref{ShapeSRs} show the triaxiality, stellar velocity anisotropy, $f_{\rm ex-situ}$, r-band half-light radius and BH-to-stellar mass ratio, as a function of stellar mass of SRs at $z=0$ in \eagle\ classified based on their merger history. The first two quantities above come from the \eagle\ analysis of \citet{Thob19}, which we briefly describe here. For each galaxy, all stellar particles within a spherical aperture of radius $30$~pkpc are used to measure the tensor of the quadrupole moments of the mass distribution (which share eigenvectors with the inertia tensor). The axes lengths $a$ (major axis), $b$ (intermediate axis) and $c$ (minor axis) are defined by the square root of the eigenvalues of the mass distribution tensor, $\lambda_{\rm i}$ (for $i=0,\, 1,\, 2$). These axes are then used to measure a first pass for the ellipticity ($\epsilon=1-c/a$) and triaxiality ($T=(a^2-b^2)/(a^2-c^2)$). These values are then used to select stellar particles that are enclosed in the ellipsoid of axes ratios $a/b$, $a/c$ of equal volume as the sphere of $r=30$~pkpc. These particles are used to remeasure the ellipsoid axes. This iterative process continues until changes in $a$, $b$, $c$ are $<1$\%. A perfect spherical galaxy has $\epsilon=0$ and $T$ is undefined. Low and high values of $T$ correspond to oblate and prolate ellipsoids, respectively. The stellar velocity anisotropy, $\delta_{\rm stars}$, depends on the velocity dispersion parallel, $\sigma_{\parallel}$, and perpendicular, $\sigma_{\bot}$, to the stellar angular momentum vector of the galaxy (all measured with stellar particles at $r<30$~pkpc from the centre of potential), $\delta_{\rm stars}=1-(\sigma_{\bot}/\sigma_{\parallel})^2$. If $\delta_{\rm stars}>0$, then the stellar velocity dispersion is dominated by disordered motions in the disk plane.

We also make use of the stellar ex-situ fractions, $f_{\rm ex-situ}$ computed by \citet{Davison20} for \eagle\ galaxies at $z=0$. Here, $f_{\rm ex-situ}$ refers to the fraction of stars that did not form in the main progenitor branch of the $z=0$ galaxy, and hence was acquired from galaxies that merged onto the main progenitor in the past (or were acquired after close interactions). This is computed considering all stellar particles within $30$~kpc from the galaxy's centre. 
For reference, the left panels of Fig.~\ref{ShapeSRs} also show the median of these quantities as a function of stellar mass for galaxies that are considered to be main sequence (those with a $\rm sSFR> 0.01\,\rm Gyr^{-1}$; \citealt{Furlong14}). 

The samples of SRs split by their merger history can quickly become very small and hence the correlation with stellar mass can be noisy. To try to identify main trends, we also show in the right panels of Fig.~\ref{ShapeSRs} the median ratio between the quantity in the left panel for the subsample of SRs and for a control sample of fast rotators and SRs matched to have the same stellar mass distribution\footnote{If our sample of interest is A and we want to draw a subsample from B to have the same stellar mass distribution of A, we randomly choose $N$ galaxies in narrow stellar mass bins from B, where $N$ is the number of galaxies of that stellar mass in A. In our case A are the subsamples of SRs split by their merger history, and B are either all fast rotators or SRs in \eagle.}.

We find that SRs without mergers are very oblate ($T\lesssim 0.2$) compared to other SRs, and in fact similar to what we expect for main sequence galaxies and fast rotators of the same stellar mass. Even though these SRs are very compact, $r_{\rm 50}\approx 2-3$~pkpc, they are still above the resolution limit by a factor of $\approx 3-4$, and given their mass, we expect them to be resolved with $\gtrsim 3000$ particles, so we consider these measurements reliable.

There is a tendency for T to increase going from SRs that went exclusively through very minor mergers to those that went through major mergers at fixed stellar mass. Most of the prolate SRs ($T\gtrsim 0.7$) correspond to galaxies that went through major mergers, while very minor and minor mergers are preferentially associated with triaxial systems ($0.3\lesssim T\lesssim 0.7$), particularly at $10^{10}\,\rm M_{\odot}\lesssim M_{\star}<10^{10.7}\,\rm M_{\odot}$. Compared to other simulations we find some interesting differences. \citet{Pulsoni20} found that in the Illustris-TNG100 simulation there is a large fraction ($\approx 83$\%) of SRs that are triaxial ($0.3\lesssim T\lesssim 0.7$) at $r<1-2\rm r_{\rm 50}$, which we do not see in \eagle\ ($\approx 34$\% are triaxial at small radii). Most of the triaxial SRs in Illustris-TNG are in the stellar mass range $10^{10.5}-10^{11.5}\,\rm M_{\odot}$, while in \eagle, $50$\% ($80$\%) are $<10^{10.5}\,\rm M_{\odot}$ ($10^{10.9}\,\rm M_{\odot}$). The reasons for these differences are not easy to pinpoint but it is worth highlighting them for future research. 

Most SRs have $0.2\lesssim \delta_{\rm stars}\lesssim 0.6$, which is similar to the values reported for SRs in \citet{Schulze18} for the Magneticum simulations. Most of the galaxies with $\delta_{\rm stars}<0.2$ are main sequence galaxies (with $\rm sSFR\gtrsim 0.025\,\rm Gyr^{-1}$) and fast rotators ($0.2\lesssim \lambda_{\rm r_{50}}\lesssim 0.7$, where the limits correspond to the $25^{\rm th}-75^{\rm th}$ percentile range) also in agreement with the findings in \citet{Schulze18}.
We identify a weak trend of $\delta_{\rm stars}$ increasing when going from SRs that went through major mergers, minor and very minor mergers, to those that have not had mergers, at fixed stellar mass. The medians in the right panel show this trend more clearly. Interestingly, most galaxies, even main sequence galaxies, show $\delta_{\rm stars}>0$, indicating $\sigma_{\rm \parallel}>\sigma_{\bot}$. \citet{Thob19} found that the most flattened systems are also the ones with the highest $\delta_{\rm stars}$ due to the fact that in a flat system you expect little vertical stellar velocity dispersion, which leads to a smaller scale-height. Major mergers therefore act to dynamically heat the galaxies making $\sigma_{\bot}$ approach $\sigma_{\parallel}$. 

{The third panels of Fig.~\ref{ShapeSRs} show the 3D S\`ersic index, $n_{\rm Sersic}$ (measured from the $3$-D stellar mass distributions). There is a trend between $n_{\rm Sersic}$ and the assembly history of a SR galaxy, whereby galaxies that have had major/minor mergers tend to have higher $n_{\rm Sersic}$ than those that had only very minor mergers or no mergers at all. Note that SRs with no mergers or very minor mergers have lower $n_{\rm Sersic}$ than even main sequence galaxies. \citet{Lagos18b} showed that in \eagle, galaxies that have had dry or wet mergers had a higher $n_{\rm Sersic}$ than galaxies without mergers. Here, we show that trends with merger history remain even when we select slow rotators only.} 

The fourth panels of Fig.~\ref{ShapeSRs} show that the ex-situ fraction strongly increases going from SRs without mergers, to SRs that have had $\ge 1$ major mergers, at fixed stellar mass. The subsample of SRs without mergers has an even smaller $f_{\rm ex-situ}$ than main sequence galaxies of the same stellar mass, while the subsample of SRs with exclusively very minor mergers appears similar to main sequence galaxies. Interestingly, at $M_{\star}\gtrsim 10^{11}\,\rm M_{\odot}$, SRs that went through $N\ge1$ major merger have as much $f_{\rm ex-situ}$ as those that went only through minor mergers.

The fifth and bottom panels of Fig.~\ref{ShapeSRs} show a tendency for SRs without mergers to be more compact and have a lower BH-to-stellar mass ratio than the rest of the SRs at fixed stellar mass. The half-light radius increases from SRs that exclusively had very minor mergers to those that had major or minor mergers. The latter are also the ones with the highest BH-to-stellar mass ratio. These differences are suggestive of different quenching mechanisms between the subsamples. This is further discussed in the next section.

\begin{figure}
        \begin{center}
                \includegraphics[trim=3mm 7mm 0mm 0mm, clip,width=0.49\textwidth]{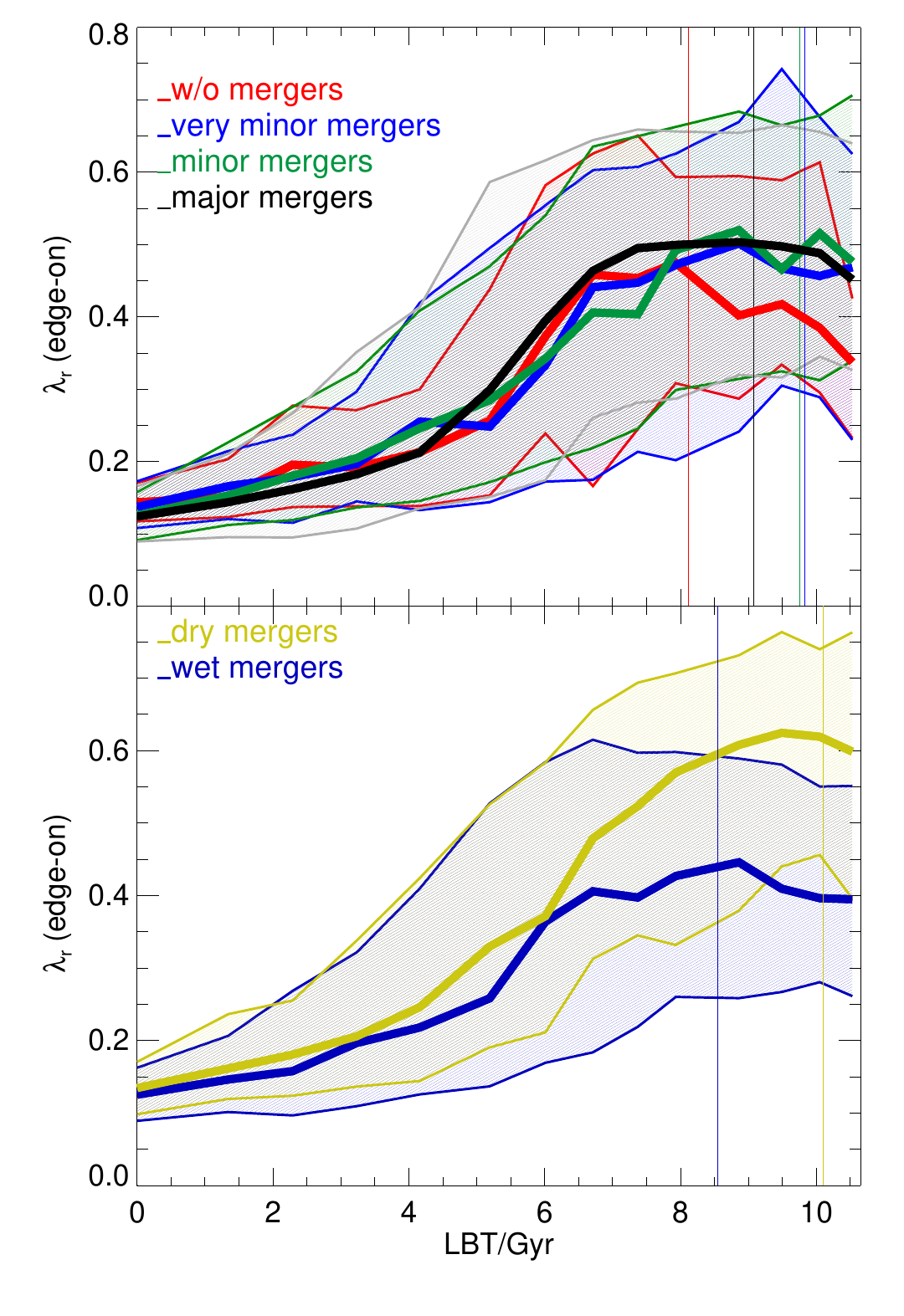}
                \caption{{\it Top panel:} The $\lambda_{\rm r_{50},edge-on}$ (measured orienting galaxies edge-on) history of $z=0$ SRs
                 that had $\ge 1$ major mergers, $= 0$ major but $\ge 1$ minor mergers,
                $= 0$ major/minor mergers but $\ge 1$ very minor mergers, and $= 0$ mergers
                in the last $10$~Gyrs, as labelled. 
                Lines with shaded regions show the median and $25^{\rm th}-75^{\rm th}$
                percentile range, respectively. Vertical lines show the median r-band weighted stellar age of the samples. The latter was computed with all stellar particles within $r_{50}$ at $z=0$. {\it Bottom panel:} as in the top panel but for SRs that had $\ge 1$ wet or dry merger (of any mass ratio) in the last $10$~Gyrs, as labelled.} 
                \label{SRsHist}
        \end{center}
\end{figure}

\begin{figure}
        \begin{center}
                \includegraphics[trim=3mm 7mm 0mm 0mm, clip,width=0.49\textwidth]{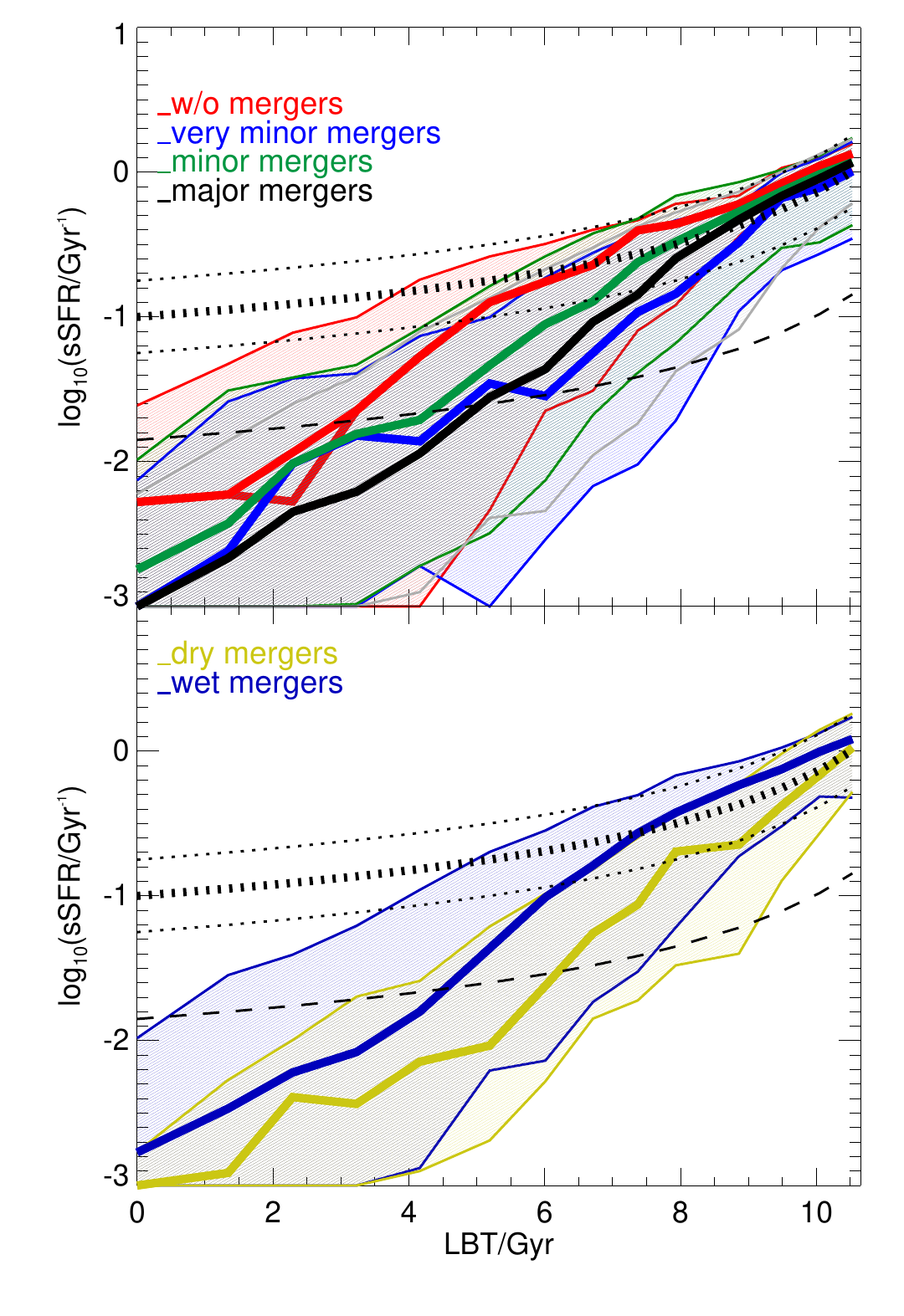}
                \caption{As in Fig.~\ref{SRsHist} but for the sSFR history. Here, the thick and thin dotted lines show the position of the main sequence in \eagle\ at $M_{\star}\approx 10^{10}\,\rm M_{\odot}$, and a scatter of $\pm 0.25$~dex, respectively, which is approximately the value measured by \citet{Furlong14} in \eagle. The dashed line shows a distance to the main sequence of $-0.85$~dex, which we use to define quenched galaxies.} 
                \label{SFsHist}
        \end{center}
\end{figure}

\begin{figure}
\begin{center}
\includegraphics[trim=3mm 2mm 3mm 0mm, clip,width=0.4\textwidth]{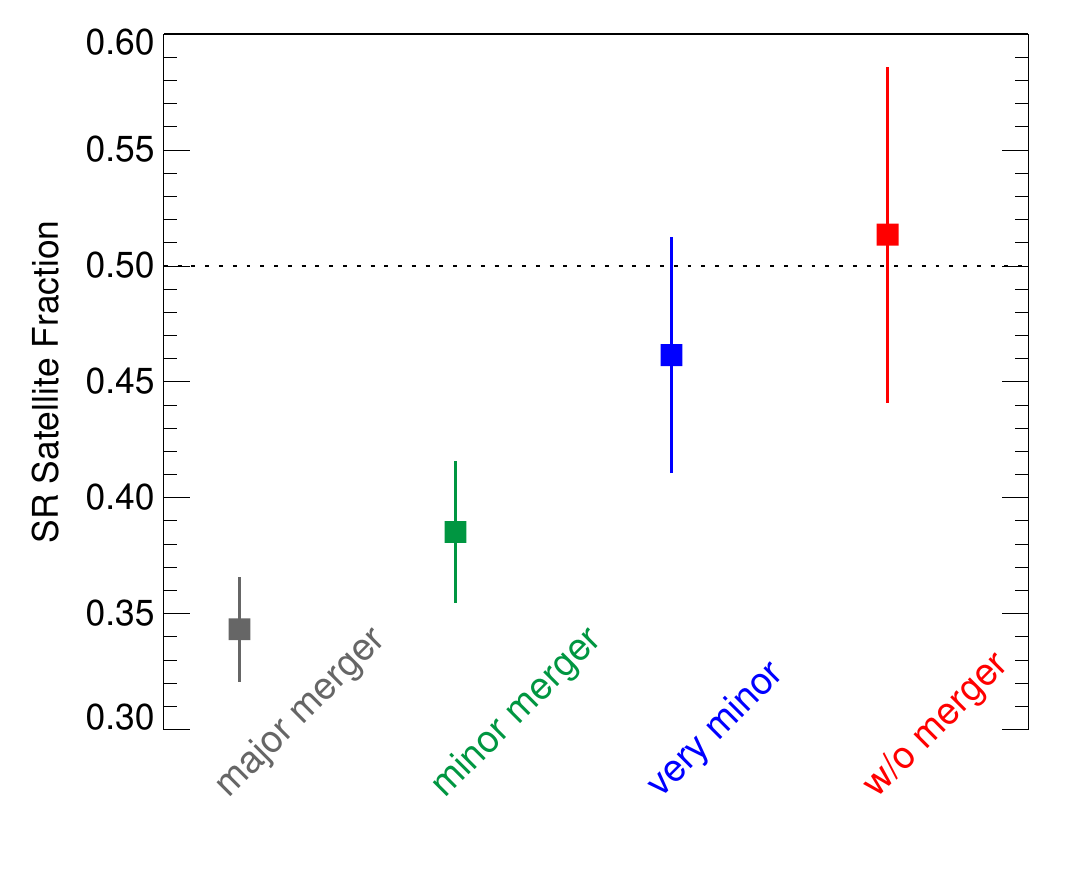}
\caption{The fraction of $z=0$ galaxies that are classed as satellites, for SRs that had $\ge 1$ major mergers (grey shaded region), no major mergers but $\ge 1$ minor mergers  (green), no minor or major mergers, but $\ge 1$ very minor mergers (blue) and no mergers (red) over the last $10$~Gyr (solid lines). Errorbars show Poisson errors. SRs that had exclusively very minor mergers or no mergers have a much higher probability of being a satellite galaxy than other SRs.} 
\label{SatelliteFractions}
\end{center}
\end{figure}

\subsection{Kinematic transformation and quenching of slow rotators}\label{kinandquenching}

The top panel of Fig.~\ref{SRsHist} shows the evolution of $\lambda_{\rm r_{\rm 50},edge-on}$ of $z=0$ SRs classified by their merger history. We find that they follow similar $\lambda_{\rm r_{50},edge-on}$ evolutionary tracks, in which most of the transformation happens in the last $6$~Gyrs, on average, and by $<2$~Gyr they are almost all completed. The main difference between SRs that went through different merger histories is when they formed their stars. SRs without mergers are the youngest ones, while those that went exclusively through very minor or minor mergers are the oldest, again suggesting different quenching mechanisms. The top panels of Fig.~\ref{SFsHist} shows this more clearly. The sSFR of SRs without mergers deviated from the main sequence at later times than other SRs, and by $z=0$ some of them continue to have low levels of star formation ($\approx 10-20$ times below the main sequence, on average). On the other hand, the rest of the SRs are much more quenched by $z=0$, with those having only very minor mergers being the first to deviate from the main sequence at $\approx 9$~Gyrs of lookback time. We find that in the sample of SRs without mergers, deviations from the main sequence are accompanied by changes in $\lambda_{\rm r_{50}}$, which start happening on average at a lookback time of $6$~Gyrs. This is not the case for the other SRs, where the kinematic transformation is disconnected from the star formation quenching.
Even though the medians of the evolutionary tracks of $\lambda_{\rm r_{50}}$ are smooth, by visual inspection of individual tracks, we find that most galaxies tend to display a sharp decrease in $\lambda_{\rm r_{50}}$ (suggestive of mergers). The timing of galaxy mergers average out to give a smooth average track but leading to a large scatter around the median at lookback times $\gtrsim 1.5-2$~Gyrs; the quick transformation of $\lambda_{\rm r_{50}}$ in individual galaxies is happening mostly throughout lookback times $\approx 2-6$~Gyr. This agrees with the low fraction of SRs found in observations at $z\approx 0.6$, which increases rapidly to $z=0$ \citep{Cole20}.

The bottom panels of Fig.~\ref{SRsHist} and \ref{SFsHist} focus on SRs that had $N\ge 1$ mergers (of any mass ratio), but we separate them between wet and dry mergers (based on whether $M_{\rm SFgas,total}/M_{\star,\rm total}$ is $>$ or $<0.1$, respectively; see $\S$~\ref{galmergerssec} for details). SRs that had $N\ge 1$ dry mergers are older and have progenitors with higher $\lambda_{\rm r_{50},edge-on}$ than the counterparts with $N\ge1$ wet mergers {and no dry mergers}. This happens because dry mergers are more effective at decreasing $\lambda_{\rm r_{50},edge-on}$, as shown by \citet{Lagos18b}. Although by $z=0$ both types of SRs have similarly low sSFRs, SRs that had $N\ge1$ dry merger started deviating from the main sequence earlier than those that had $N\ge1$ wet merger, explaining their older age. 

One key question is how these SRs quenched - or similarly, what led them to start deviating from the main sequence. Because of the high stellar masses of these galaxies ($\ge 10^{10}\,\rm M_{\odot}$), there are only two plausible pathways in which they could have quenched in \eagle: due to AGN feedback or environmental effects. The latter mostly happens in \eagle\ due to tidal interactions between galaxies, tidal stripping and ram pressure stripping \citep{Marasco16,Bahe17b}. We explore this by separating the merger-samples of SRs of Fig.~\ref{SFsHist} into centrals and satellites. Fig.~\ref{SatelliteFractions} shows the fraction of galaxies among SRs that had different merger histories that are satellites by $z=0$.  SRs that have not had mergers or have had exclusively very minor mergers have a clear preference for being satellite galaxies compared to other slow and fast rotators of the same stellar mass. 
Within SRs, those that have had major or minor mergers make the vast majority of central galaxies. 
These results already indicate environment has likely played an important role in quenching SRs mostly for the subsamples that had no mergers or exclusively very minor mergers. However, the excess in satellites does not uniquely point to environment as a source of quenching.

\begin{figure}
        \begin{center}
                \includegraphics[trim=2mm 1mm 0mm 0mm, clip,width=0.45\textwidth]{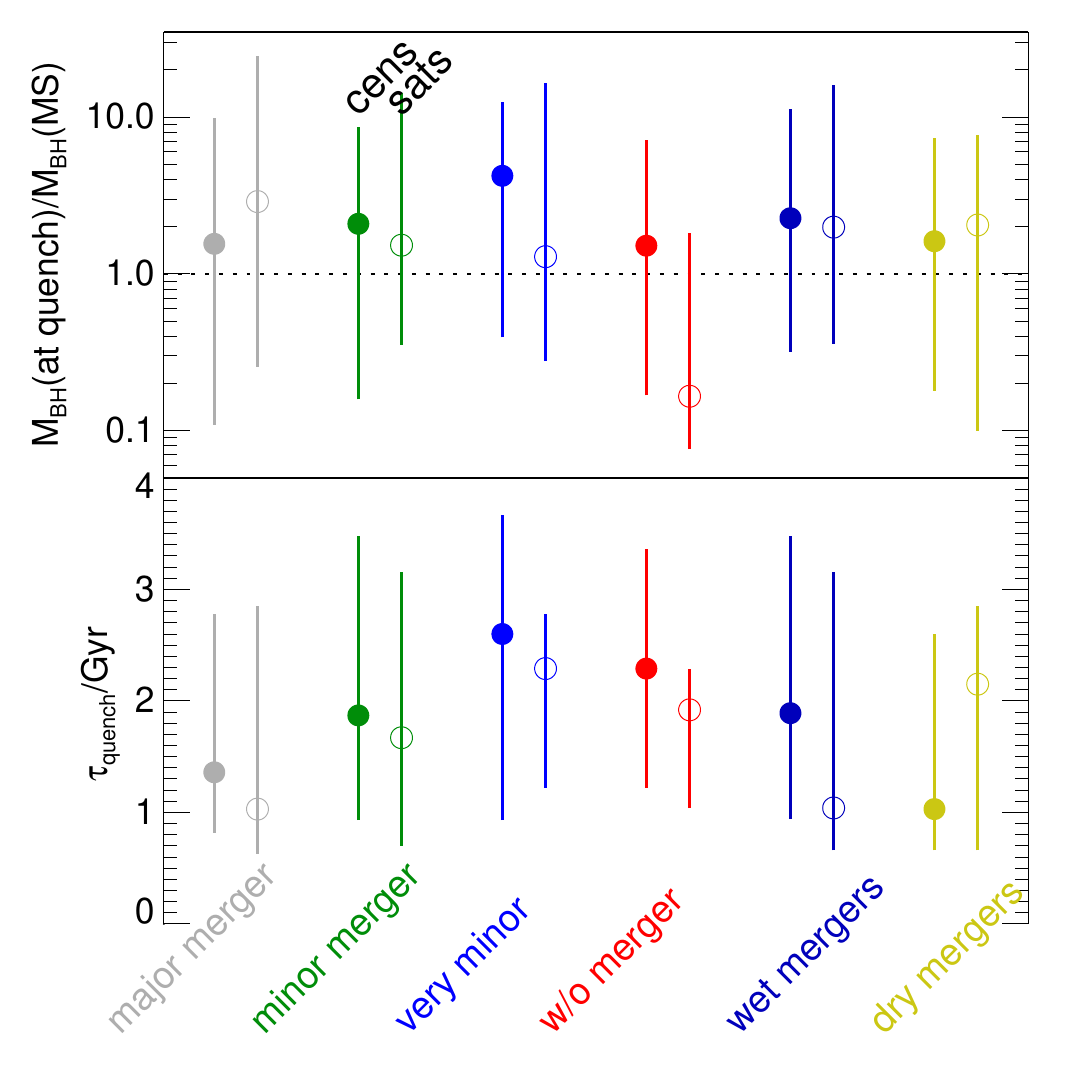}
                \caption{{\it Top panel:} The ratio between the BH mass of the SRs in Fig.~\ref{SFsHist} and the median BH mass of main sequence galaxies of the same stellar mass at the time the SRs leave the main sequence for the first time (lower dotted line in Fig.~\ref{SFsHist}). We show this for central (filled circles) and satellite (open circles) galaxies separately. Symbols with errorbars show the median and $25^{\rm th}-75^{\rm th}$ percentile range, respectively. Ratios $\ge 1$ correspond to galaxies that leave the main sequence with an overly massive BH compared to galaxies on the main sequence of the same stellar mass. {\it Bottom panel:} Quenching timescale of the SRs in the top panel, defined as the time galaxies take to transition from the lower dotted to the dashed lines in Fig.~\ref{SFsHist}.}
                \label{MBHAtQuench}
        \end{center}
\end{figure}

To isolate environment from AGN feedback as potential sources of quenching, we track back the time at which each SR departed the main sequence for the first time ($\tau_{\rm depart}$; which corresponds to the first time the sSFR of the SRs crossed the lower dotted line in Fig.~\ref{SFsHist}) and measure their central BH mass, $M_{\rm BH(at\, quench)}$. \citet{Bower17} showed that in \eagle, galaxies being quenched by AGN feedback are characterised by a strongly non-linear BH growth phase, which makes the relative BH-to-stellar mass ratio a good indicator of AGN feedback in action. We normalise $M_{\rm BH(at quench)}$ by the median central BH mass of main sequence galaxies of the same stellar mass of the SR's progenitor at $\tau_{\rm depart}$, $M_{\rm BH(MS)}$, and save the ratio, $M_{\rm BH(at\, quench)}/M_{\rm BH(MS)}$. The top panel of Fig.~\ref{MBHAtQuench} shows the median and $25^{\rm th}-75^{\rm th}$ percentiles of the distribution of $M_{\rm BH(at\, quench)}/M_{\rm BH(MS)}$ for $z=0$ SRs selected based on their merger histories. We show this separately for SRs that by $z=0$ are centrals and satellites. Overall we see a tendency for SRs to have overly massive BHs compared to main sequence galaxies of the same stellar mass at $\tau_{\rm depart}$. The only exceptions are SRs that have not had mergers and end up as satellite galaxies by $z=0$; this population has light black holes compared to main sequence galaxies at $\tau_{\rm depart}$. 

\citet{Trayford16} showed that excess BH mass is a strong indicator of colour transformation and quenching; \citet{Trayford16,Wright19} quantified that and showed that galaxies with overly massive BHs or high specific BH growth rates quench much more rapidly than those with lighter BHs (relative to their stellar mass), both in terms of colour transformation, as well as departures from the sSFR main sequence. 
Hence, the top panel of Fig.~\ref{MBHAtQuench} suggests that the vast majority of SRs quenched due to AGN feedback, with the exception of $z=0$ satellite SRs that have not had mergers. We also highlight that AGN are likely to be the source of quenching even for satellite SRs that have had major, minor, or very minor mergers. 
{We caution that this interpretation of overly sized BHs being an indicator of AGN feedback quenching applies to \eagle\ (see \citealt{Bower17}). However, this may not work in simulations implementing different models of AGN feedback that are not tight to rapid BH growth phases.}
Note that the trends of $M_{\rm BH(at quench)}/M_{\rm BH(MS)}$ at $\tau_{\rm depart}$ for these different SRs continues to hold at later times, as shown in the bottom panels of Fig.~\ref{ShapeSRs} for $z=0$. 
The top panel of Fig.~\ref{MBHAtQuench} also shows $M_{\rm BH(at quench)}/M_{\rm BH(MS)}$ at $\tau_{\rm depart}$ for $z=0$ SRs that had $\ge 1$ dry or wet merger in the last $10$~Gyrs. Overall these are similar to the overall major/minor merger SRs subsample.

The bottom panel of Fig.~\ref{MBHAtQuench} shows the quenching timescale of these SRs defined as the time it took to transition from the main sequence down to an arbitrary low level of star formation, $\tau_{\rm quench}$. In this paper we use the method of \citet{Wright19}, which consists of measuring the time it took for a galaxy to change its sSFR from $\rm MS(M_{\star})-c_{\rm high}$ to $\rm MS(M_{\star})-c_{\rm low}$, where $\rm MS(M_{\star})=log_{10}(sSFR_{\rm MS}(M_{\star})/Gyr^{-1})$ is the sSFR of the main sequence at $M_{\star}$. Here, we adopt $c_{\rm high}=0.25$ and $c_{\rm low}=0.85$ (lower thin dotted and dashed lines in Fig.~\ref{SFsHist}). Note that these values are slightly different to those adopted in \citet{Wright19}, but they give us the best statistics for $\tau_{\rm quench}$, as the value of $c_{\rm low}=1.3$~dex adopted in \citet{Wright19} leads to about half of the SRs in the ``no merger'' subsample to have undefined $\tau_{\rm quench}$. We note that typical values for $c_{\rm low}$ adopted in the literature range from $\approx 1.6$ to $0.3$~dex \citep{Bethermin14,Davies18,Wang18}.

Most SRs have $\tau_{\rm quench}\approx 1-2$~Gyr, in agreement with quenching timescales derived in observations \citep{Smethurst18}.
Note that satellite SRs in the ``no merger'' sample quench fast, $\tau_{\rm quench}\approx 1.9$~Gyr, despite them having overly light BHs. This is the only subsample in which this happens. If we focus on central galaxies that by $z=0$ are SRs and had galaxy mergers, we find that changes in $\lambda_{\rm r_{50},edge-on}$ happen after galaxies quench, on average (galaxies start deviating from the main sequence earlier than they start showing $\lambda_{\rm r_{50}, edge-on}\lesssim 0.5$). We quantify this by comparing the lookback time at which the SRs' progenitors leave the main sequence ($c_{\rm high}=0.25$) with the time at which progenitors have a $\lambda_{\rm r_{50},edge-on}$ that is $0.8$ times the maximum $\lambda_{\rm r_{50},edge-on}$ they had. We find that for SRs that have had mergers, quenching starts happening $\approx 1-2$~Gyrs before $\lambda_{\rm r_{50},edge-on}$ drops below $80$\% its maximum historical value, while for SRs that have not had mergers, this happens roughly at the same time (within $0.2$~Gyrs).
The emerging picture is that AGN feedback quenches these centrals galaxies at $z\gtrsim 1$, and subsequent mergers are responsible for the kinematic transformation leading them to become SRs.

\begin{figure}
\begin{center}
\includegraphics[trim=4mm 2mm 3mm 0mm, clip,width=0.235\textwidth]{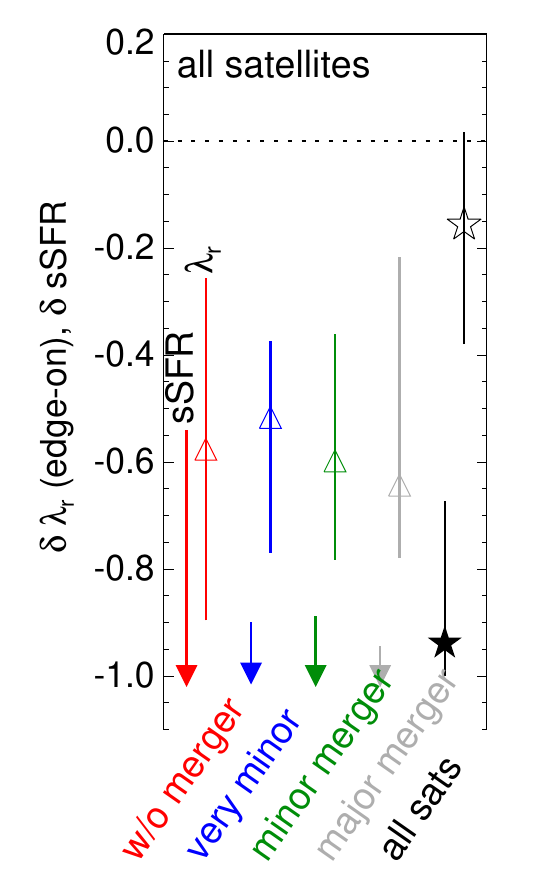}
\includegraphics[trim=4mm 2mm 3mm 0mm, clip,width=0.235\textwidth]{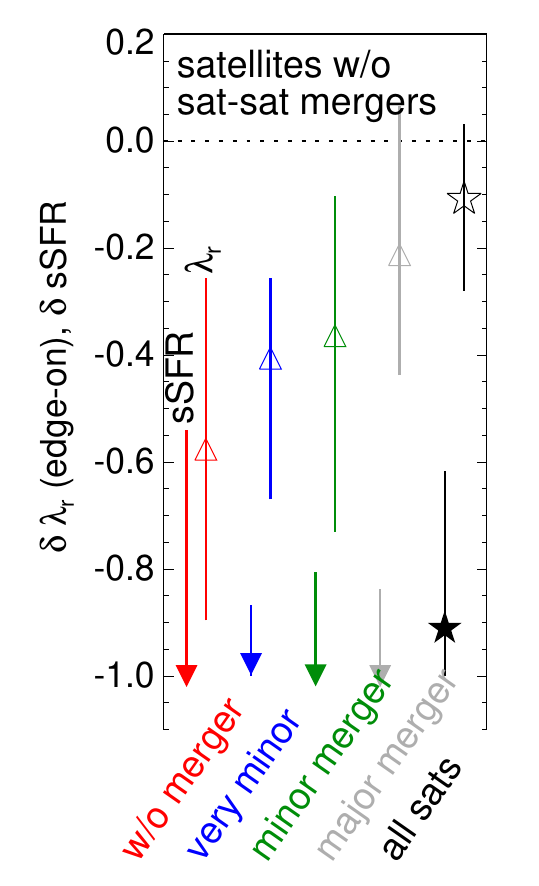}
\caption{{\it Left panel:} Relative change in $\lambda_{\rm r_{50}}\,\rm (edge-on)$ (up-pointing triangles) and sSFR (down-pointing triangles) for the subsample of satellite $z=0$ SRs of Fig.~\ref{SatelliteFractions}, between $z=0$ and the time they were last central. Symbols with errorbars show the median and $25^{\rm th}-75^{\rm th}$ percentile range. We also show for reference the relative change in $\lambda_{\rm r_{50}}\,\rm (edge-on)$ and sSFR for all $z=0$ satellite galaxies in \eagle\ as black stars. {\it Right panel:} As in the left panel but only for satellites that have not had a galaxy merger since becoming a satellite (i.e. no satellite-satellite mergers).} 
\label{SatelliteLambdaRChange}
\end{center}
\end{figure}

To address the effect of environment in quenching and potential transformation of the kinematics of satellite galaxies, we show in the left panel of Fig.~\ref{SatelliteLambdaRChange} the relative change of $\lambda_{\rm r_{50},edge-on}$, $\delta \lambda_{\rm r_{50},edge-on}$, and sSFR, $\delta \rm \, sSFR$, for satellite SRs in our four samples split by their merger history. This relative change is computed between $z=0$ and the last time the galaxy was a central, 

\begin{equation}
\delta i=\frac{i (z=0) - i (\rm last\, central)}{i (\rm last\, central)},
\end{equation}

\noindent with $i=\lambda_{\rm r_{50},edge-on}$ or $\rm sSFR$. By definition, $\delta i\ge -1$, with a value $\delta i=-1$ indicating the quantity of interest at $z=0$ is $=0$ (which is often the case for sSFR). {The lookback time to when satellite SRs were last central has a median of $\approx 6$~Gyr, and a $16^{\rm th}$ and $84^{\rm th}$ percentiles of $\approx 2.5$~Gyr and $\approx 8.6$~Gyr, respectively (most of them became satellites at $z<1$, as expected).} The right panel of Fig.~\ref{SatelliteLambdaRChange} also shows $\delta \lambda_{\rm r_{50},edge-on}$ and $\delta \rm \, sSFR$ for the whole population of $z=0$ satellite galaxies with $M_{\star}\ge 10^{10}\,\rm M_{\odot}$. We find that in all cases, satellites that by $z=0$ are visually classified as SRs suffered significant quenching and kinematic transformation since becoming a satellite galaxy. 
The case of the whole population of satellites is very different; quenching here is unaccompanied by kinematic transformations. Even though most satellites suffer an overall decrease in $\lambda_{\rm r_{50},edge-on}$, this is small compared to the change in sSFR. The latter is consistent with what \citet{Cortese19} inferred for satellite galaxies in the SAMI survey, and agree with the analysis presented there for \eagle\ in which net changes in stellar rotation-to-dispersion velocity ratio were compared to net changes in SFR to find that the two were decoupled. 
We investigate the effect of environment further by studying how many of these $z=0$ SR satellites suffered their last merger {\it after} becoming satellites. We find that this happens frequently: $\approx 50$\% of SR satellites had their last merger {\it after} becoming satellites (for the general population of satellites with $M_{\star}\ge 10^{10}\,\rm M_{\odot}$ this is much lower, $\approx 20$\%). To isolate the environment effect (which we associate with interactions with the tidal field of the halo and the central galaxy) from those of mergers with other satellite galaxies, the right panel of Fig.~\ref{SatelliteLambdaRChange} shows the subsample of satellites that had their last merger {\it prior} to becoming satellites. We see significant differences with the left panel of Fig.~\ref{SatelliteLambdaRChange}. Overall, the relative change in $\lambda_{\rm r_{50}, edge-on}$ is significantly smaller for satellites that had mergers (particularly major or minor mergers) prior to becoming satellites. {This shows that satellite-satellite mergers are at least as effective (or more so) in reducing $\lambda_{\rm r}$ as the environment (as defined here).} In the sample of SRs of the right panel of Fig.~\ref{SatelliteLambdaRChange}, there is a trend of the kinematic transformation being weaker when going from SRs that had exclusively very minor, minor to major mergers prior to becoming satellites. The weak environmental effect on $\lambda_{\rm r_{50}, edge-on}$ in SRs that have had mergers is due to the fact that by the time they become satellites they already have low $\lambda_{\rm r_{50},edge-on}\approx 0.19-0.29$, with the lower (higher) value corresponding to the median for SRs at the time of accretion that had major (only very minor) mergers prior to becoming satellites. In comparison, satellites that have not had mergers by $z=0$ and are SRs had a median $\lambda_{\rm r_{50},edge-on}\approx 0.47$ by the time they became satellites. For reference, satellite fast rotators at $z=0$ had a median $\lambda_{\rm r_{50},edge-on}\approx 0.6$ at the time they became satellites. 

In contrast, SRs that had mergers after becoming satellites were accreted with much higher $\lambda_{\rm r_{50},edge-on}\approx 0.5$, on average, showing that those that do not experience mergers during their lifetime as satellites suffer from strong progenitor bias\footnote{Their progenitors have sufficiently different properties as to cause the differences seen at $z=0$ with the SRs that had satellite-satellite mergers.}.
The comparison between the left and right panels of Fig.~\ref{SatelliteLambdaRChange} also shows that the large kinematic transformation seen in the left panel for SRs that have had mergers is driven in great part by satellite-satellite mergers (particularly for those that go through minor/major mergers with other satellites) rather than by interactions with the central galaxy or the tidal field of the host halo (which are the main mechanisms of kinematic transformation due to the environment; \citealt{Choi17}). 

\citet{Choi17} analysed satellite SRs in the Horizon-AGN hydrodynamical simulations and found that only $22$\% of their satellite SRs appeared to have low spins due to galaxy mergers. This appears to be in contradiction with our findings in \eagle. Part of that can be related to the high contamination parametric selections used in \citet{Choi17} have in distinguishing unambiguous SRs (see $\S$~\ref{CompWithSAMISec}), but more likely is that there are significant differences in the properties of satellite galaxies between the two simulations that allow environment to play a more significant role in the spin down of galaxies in Horizon-AGN compared to \eagle.

Despite the clear trends found in \eagle\ between the properties of SRs and their assembly history, it is important to highlight that having had mergers of some sort does not guarantee the formation of a SR. In fact, many fast rotators have also gone through galaxy mergers of different mass ratios and gas content (as seen from the difference in the number of visually classified SRs, $479$, and the number of galaxies that went through different merger histories in Table~\ref{mergers}). The emerging picture from \eagle\ is that {\it the required condition to form a SR is the process of quenching prior to or simultaneously with the kinematic transformation}.

\begin{figure}
\begin{center}

\label{SatelliteFractions2}
\end{center}
\end{figure}

\subsection{The stellar populations of slow rotators}\label{subsect_stellarpops}

\begin{figure}
        \begin{center}
                \includegraphics[trim=2mm 5mm 0mm 0mm, clip,width=0.47\textwidth]{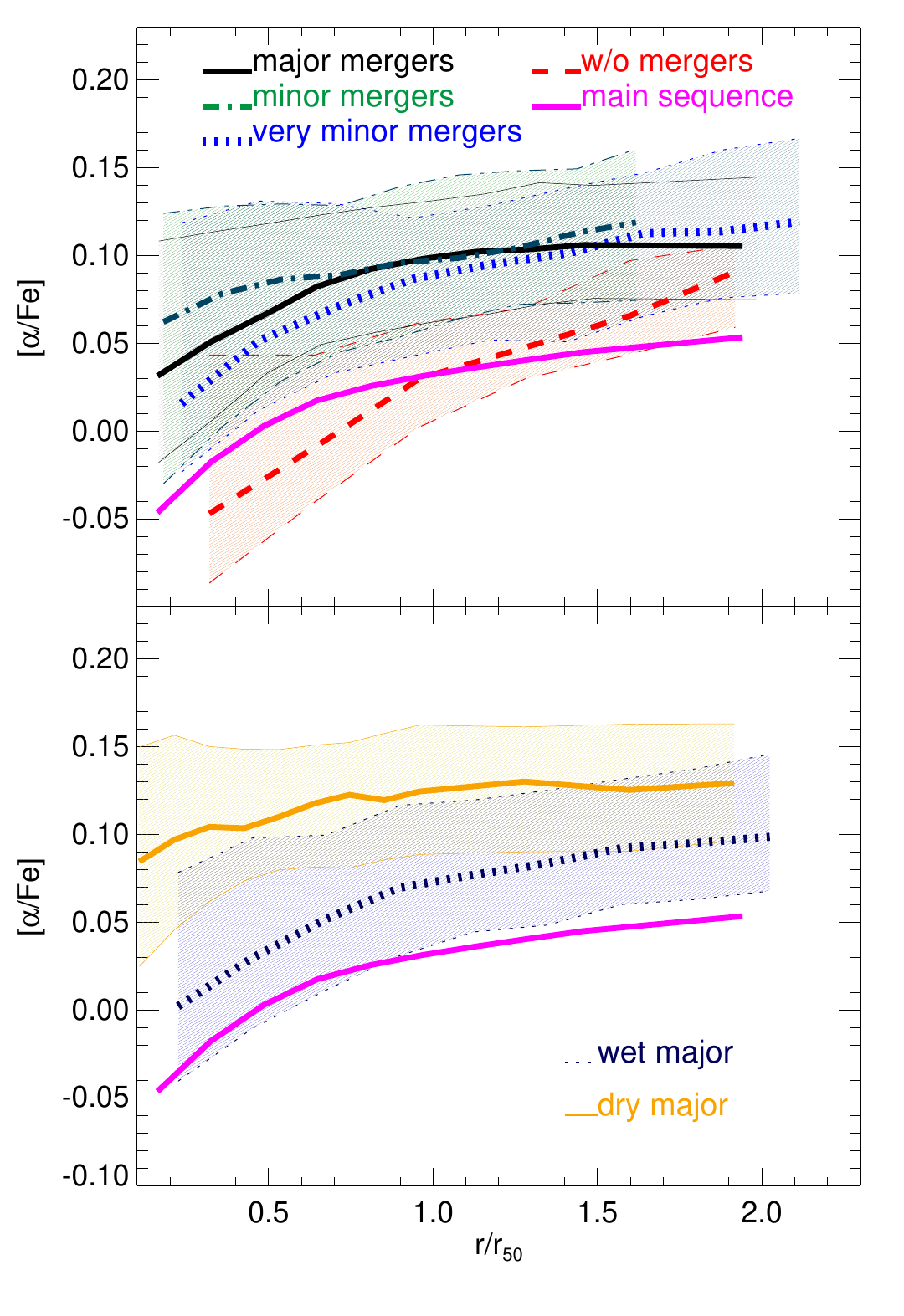}
                \caption{{\it Top panel:} Radial profiles of the stellar abundance of $\alpha$-elements over iron, in units of the solar abundance, [$\alpha$/Fe], for SRs at $z=0$ that have had $\ge 1$ major mergers (solid lines), $= 0$ major but $\ge 1$ minor mergers (dot-dashed lines),
                $= 0$ major/minor mergers but $\ge 1$ very minor mergers (dotted lines), and $= 0$ mergers (dashed lines)
                in the last $10$~Gyrs. 
                {Thick} lines show the median and {the shaded regions plus thin lines show the $25^{\rm th}-75^{\rm th}$}
                percentile range, respectively. The magenta lines shows the median for galaxies with $M_{\star}>10^{10}\,\rm M_{\odot}$ and $\rm sSFR>0.01\, Gyr^{-1}$ (considered to be representative of the main sequence). Radii are normalised by the half r-band luminosity radii of galaxies. {\it Bottom panel:} as in the top panel but for SRs that had $\ge 1$ wet (dotted line) or dry (solid line) merger (of any mass ratio) in the last $10$~Gyrs, as labelled.}
                \label{MetalsSRs}
        \end{center}
\end{figure}

\begin{figure}
        \begin{center}
                \includegraphics[trim=3mm 5mm 0mm 0mm, clip,width=0.47\textwidth]{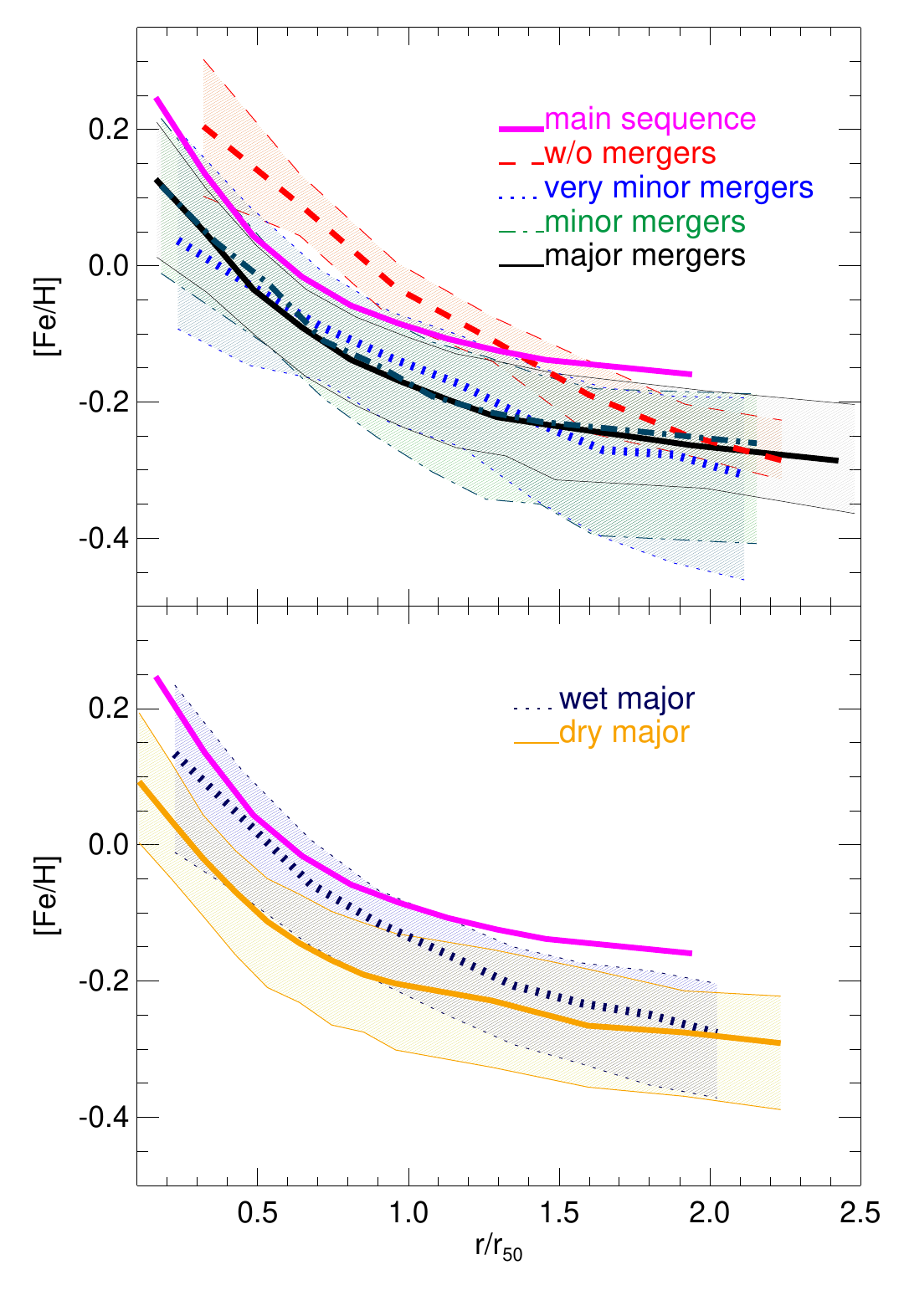}
                \caption{As in Fig.~\ref{MetalsSRs} but for radial profiles of [Fe/H].}
                \label{MetalsSRs2}
        \end{center}
\end{figure}

The different star formation and assembly histories of SRs in \eagle\ should leave imprints on the stellar populations of these galaxies that are potentially observable. Here, we focus on the metal abundance and stellar ages of $z=0$ SRs in \eagle.

Fig.~\ref{MetalsSRs} shows the radial profile of the abundance of $\alpha$ elements relative to Fe. We compute $\rm [\alpha/Fe]=\rm log_{10}(M_{\alpha}/M_{\rm Fe})-log_{10}(\alpha/Fe)_{\odot}$, where $M_{\alpha}$ is the mass contributed by $\alpha$ elements (the sum of the masses contained in Si, O, Mg, Ne and C), $M_{\rm Fe}$ the mass in iron and $\rm log_{10}(\alpha/Fe)_{\odot}=13.1206$ \citep{Asplund05}. The top panel of Fig.~\ref{MetalsSRs} shows $z=0$ SRs split by their assembly history, as labelled; also shown is the median $\rm [\alpha/Fe]$ of main sequence galaxies ($\rm sSFR>0.01\,\rm Gyr^{-1}$) with $M_{\star}>10^{10}\,\rm M_{\odot}$. All SRs that have had mergers are $\alpha$-enhanced relative to the sun across the whole radial range investigated. This is not the case for SRs that have not had mergers, in which $\rm [\alpha/Fe]<0$ at $r<r_{50}$, on average - $\approx 0.1$~dex lower than $\rm [\alpha/Fe]$ of other SRs. Interestingly, SRs in the ``no merger'' group are even less $\alpha$-enhanced than main sequence galaxies at $r<r_{\rm 50}$. This shows that low $\rm [\alpha/Fe]$ SRs are more likely to belong to the ``no merger'' sample than the other ones. The bottom panel of Fig.~\ref{MetalsSRs} shows $\rm [\alpha/Fe]$ radial profiles this time for SRs that had wet or dry mergers of any mass ratio (see $\S$~\ref{galmergerssec}). Those that had dry mergers are the ones with the flattest and most $\alpha$-enhanced $\rm [\alpha/Fe]$ radial profiles. 

\begin{figure*}
        \begin{center}
                \includegraphics[trim=2mm 2mm 0mm 0mm, clip,width=0.95\textwidth]{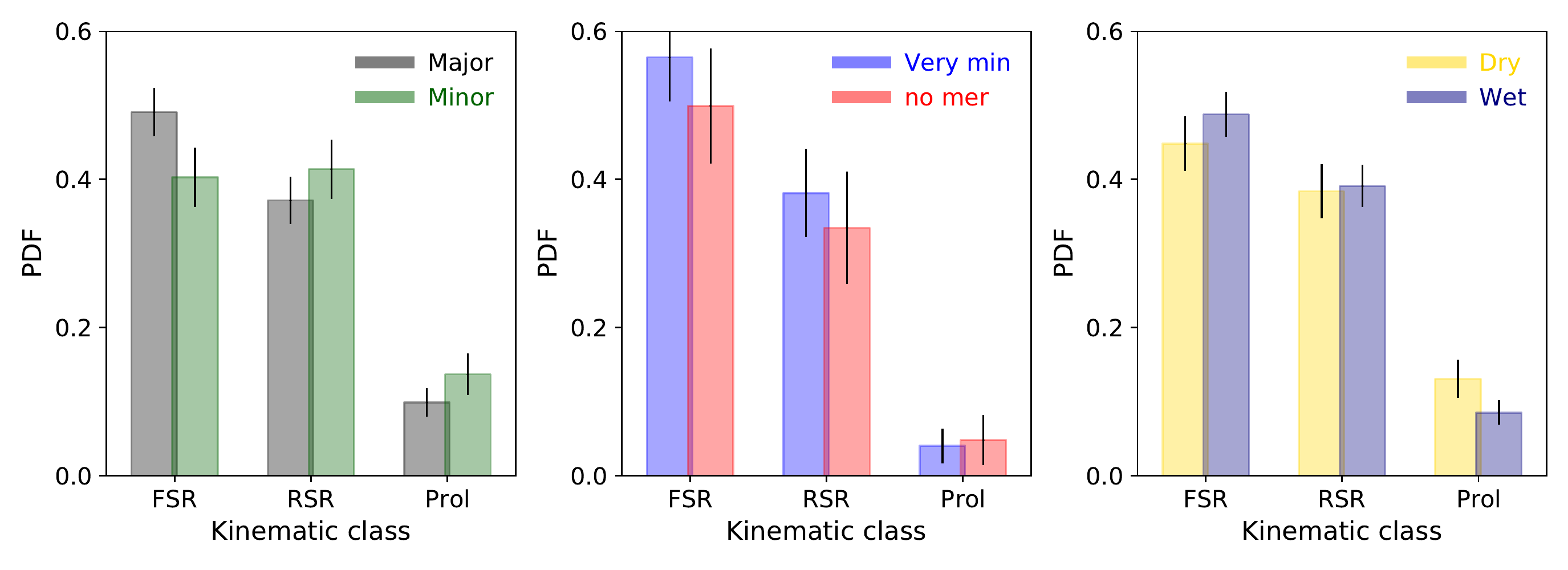}
                \caption{PDF of the kinematic classes of $z=0$ \eagle\ galaxies in the visually classified sample split by their merger history. {\it The left panel} shows those that had $\ge 1$ major mergers in the last $10$~Gyr and those that had $= 0$ major mergers but $\ge 1$ minor mergers, as labelled. {\it The middle panel} shows those that had $=0$ major/minor mergers but $\ge 1$ very minor mergers and $=0$ mergers. {\it The right panel shows} those that had $\ge 1$ dry mergers and those with $=0$ dry mergers but $\ge 1$ wet mergers. Errorbars were computed from jackknife resampling and are displaced arbitrarily from the centre of the bin to aid visualisation.}
                \label{KinematicMergers}
        \end{center}
\end{figure*}

A general feature is that most galaxies in \eagle\ tend to exhibit inverted $\rm [\alpha/Fe]$ profiles in which the central parts are less $\alpha$-enhanced than the outer parts. This happens because, on average, the stellar age radial profiles are inverted in \eagle, with the central parts being younger than the outer parts. Although early-type and passive galaxies in observations are consistent with flat (or even inverted) stellar age and $\rm [\alpha/Fe]$ radial profiles (e.g. \citealt{Kuntschner10,Greene15,Li18c,Bernardi19,Barsanti20,Santucci20}), late-type galaxies tend to have stellar age profiles consistent with the inner parts being older (e.g. \citealt{Gonzalez-Delgado15,Barsanti20}). For star-forming galaxies with $M_{\star}>10^{10}\,\rm M_{\odot}$ and $\rm sSFR > 0.01\,\rm Gyr^{-1}$ at $z=0$ in \eagle, we find $\approx 88$\% have inverted stellar age profiles (younger central parts; a similar percentage is found for those with $\rm sSFR < 0.01\,\rm Gyr^{-1}$), which disagrees with observational evidence. We do note, however, that the integrated $\rm [\alpha/Fe]$ ratios in \eagle\ galaxies agrees well with observations \citep{Segers16}.
This shows that even though feedback in \eagle\ is sufficient to quench star formation to reproduce the correct stellar mass function and other global properties related to metallicities and element abundances, the predicted radial properties of the stellar populations in galaxies has some important discrepancies with observations. The excess star formation in the centre is then the likely culprit of many of \eagle\ galaxies exhibiting inverted stellar velocity dispersion radial profiles, where the central velocity dispersion is lower (see for example the top three and bottom panels of Fig.~\ref{ExampleImages}). 
$\S$~\ref{buildingmocks} reported that $\approx 55$\% of galaxies with $M_{\star}>10^{10}\,\rm M_{\odot}$ have $\sigma_{\star}(0.5\,\rm r_{50})<\sigma_{\star}(r_{50})$. This percentage reduces to $43$\% for passive galaxies or SRs in \eagle. This is much larger than what is reported in observations. \citet{Falcon-Barroso17} found that in the sample of early-type galaxies in CALIFA, only 1 or 2 galaxies (out of $47$) have $\sigma_{\star}(0.5\,\rm r_{50})<\sigma_{\star}(r_{50})$.

In the galaxies with $\sigma_{\star}(0.5\,\rm r_{50})<\sigma_{\star}(r_{50})$, the r-band weighted stellar ages increase from $7.7$~Gyr at $\rm r<0.5\,r_{50}$ to $8.2$~Gyr at $\rm r<r_{50}$, on average, showing the connection between the lower central stellar velocity dispersions and the inverted stellar age profiles. We measure the Pearson correlation coefficient between $\rm log_{10}(\sigma_{\star}(0.5\,\rm r_{50})/\sigma_{\star}(r_{50}))$ and $\rm log_{10}(age_{\star}(0.5\,\rm r_{50})/age_{\star}(r_{50}))$, for all galaxies with $M_{\star}>10^{10}\,\rm M_{\odot}$ and obtained $P=0.43$. This correlation becomes stronger, $P=0.6$, for SRs in \eagle, which shows that the more strongly inverted the stellar age profile, the more strongly inverted the $\sigma_{\star}$ profile. This is an important shortcoming of \eagle\ that upcoming hydrodynamical simulations need to address.

Fig.~\ref{MetalsSRs2} shows radial profiles of $\rm [Fe/H]=\rm log_{10}(M_{\rm Fe}/M_{\rm H})-log_{10}(Fe/H)_{\odot}$ for $z=0$ SRs in \eagle, with $\rm (Fe/H)_{\odot}=0.001798$ \citep{Asplund05}. SRs in the ``no merger'' sample have the highest metallicities due to their delayed quenching times compared to other SRs (see Fig.~\ref{SFsHist}). Again, we see that these SRs have even higher $\rm [Fe/H]$ than main sequence galaxies, and display the steepest radial profiles. SRs in the ``very minor merger'' sample have the lowest and flattest $\rm [Fe/H]$ radial profiles due to their early quenching (see Fig.~\ref{SFsHist}). The bottom panel of Fig.~\ref{MetalsSRs2} separates SRs between dry and wet mergers. Dry mergers lead to SRs that have flatter $\rm [Fe/H]$ profiles due to the effective redistribution of stellar mass during dry mergers \citep{Lagos18a}. \citet{Krajnovic20} found that classical slow rotators (which they linked to the dissipation-less galaxy mergers; i.e. dry mergers) have flatter metallicity gradients than other slow rotators in ATLAS$^{\rm 3D}$, which is in qualitative agreement to what we find in \eagle.

The trends shown in Figs.~\ref{MetalsSRs}~and~\ref{MetalsSRs2} show that observations of the stellar populations in SRs can provide a broad indication of the most likely merger history. However, as these are trends, application on a one-to-one basis is not advised.

\section{Kinematic classes of slow rotators}\label{KinClassesSec}

In this section we use the visual kinematic classification of \eagle\ galaxies of $\S$~\ref{secclassification} and analyse their connection to the galaxy merger history of to understand the effect the galaxy mass ratio and gas ratio involved in mergers have on the kinematic class.  We focus on the latter two merger parameters as \citet{Lagos18a,Lagos18b} have shown that those have the most effect in modifying the kinematics of galaxies. We also study possible connections between environment and stellar mass with the different SR's kinematic classes. In this section we only study the kinematic classes FSR, RSR and prolates (overall classed as SRs), ignoring the $2\sigma$, unclear and rotator classes.

\subsection{The relation between kinematic class and galaxy merger history}

\begin{figure}
        \begin{center}
                \includegraphics[trim=14mm 10mm 7mm 10mm, clip,width=0.5\textwidth]{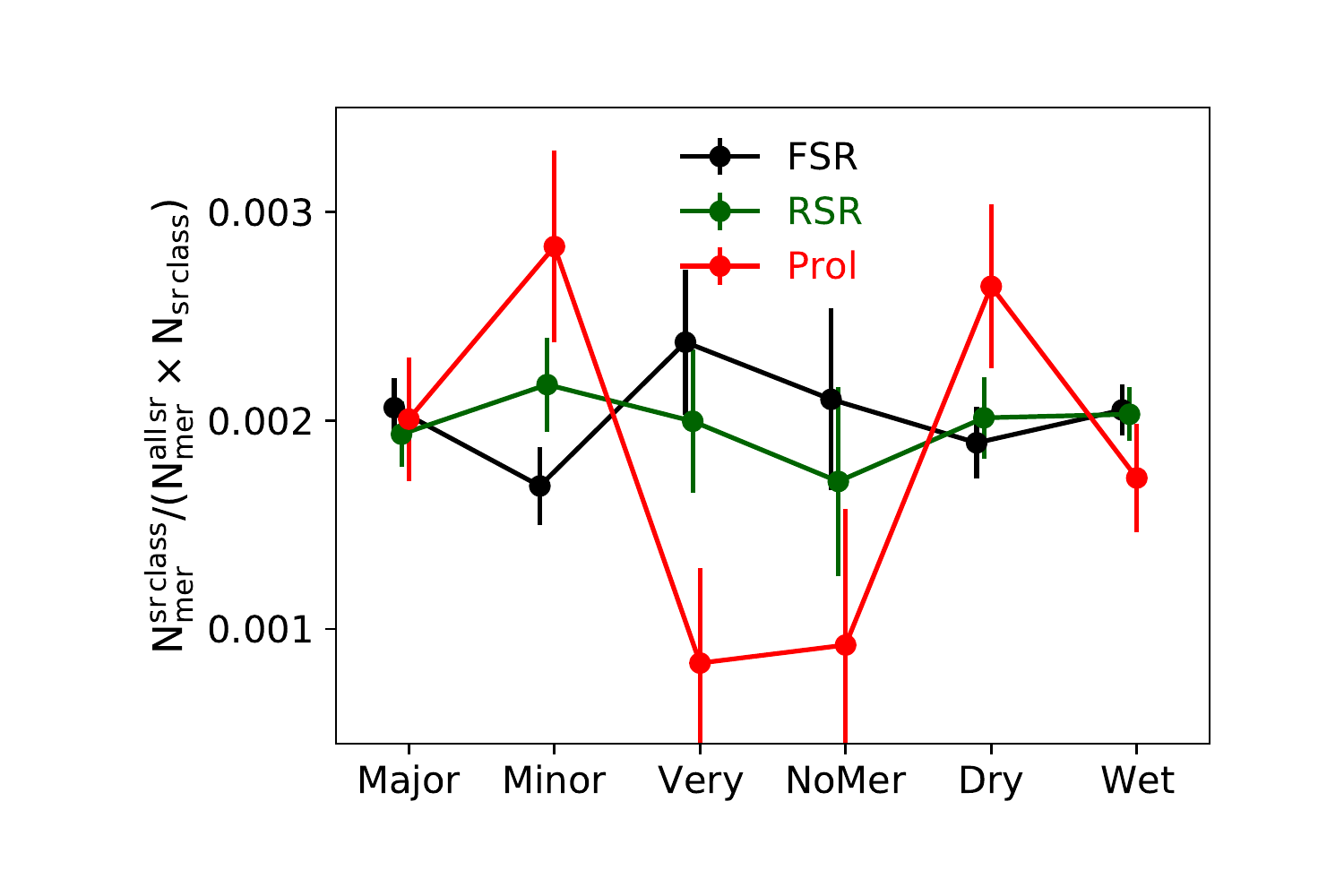}
                \caption{Relative frequency of different types of mergers for a given kinematic class. We show this for $6$ different galaxy merger types: major mergers, minor mergers (with no major mergers), exclusively very minor mergers, no mergers, dry and wet mergers, counted over the last $10$~Gyr of evolution of these galaxies that at $z=0$ are classified as belonging to the four kinematic classes shown, as labelled. Errorbars were computed from jackknife resampling.} 
                \label{KinematicMergersByClass}
        \end{center}
\end{figure}

\begin{figure*}
        \begin{center}
                \includegraphics[trim=4mm 5mm 4mm 3mm, clip,width=1\textwidth]{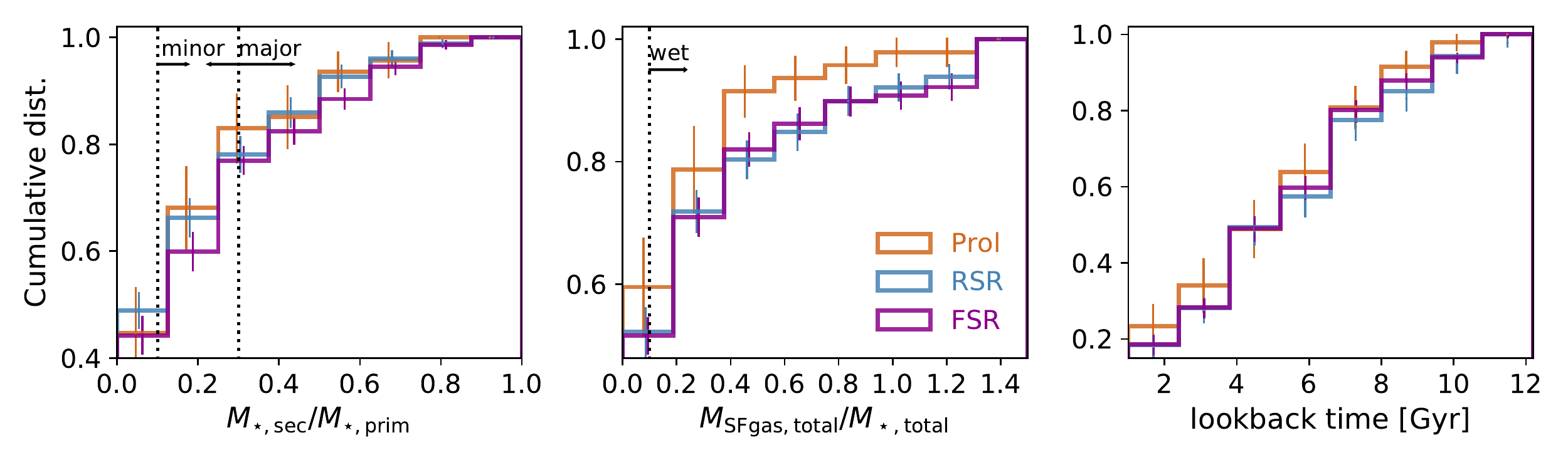}
                \caption{Cumulative distribution of the stellar mass ratio (left), SF gas ratio (middle) and lookback time (right) of the last merger SRs in the four kinematic classes, labelled in the middle panels, went through. Vertical lines in the top and middle panels show what we classify as major/minor/very minor mergers, and wet/dry mergers, respectively. Errorbars were computed from jackknife resampling and are displaced from the centre of the bin by arbitrary amounts to aid their visualization. Prolate SRs have a preference of lower gas ratios. Flat SRs prefer high stellar mass ratios compared to round SRs.}
                \label{PropsMergers}
        \end{center}
\end{figure*}

\begin{figure}
        \begin{center}
                \includegraphics[trim=4mm 96mm 6mm 95mm, clip,width=0.24\textwidth]{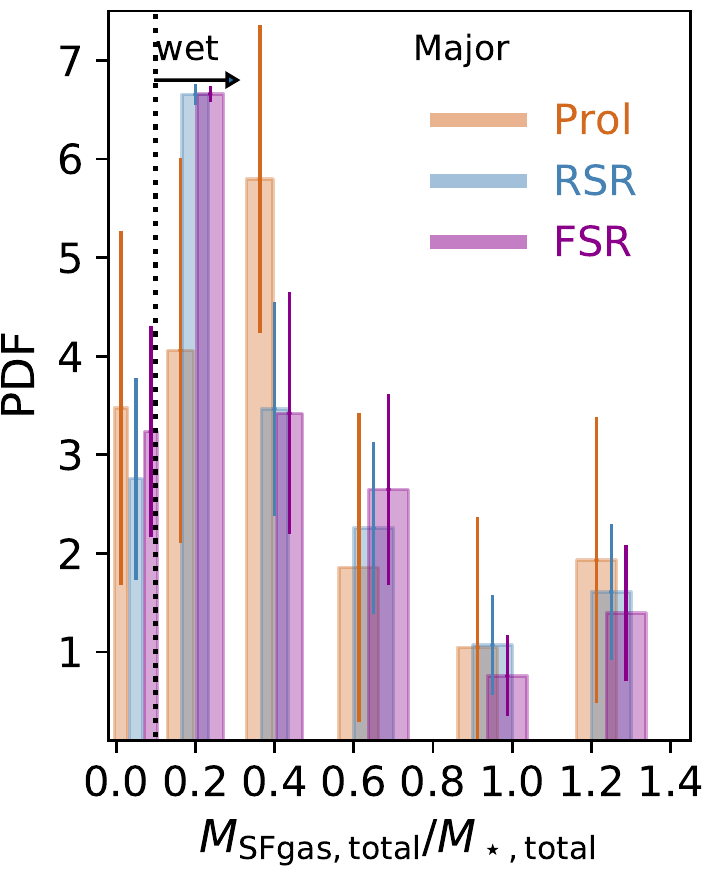}
                \includegraphics[trim=8mm 96mm 6mm 95mm, clip,width=0.225\textwidth]{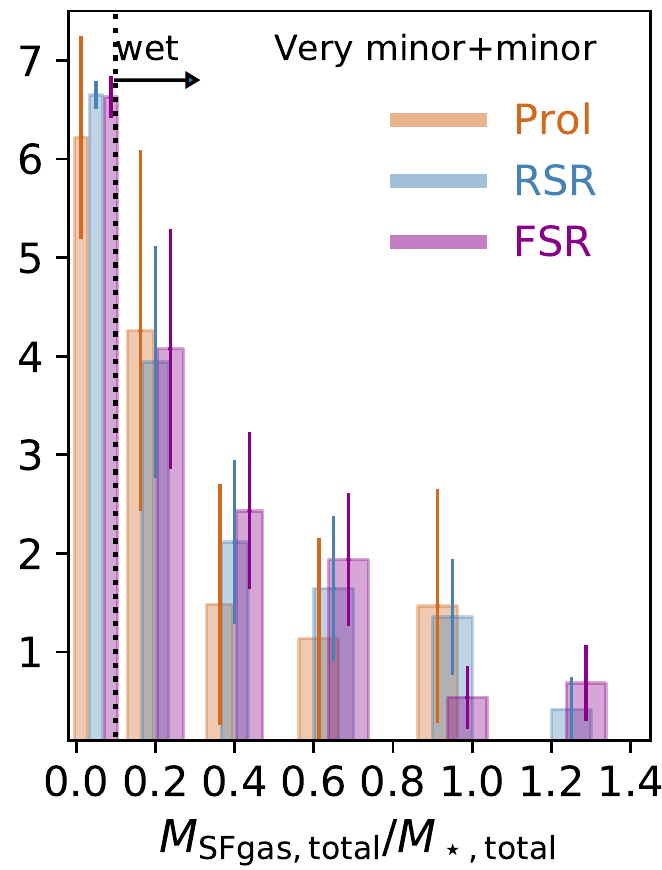}
                \caption{PDF of the SF gas ratio of the last major merger (left) and very minor plus minor mergers (right) SRs went through. We show this for 3 kinematic classes, as labelled. In the right panel we show SRs that experienced very minor or minor mergers (no major mergers in their history), while the SRs in the left panel has gone through major mergers. Errorbars were computed from jackknife resampling and are displaced from the centre of the bin by arbitrary amounts to aid their visualization. Note that we use arbitrary bin widths that are narrower in the ranges where there are more galaxies. Very minor/minor mergers that lead to SRs are preferentially low gas ratios compared to the major mergers that lead to SRs.}
                \label{PropsMergers2}
        \end{center}
\end{figure}
Fig.~\ref{KinematicMergers} shows the PDF of the SRs kinematic classes described in $\S$~\ref{secclassification} for all the galaxies visually classified that have a confidence $\ge 60$\%, split by their merger history. By comparing galaxies that had $\ge 1$ major mergers and those that had $\ge 1$ minor mergers but no major mergers (left panel), we see that the former tend to be associated with more FSRs, while prolate galaxies appear to have a preference for minor mergers. 
The overall distribution of galaxies that had exclusively very minor mergers is qualitatively similar to those that had no mergers (middle panel).
The distribution of minor and very minor mergers are similar, and we see that a similar fraction of those are associated with FSRs and RSRs. Although prolates do happen in these SRs, their relative fraction is small compared to what is seen for SRs that had major/minor mergers.

The right panel of Fig.~\ref{KinematicMergers} shows dry and wet mergers (see $\S$~\ref{galmergerssec} for the criterion to define wet and dry mergers). We remind the reader that the sample of dry mergers can also contain wet mergers, while the sample of wet mergers excludes dry mergers. We find that FSRs are overly represented in the sample of wet mergers compared to the dry merger sample. If we split wet mergers in two bins of gas fraction we obtain similar distributions (not shown).
Although about half of the prolate galaxies formed via wet mergers and the other half via dry mergers in bulk numbers, we see that the normalised distributions of dry mergers have a higher incidence of prolate galaxies than wet mergers. 

Fig.~\ref{KinematicMergersByClass} shows the relative frequency of different types of galaxy mergers for a given kinematic class. We define this relative frequency as $=N^{\rm sr\,class}_{\rm mer}/(N^{\rm all\, sr}_{\rm mer}\times N_{\rm sr\, class})$, where $N^{\rm sr\,class}_{\rm mer}$, $N^{\rm all\, sr}$ and $N_{\rm sr\, class}$ 
are the number of SRs in a given kinematic class that went through the corresponding type of merger, the total number of SRs that went through that same type of merger, and the number of SRs in the kinematic class, respectively. This way we normalise by the different number of galaxies in each kinematic class and in each merger history type. We confirm that FSRs have a preference for major mergers compared to minor ones, while RSRs appear to be similarly represented for the minor, very minor and no merger cases (considering the relative numbers of those). Prolate galaxies have a preference for dry mergers, and are also overly represented in the case of minor mergers. {However, we caution that the latter is highly uncertain due to the small number of prolate galaxies in our sample (see Table~\ref{kinclassesnum}).}

In order to get a better understanding of the connection between the merger parameters and the different kinematic classes of SRs, we study the distribution of the stellar mass and gas ratios, and lookback time of the last galaxy merger each SR in our sample had (in this case we remove the sample of SRs that have not had mergers). This is shown in Fig.~\ref{PropsMergers}. Below we discuss the main trends in the three quantities shown in Fig.~\ref{PropsMergers}:

(i) \noindent {\it Merger mass ratio.} Comparing FSRs and RSRs, we see a preference of RSRs for smaller stellar mass ratios, even within the major merger band ($M_{\rm \star,sec}/M_{\rm \star,prim}>0.3$), compared to FSRs. FSRs are the sample that is most skewed towards high $M_{\rm \star,sec}/M_{\rm \star,prim}$. Prolate galaxies seem to be associated with either very low or intermediate mass ratios,  
$0.1\lesssim M_{\rm \star,sec}/M_{\rm \star,prim}\lesssim 0.45$. We will show later that the lower stellar mass ratios are mostly associated with gas-poor mergers, while the higher ratios to gas-richer mergers. 

(ii) \noindent {\it Merger gas ratio.} FSRs and RSRs show comparable distributions of gas ratios
Prolates, on the other hand, prefer lower gas ratios compared to both FSRs and RSRs, that is most clear in the regime of dry mergers. 

(iii) \noindent {\it Lookback time to last merger.} FSRs, RSRs and prolates have a similar distribution of lookback time to their last galaxy mergers. Even though the errors are large there is a small preference for prolates to have had their last merger at later times. The latter would be expected given that gas poorer mergers happen preferentially at later times in \eagle\ \citep{Lagos18a}. 

In the case of prolates, \citet{Li18} found that in the Illustris simulations they were predominantly associated with late, dry major mergers. In \eagle\ we find a clear preference for dry mergers, but find that in bulk numbers a similar percentage of prolates are associated to major and minor mergers; i.e. $\approx 47$\% to major mergers and $\approx 43$\% to minor mergers. When the distributions are normalised by the relative numbers of these mergers in SRs, we find that minor mergers have a higher incidence of prolates. Furthermore, the remaining $10$\% are associated to very minor mergers or no mergers. Hence, it appears like the formation mechanisms of prolates in \eagle\ are more diverse than in Illustris. 

To connect the stellar mass and gas ratios of the mergers of the different kinematic classes of SRs in \eagle, we show in Fig.~\ref{PropsMergers2} the distribution of the gas ratio of the last merger SRs went through, split into major and very minor plus minor mergers. Generally, we find that in all three kinematic SR classes, minor and very minor mergers that lead to remnant SRs tend to be gas poorer than major mergers leading to SRs. FSR and RSR show similar distributions of merger gas ratios in both panels, which means that the main difference between these two subclasses is the higher stellar mass ratios of the FSRs (Fig.~\ref{PropsMergers}). Prolates behave similarly, with the minor/very minor mergers being heavily skewed towards gas-poor mergers, and major mergers having a wider range of gas ratios. Connecting to the left panel of Fig.~\ref{PropsMergers}, we find that for prolates, the lower (higher) stellar mass ratios are primarily associated with low (high) gas ratios.

Although the trends above are connected to possible physical drivers, it is important, however, to highlight that Poisson noise is quite significant in these trends due to low number statistics. Ideally we would like to study the three dimensional space between kinematic classes, stellar mass and gas ratios but the current statistics in \eagle\ are prohibiting. We are then forced to marginalise over one of these properties. Upcoming simulations of much larger cosmological volume but comparable or even higher resolution than \eagle\ are required to consolidate some of the trends reported here and to open the possibility to a much finer connection between merger parameters and the kinematic properties of SRs (including the extension to more mergers parameters associated with the orbits of satellite galaxies).  

\subsection{The relation between kinematic class, stellar mass and environment}

\begin{figure}
        \begin{center}
                \includegraphics[trim=4mm 5mm 3mm 3mm, clip,width=0.5\textwidth]{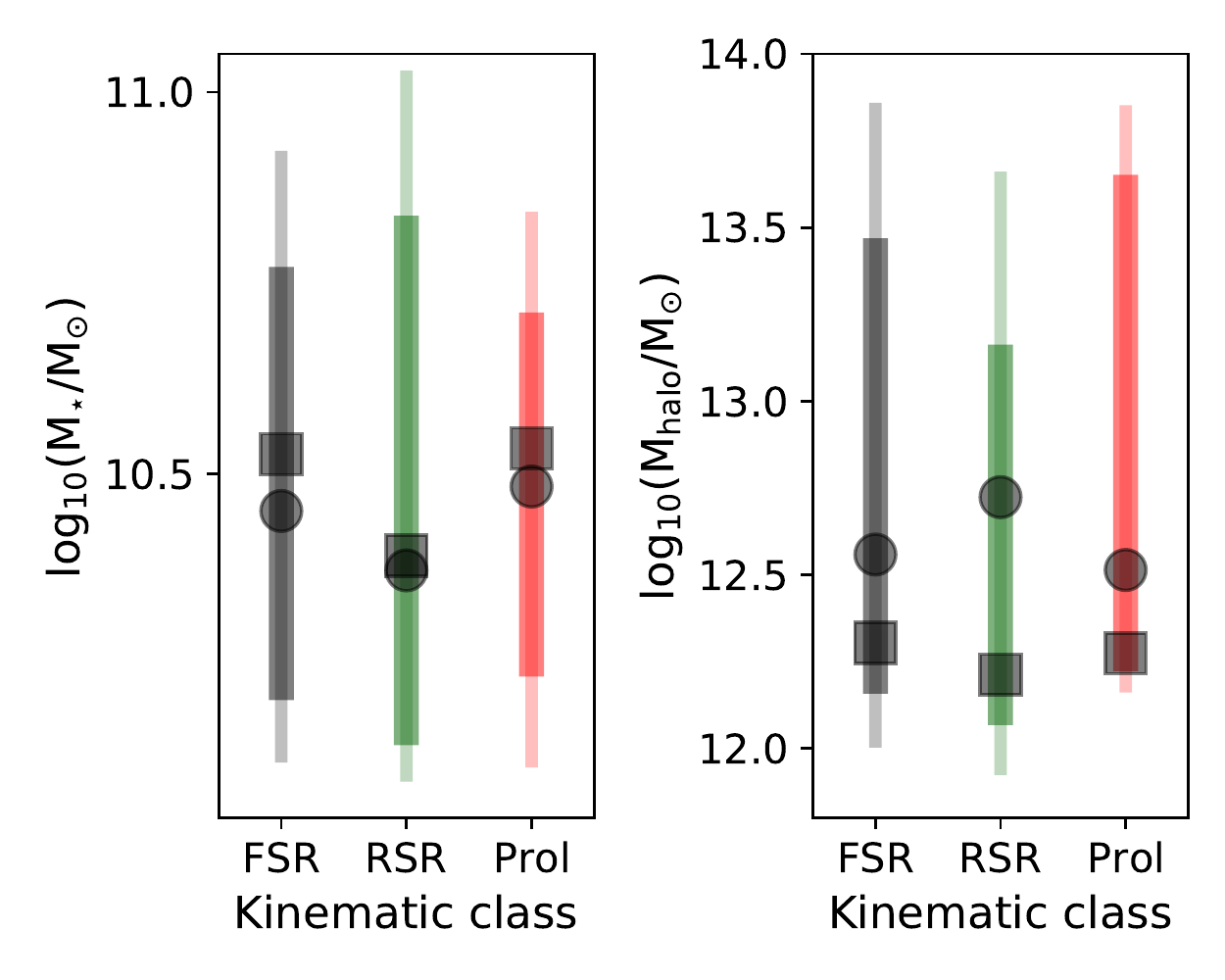}
                \caption{Median (filled circles) stellar (left) and host halo (right) mass of the SRs in the kinematic classes labelled in the x-axis. The errorbars show the $25^{\rm th}-75^{\rm th}$ (thicker lines) and $16^{\rm th}-84^{\rm th}$ (thinner lines) percentile ranges. Squares show the median of the subsample of central SRs in each kinematic class.}
                \label{KinClass_MsMh}
        \end{center}
\end{figure}

Fig.~\ref{KinClass_MsMh} shows the median stellar and halo masses of SRs in \eagle\ split by their kinematic class. SRs of different kinematic classes have a similar median stellar masses, but prolate SRs tend to be skewed towards higher masses. The typical stellar masses of prolate galaxies in \eagle\ agree well with those reported in \citet{Schulze18} in the Magneticum simulations, but differ significantly from the ones in Illustris reported in \citet{Li18}. \citet{Li18} found that prolates in Illustris are almost exclusively galaxies with $M_{\star}\gtrsim 3\times 10^{11}\,\rm M_{\odot}$. Part of this discrepancy may come from the fact that at halo masses $>10^{12.2}\,\rm M_{\odot}$, Illustris produces galaxies too massive in stars compared to observational inferences by a factor of $\approx 7-10$ (see Fig.~$4$ in \citealt{Pillepich17}). \eagle\ on the other hand produces a stellar-halo mass relation in better agreement with observations \citep{Schaye14}. If one was to instead analyse the host halo masses of prolate galaxies in \eagle\ and Illustris, the difference above would be largely alleviated.

For the halo masses of SRs in \eagle\ (right panel in Fig.~\ref{KinClass_MsMh}), we find larger variations than for stellar mass. Interestingly, FSRs are more massive in stars but are hosted by lower mass halos than RSRs, on average, and prolates have the highest stellar-to-halo mass ratio. 
This is the result of two factors: the fact that Fig.~\ref{KinClass_MsMh} includes both centrals and satellites (and for the latter we expect no correlation between stellar and host halo mass), and the fact that at fixed halo mass, the scatter in the stellar-halo mass relation is correlated with the assembly history of galaxies (e.g. \citealt{Correa20}). To disentangle these effects, we also show in Fig.~\ref{KinClass_MsMh} the median stellar and halo mass of central SRs only in the same three kinematic classes (squares). The most striking trend is that central prolates tend to have a slightly higher stellar-to-halo mass ratios (median $0.02$) compared to the other SRs by $\approx 12$\%. Given how tight the stellar-halo mass relation is in \eagle\ (median stellar-to-halo mass ratio of all central SRs is $0.0179\pm 0.006$), this difference is significant. \citet{Correa20} found that at fixed halo mass in \eagle, higher stellar mass galaxies formed in halos that assembled earlier than lower stellar mass galaxies. Galaxies in halos that assemble earlier also tend to have higher BH-to-stellar mass ratios, indicating that AGN feedback can be more effective there \citep{Bower17}. Since prolates tend to prefer gas-poor galaxy mergers (Figs.~\ref{PropsMergers}~and~\ref{PropsMergers2}), the more efficient AGN feedback can help promote their formation. 

\begin{figure}
        \begin{center}
                \includegraphics[trim=36mm 2mm 20mm 0mm, clip,width=0.5\textwidth]{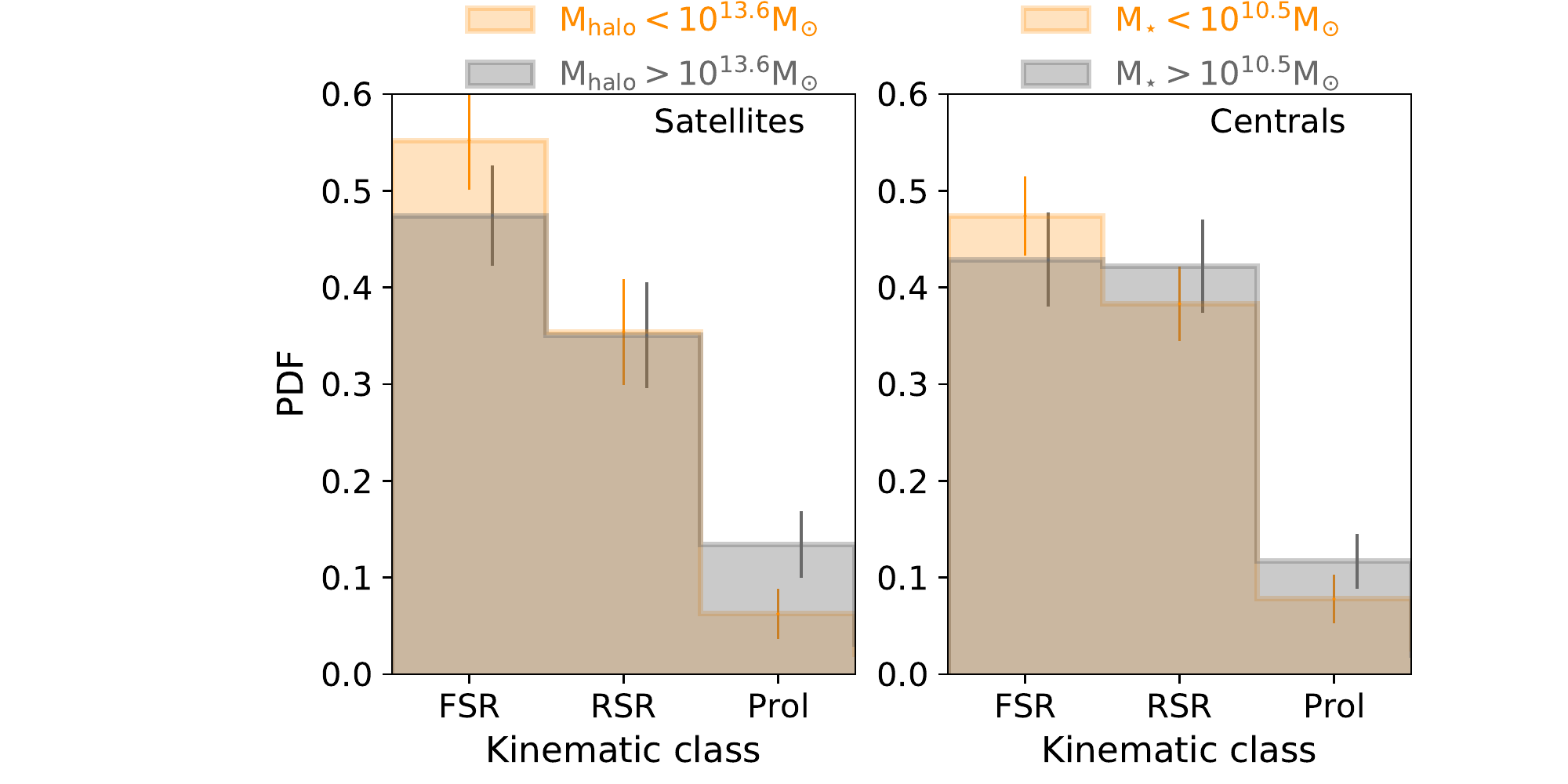}
                \caption{PDF of the kinematic classes of $z=0$ SRs separating satellite (left) and central (right) galaxies. For satellites, we show separately the distribution for galaxies above and below the median hot halo mass, $10^{13.6}\,\rm M_{\odot}$. For centrals, we separate them by their stellar mass, above and below the median, $10^{10.5}\,\rm M_{\odot}$, as labelled {(the darker region is where histograms overlap)}. Errorbars were computed from jackknife resampling and are displaced from the centre of the bin by arbitrary amounts to aid their visualization.}
                \label{KinClass_CensSats}
        \end{center}
\end{figure}

In $\S$~\ref{kinandquenching} we showed the connection between environment and kinematic transformation in satellite galaxies that end up as SRs, as well as the connection between AGN feedback and quenching of central SRs. Here, we explore the connection between the kinematic class of satellites and centrals with their environment and stellar mass. 
Because our sample of SRs is small, we only split the subsample of satellites in two halo masses (left panel of  Fig.~\ref{KinClass_CensSats}). We split SR satellites by their host halo mass, below and above the median host halo mass, $\approx 10^{13.6}\rm \, M_{\odot}$. The median stellar masses of these two samples of satellites are very similar ($\approx 2.4\times 10^{10}\,\rm M_{\odot}$ and $\approx 2.7\times 10^{10}\,\rm M_{\odot}$, respectively), hence yielding no bias in stellar mass when splitting by halo mass.
We find an environmental trend, with SR satellites in low-mass halos showing a preference for being FSRs, while those in high-mass halos have a preference for being RSRs. 
Although prolate are found in both halo mass samples, their frequency is higher in high-mass halos. Given that we find environment has a differential effect on $\lambda_{\rm r_{50},edge-on}$ depending on the type of merger suffered by the SR prior to being accreted, it is likely that the environmental trend of Fig.~\ref{KinClass_CensSats} is at least in part driven by progenitor bias. To asses this, we study the incidence of different types of mergers between the satellite SRs in halos of mass above and below $\approx 10^{13.6}\rm \, M_{\odot}$ and find that satellites hosted in halos of mass $M_{\rm halo}\lesssim 10^{13.6}\rm \, M_{\odot}$ are more (less) likely to have had major (minor) mergers compared to those hosted in more massive halos, $M_{\rm halo}\gtrsim 10^{13.6}\rm \, M_{\odot}$. The relative frequency of major and minor mergers in satellite SRs below and above the median host halo mass is $47\pm 3$\% vs. $37\pm 2.8$\% (major mergers) and $24\pm 2$\% vs. $32\pm 3$\% (minor mergers), respectively, showing a significant effect of progenitor bias (uncertainties in percentages were computed from jackknife resampling). 
As Fig.~\ref{PropsMergers} shows, FSRs are preferentially associated with major mergers, hence explaining why satellite FSRs are more common at $M_{\rm halo}\lesssim 10^{13.6}\rm \, M_{\odot}$. We find no difference between the fraction of SRs that have exclusively very minor mergers or no mergers for satellites in halos below/above $\approx 10^{13.6}\rm \, M_{\odot}$.
The persistence of major mergers in satellites hosted by halos of low masses partially explains the finding in \citet{Davison20} of the ex-situ stellar fraction being {\it higher} in satellites at lower halo masses in \eagle.
Related to the difference between prolate satellite galaxies above/below $\approx 10^{13.6}\rm \, M_{\odot}$, the likely cause is the fact that galaxies tend to be gas-richer in lower mass halos, and hence less likely to lead to a prolate galaxy, give the trends of Fig.~\ref{PropsMergers}.

The right panel of Fig.~\ref{KinClass_CensSats} focuses now on central SRs, showing the distribution of kinematic classes in two bins of stellar mass, above/below the median of the sample, $\approx 10^{10.5}\,\rm M_{\odot}$. We see that massive central SRs have a preference for being FSRs, while at lower mass, FSRs and RSRs are similarly common. Prolate centrals tend to be similarly common below/above a stellar mass of $\approx 10^{10.5}\,\rm M_{\odot}$.
In \eagle, the fraction of galaxies that by $z=0$ have experienced a galaxy merger increases with stellar mass \citep{Lagos18a}, and the bottom panel of Fig.~\ref{LambdaREllipSRs2} shows that in particular the incidence of major mergers increases with stellar mass. Given the preference for FSRs to be associated with galaxy mergers with higher stellar mass ratios compared to RSRs (Fig.~\ref{PropsMergers}) it is not surprising that the massive central SRs have a preference for being FSRs.

\section{Discussion and conclusions}\label{Conclusions}

The possible formation paths of fast and slow rotators has been an area of intense research since the advent of IFS surveys, sampling hundreds of galaxies, revealed the existence of these populations \citep{Emsellem07,Emsellem11}. Their connection to galaxy mergers has been explored in hydrodynamical simulations, which have found varied and often contradictory results in the types of mergers that lead to the formation of SRs  \citep{Naab14,Choi17,Penoyre17,Lagos18b,Schulze18}. 
Generally to distinguish these two populations of rotators, both observations and simulations have generally employed parametric selections in the $\lambda_{\rm r}-\epsilon$ plane, motivated by the results of the ATLAS$^{\rm 3D}$ survey (see \citealt{Cappellari16} for a review). However, a significant problem that has become more evident is that these parametric selections lead to significant contamination (i.e. a high ratio of visual rotators to non-rotators) in the retrieved samples of SRs \citep{vandeSande20}. This hinders the study of the formation paths of SRs in both observations and simulations, and could in part be responsible for the difficulty in isolating the main formation paths of SRs experienced in the latter. 

To remedy this low SRs purity, here we employ visual classification of the stellar kinematic maps of galaxies in the \eagle\ hydrodynamical simulations suite. We focused on galaxies with $M_{\star}\ge 10^{10}\,\rm M_{\odot}$ which have well resolved kinematics \citep{Lagos16b,Lagos18b}. We place our galaxies in the context of the SAMI survey by creating kinematic maps using the SAMI specifications, which are readily available in the {\sc SimSpin} package (see $\S$~\ref{buildingmocks}). We had $5$ classifiers separating galaxies into $6$ kinematic classes: flat SRs, round SRs, $2\sigma$, prolates, unclear and rotators. We found that $\ge 60$\% agreement among classifiers is reached in $90$\% of the classified galaxies, showing that the vast majority of galaxies can be cleanly separated into these kinematic classes. We use this sample to select unambiguous SRs in \eagle, which correspond to flat or round SRs and prolate galaxies. 
Using the unambiguous SRs in \eagle, we find that parametric classifications have at best a purity of $65$\% (i.e. $65$\% of the galaxies that comply with the parametric selection of SRs are not considered as such by our visual classification), showing the requirement of the visual classification to isolate unambiguous SRs.

For the sample of unambiguous SRs in \eagle, we study the connection to galaxy mergers, differences in intrinsic galaxy properties and the connection between quenching and kinematic transformation. We summarise our findings below:

\begin{itemize}
    \item SRs with $M_{\star}\gtrsim 10^{10.8}\,\rm M_{\odot}$ have triaxiality, $T$, consistent with being prolates, $T>0.7$, while lower mass SRs span the whole range of $T$. Major and minor mergers lead to triaxial or prolate SRs ($T\gtrsim 0.5$), while exclusively very minor mergers are largely associated with triaxial systems ($0.3\lesssim T\lesssim 0.7$). SRs that formed in the absence of mergers are oblate ($T\lesssim 0.2$). These classes of SRs are clearly linked to different ex-situ stellar fractions, with SRs that had minor/major mergers or exclusively very minor mergers typically having $f_{\rm ex-situ}\gtrsim 0.4$ and $f_{\rm ex-situ}\gtrsim 0.1$, respectively. SRs in the ``no merger'' category have $f_{\rm ex-situ}\lesssim 0.05$ (Fig.~\ref{ShapeSRs}). This clearly shows that there is a class of SRs that forms in the absence of mergers in simulations (see also \citealt{Choi17}). A higher fraction of galaxies in this class are satellites compared to SRs associated with galaxy mergers (Fig.~\ref{SatelliteFractions}). 
    \item SRs in the ``no merger'' class tend to be more compact and have lower BH-to-stellar mass ratios than other SRs at fixed stellar mass (Fig.~\ref{ShapeSRs}). They also tend to quench later (starting to drop below the main sequence at a lookback time $\approx 4.5$~Gyr, compared to $\gtrsim 6$~Gyrs for other SRs; Fig.~\ref{SFsHist}). This leaves imprints on their stellar populations, with the ``no merger'' SRs having lower $\rm \alpha/Fe$ (even below solar; Fig.~\ref{MetalsSRs}) and higher $\rm Fe/H$ ratios than other SRs (Fig.~\ref{MetalsSRs2}). 
    \item We find that in most SRs quenching happens {\it before} kinematic transformation by $\approx 2$~Gyrs (Figs.~\ref{SRsHist}~and~\ref{SFsHist}). Most SRs quenched due to AGN feedback as evidenced by their overly massive BHs at the time they left the main sequence of SF (Fig.~\ref{MBHAtQuench}). These SRs tend to have quenching timescales between $1.5-2.5$~Gyrs. The exception are satellite SRs that have not had mergers, which are quenched by the effect of environment. In the latter, quenching and kinematic transformation appear to happen in tandem (to within $0.2$~Gyrs), with the likely mechanism of kinematic transformation being interactions with the tidal field of the halo and central galaxy. The emerging picture is that in most SRs, quenching is required for galaxy mergers to more effectively decrease $\lambda_{\rm r}$.
    \item We find that $\approx 50$\% of $z=0$ satellite SRs experienced satellite-satellite mergers, which were largely responsible for their SR fate (for reference only $20$\% of the general satellite population with $M_{\star}\ge 10^{10}$ experience satellite-satellite mergers). When focusing solely on $z=0$ satellite SRs that have not had satellite-satellite mergers, we find that environment was clearly responsible for the kinematic transformation of the subsamples of ``no merger'' or ``exclusively very minor mergers" satellite SRs (Fig.~\ref{SatelliteLambdaRChange}). Nevertheless, we see an important effect of progenitor bias, with $\lambda_{\rm r}$ of the progenitors of $z=0$ satellite SRs that formed in the absence of mergers being higher than those that had mergers, regardless of whether these happen prior or after becoming satellites.
    \item Flat SRs are overly represented in the sample of SRs that had major mergers, while round SRs tend to prefer galaxies that had exclusively minor or very minor mergers (Figs.~\ref{KinematicMergers}~and~\ref{KinematicMergersByClass}). Prolate galaxies are predominantly connected to gas-poor galaxy mergers (Fig.~\ref{PropsMergers}). In the sample of flat and round SRs, we find that major and minor mergers associated with their formation tend to be gas-rich and gas-poor, respectively (Fig.~\ref{PropsMergers2}).  
    \item Flat SRs tend to be more common in satellites hosted by massive halos ($>10^{13.6}\,\rm M_{\odot}$) and centrals of high stellar mass ($>10^{10.5}\,\rm M_{\odot}$) due to the higher incidence of major mergers in these populations (Fig.~\ref{KinClass_CensSats}). Prolates are also more common in these populations due to the higher incidence of gas-poor mergers. Prolate centrals have the highest stellar-to-halo mass ratios of all the SRs (Fig.~\ref{KinClass_MsMh}), which we connect to those halos preferentially forming earlier and having more AGN activity, as indicated in \citet{Correa20}.
\end{itemize}

Although we find several trends between different types of mergers and SR's kinematic classes, we could not identify a single galaxy feature that can unambiguously indicate a given assembly history. This may not be surprising given the complexity of galaxy formation and the many physical processes that simultaneously take place. However, it does mean that the trends exposed here cannot be applied on a single galaxy basis.

There are some important limitations of our study. By visually inspecting many kinematic maps of \eagle\ galaxies, we found a common feature of the stellar $\sigma$ being smaller at the centre and increasing towards the outskirts in $\approx 50$\% of galaxies with $M_{\star}>10^{10}\,\rm M_{\odot}$ (see the top three panels of Fig.~\ref{ExampleImages} for examples). The latter was found to be related to inverted stellar age radial profiles (where the central parts of galaxies are younger than the outer parts). We concluded that although feedback in \eagle\ is sufficient to lead to integrated galaxy properties that agree well with observations (such as colour distribution, stellar mass function, global metallicities and oxygen abundance profiles, etc.; e.g. S15, \citealt{Trayford15,Trayford16,Wright19,Katsianis17,Segers16,derossi17,Tissera19,Collacchioni20}), the imprints it leaves on the internal kinematic properties of galaxies is not always realistic. This physical limitation of the simulation needs to be addressed in upcoming realisations. 
A second limitation is inherent to the cosmological volume of \eagle. After visual classification to find the unambiguous SRs in \eagle, we are left with 479 galaxies. Although this sample is sufficient to provide us with the trends presented here, we often had to resort to studying the effect of a single quantity (e.g. stellar or halo mass), without controlling for others, making it difficult to disentangle (in some cases) the primary drivers of the trends above. Larger cosmological volumes, but retaining the sub-kpc resolution are required to address this limitation. 

From the simulation's perspective, the future is promising. The advent of large cosmological volumes ($\gtrsim 300$~cMpc) at high enough spatial resolution (sub-kpc) will open the way to much more thorough studies connecting SRs and their diverse kinematic classes to a large range of merger parameters (not only mass and gas ratio, but also orbital parameters) as well as stellar mass and environment. In addition, small volume, but much higher resolution simulations, as to resolve the cold interstellar medium, will allow a better understanding of the formation of thin, flat disk galaxies, as well as how instabilities and galaxy mergers can lead to the formation of early-type, fast rotator galaxies.

Observations also promise significant progress over the next years. The fact that the kinematic transformation experienced by SRs in \eagle\ happens at lookback times $\approx 2-6$~Gyrs implies that the upcoming MUSE survey Middle Ages Galaxy Properties with Integral Field Spectroscopy (MAGPI; \citealt{Foster20}) is ideally placed to unveil these transformations. The connection to $z=0$ surveys, such as SAMI, MaNGA and Hector \citep{Bryant16}, will complete this picture. We expect that in the next 3-5 years IFS surveys observations will be able to place stringent constraints on the epoch of kinematic transformation and the (lack of) connection to star formation quenching.

\section*{Data availability}

The {\sc EAGLE} simulations are publicly available; see \citet{McAlpine15,EAGLE17} for how to access {\sc EAGLE} data.

\section*{Acknowledgements}

CL, KEH and CF have received funding from the ARC Centre of
Excellence for All Sky Astrophysics in 3 Dimensions (ASTRO 3D), through project number CE170100013. EE thanks ASTRO 3D support during his visit to Australia. CL and CF are the receipts of an Australian Research Council Discovery Project (DP210101945) funded by the Australian Government.
CL also thanks the MERAC Foundation for a Postdoctoral Research Award.
JvdS acknowledges support of an Australian Research Council Discovery Early Career Research Award (project number DE200100461) funded by the Australian Government.
LC is the recipient of an Australian Research Council Future Fellowship (FT180100066) funded by the Australian Government.
This work used the DiRAC Data Centric system at Durham University, operated by the Institute for Computational Cosmology on behalf of the STFC DiRAC HPC Facility ({\tt www.dirac.ac.uk}). 
This equipment was funded by BIS National E-infrastructure capital grant ST/K00042X/1, STFC capital grant ST/H008519/1, and STFC DiRAC Operations grant ST/K003267/1 and Durham University. DiRAC is part of the National E-Infrastructure.
We acknowledge the Virgo Consortium for making
their simulation data available. The \eagle\ simulations were performed using the DiRAC-2 facility at
Durham, managed by the ICC, and the PRACE facility Curie based in France at TGCC, CEA, Bruyeres-le-Chatel.

\bibliographystyle{mn2e_trunc8}
\bibliography{SRs}

\begin{thebibliography}{21}
\expandafter\ifx\csname natexlab\endcsname\relax\def\natexlab#1{#1}\fi

\bibitem[{{Cappellari}(2016)}]{Cappellari16}
{Cappellari} M., 2016, \araa, 54, 597

\bibitem[{{Crain} {et~al.}(2015){Crain}, {Schaye}, {Bower}, {Furlong},
  {Schaller}, {Theuns}, {Dalla Vecchia}, {Frenk}, {McCarthy}, {Helly},
  {Jenkins}, {Rosas-Guevara}, {White}, \& {Trayford}}]{Crain15}
{Crain} R.~A., {Schaye} J., {Bower} R.~G., {Furlong} M., {Schaller} M.,
  {Theuns} T., {Dalla Vecchia} C., {Frenk} C.~S. {et~al}, 2015, \mnras, 450,
  1937

\bibitem[{{Dalla Vecchia} \& {Schaye}(2012)}]{DallaVecchia12}
{Dalla Vecchia} C., {Schaye} J., 2012, \mnras, 426, 140

\bibitem[{{Dolag} {et~al.}(2009){Dolag}, {Borgani}, {Murante}, \&
  {Springel}}]{Dolag09}
{Dolag} K., {Borgani} S., {Murante} G., {Springel} V., 2009, \mnras, 399, 497

\bibitem[{{Dutton} {et~al.}(2011){Dutton}, {Bosch}, {Faber}, {Simard},
  {Kassin}, {Koo}, {Bundy}, {Huang}, {Weiner}, {Cooper}, {Newman}, {Mozena}, \&
  {Koekemoer}}]{Dutton11}
{Dutton} A.~A., {Bosch} F.~C.~V.~D., {Faber} S.~M., {Simard} L., {Kassin}
  S.~A., {Koo} D.~C., {Bundy} K., {Huang} J. {et~al}, 2011, \mnras, 410, 1660

\bibitem[{{Jiang} {et~al.}(2014){Jiang}, {Helly}, {Cole}, \& {Frenk}}]{Jiang14}
{Jiang} L., {Helly} J.~C., {Cole} S., {Frenk} C.~S., 2014, \mnras, 440, 2115

\bibitem[{{Lagos} {et~al.}(2015){Lagos}, {Crain}, {Schaye}, {Furlong}, {Frenk},
  {Bower}, {Schaller}, {Theuns}, {Trayford}, {Bah{\'e}}, \& {Dalla
  Vecchia}}]{Lagos15}
{Lagos} C.~d.~P., {Crain} R.~A., {Schaye} J., {Furlong} M., {Frenk} C.~S.,
  {Bower} R.~G., {Schaller} M., {Theuns} T. {et~al}, 2015, \mnras, 452, 3815

\bibitem[{{Lagos} {et~al.}(2018){Lagos}, {Stevens}, {Bower}, {Davis},
  {Contreras}, {Padilla}, {Obreschkow}, {Croton}, {Trayford}, {Welker}, \&
  {Theuns}}]{Lagos17}
{Lagos} C.~d.~P., {Stevens} A.~R.~H., {Bower} R.~G., {Davis} T.~A., {Contreras}
  S., {Padilla} N.~D., {Obreschkow} D., {Croton} D. {et~al}, 2018, \mnras, 473,
  4956

\bibitem[{{Lagos} {et~al.}(2017){Lagos}, {Theuns}, {Stevens}, {Cortese},
  {Padilla}, {Davis}, {Contreras}, \& {Croton}}]{Lagos16b}
{Lagos} C.~d.~P., {Theuns} T., {Stevens} A.~R.~H., {Cortese} L., {Padilla}
  N.~D., {Davis} T.~A., {Contreras} S., {Croton} D., 2017, \mnras, 464, 3850

\bibitem[{{McAlpine} {et~al.}(2015){McAlpine}, {Helly}, {Schaller}, {Trayford},
  {Qu}, {Furlong}, {Bower}, {Crain}, {Schaye}, {Theuns}, {Dalla Vecchia},
  {Frenk}, {McCarthy}, {Jenkins}, {Rosas-Guevara}, {White}, {Baes}, {Camps}, \&
  {Lemson}}]{McAlpine15}
{McAlpine} S., {Helly} J.~C., {Schaller} M., {Trayford} J.~W., {Qu} Y.,
  {Furlong} M., {Bower} R.~G., {Crain} R.~A. {et~al}, 2015, ArXiv:1510.01320

\bibitem[{{Planck Collaboration}(2014)}]{Planck14}
{Planck Collaboration}, 2014, \aap, 571, A16

\bibitem[{{Qu} {et~al.}(2017){Qu}, {Helly}, {Bower}, {Theuns}, {Crain},
  {Frenk}, {Furlong}, {McAlpine}, {Schaller}, {Schaye}, \& {White}}]{Qu17}
{Qu} Y., {Helly} J.~C., {Bower} R.~G., {Theuns} T., {Crain} R.~A., {Frenk}
  C.~S., {Furlong} M., {McAlpine} S. {et~al}, 2017, \mnras, 464, 1659

\bibitem[{{Rahmati} {et~al.}(2013){Rahmati}, {Pawlik}, {Raicevic}, \&
  {Schaye}}]{Rahmati13}
{Rahmati} A., {Pawlik} A.~H., {Raicevic} M., {Schaye} J., 2013, \mnras, 430,
  2427

\bibitem[{{Rosas-Guevara} {et~al.}(2015){Rosas-Guevara}, {Bower}, {Schaye},
  {Furlong}, {Frenk}, {Booth}, {Crain}, {Dalla Vecchia}, {Schaller}, \&
  {Theuns}}]{Rosas-Guevara13}
{Rosas-Guevara} Y.~M., {Bower} R.~G., {Schaye} J., {Furlong} M., {Frenk} C.~S.,
  {Booth} C.~M., {Crain} R.~A., {Dalla Vecchia} C. {et~al}, 2015, \mnras, 454,
  1038

\bibitem[{{Schaller} {et~al.}(2015){Schaller}, {Dalla Vecchia}, {Schaye},
  {Bower}, {Theuns}, {Crain}, {Furlong}, \& {McCarthy}}]{Schaller15b}
{Schaller} M., {Dalla Vecchia} C., {Schaye} J., {Bower} R.~G., {Theuns} T.,
  {Crain} R.~A., {Furlong} M., {McCarthy} I.~G., 2015, \mnras, 454, 2277

\bibitem[{{Schaye} {et~al.}(2015){Schaye}, {Crain}, {Bower}, {Furlong},
  {Schaller}, {Theuns}, {Dalla Vecchia}, {Frenk}, {McCarthy}, {Helly},
  {Jenkins}, {Rosas-Guevara}, {White}, {Baes}, {Booth}, {Camps}, {Navarro},
  {Qu}, {Rahmati}, {Sawala}, {Thomas}, \& {Trayford}}]{Schaye14}
{Schaye} J., {Crain} R.~A., {Bower} R.~G., {Furlong} M., {Schaller} M.,
  {Theuns} T., {Dalla Vecchia} C., {Frenk} C.~S. {et~al}, 2015, \mnras, 446,
  521

\bibitem[{{Schaye} \& {Dalla Vecchia}(2008)}]{Schaye08}
{Schaye} J., {Dalla Vecchia} C., 2008, \mnras, 383, 1210

\bibitem[{{Springel}(2005)}]{Springel05b}
{Springel} V., 2005, \mnras, 364, 1105

\bibitem[{{Springel} {et~al.}(2001){Springel}, {White}, {Tormen}, \&
  {Kauffmann}}]{Springel01}
{Springel} V., {White} S.~D.~M., {Tormen} G., {Kauffmann} G., 2001, \mnras,
  328, 726

\bibitem[{{Wiersma} {et~al.}(2009{\natexlab{a}}){Wiersma}, {Schaye}, \&
  {Smith}}]{Wiersma09b}
{Wiersma} R.~P.~C., {Schaye} J., {Smith} B.~D., 2009{\natexlab{a}}, \mnras,
  393, 99

\bibitem[{{Wiersma} {et~al.}(2009{\natexlab{b}}){Wiersma}, {Schaye}, {Theuns},
  {Dalla Vecchia}, \& {Tornatore}}]{Wiersma09}
{Wiersma} R.~P.~C., {Schaye} J., {Theuns} T., {Dalla Vecchia} C., {Tornatore}
  L., 2009{\natexlab{b}}, \mnras, 399, 574

\end{thebibliography}


\begin{thebibliography}{107}
\expandafter\ifx\csname natexlab\endcsname\relax\def\natexlab#1{#1}\fi

\bibitem[{{Asplund} {et~al.}(2005){Asplund}, {Grevesse}, \&
  {Sauval}}]{Asplund05}
{Asplund} M., {Grevesse} N., {Sauval} A.~J., 2005, in Astronomical Society of
  the Pacific Conference Series, Vol. 336, Cosmic Abundances as Records of
  Stellar Evolution and Nucleosynthesis, {T.~G.~Barnes III \& F.~N.~Bash}, ed.,
  pp. 25--+

\bibitem[{{Bah{\'e}} {et~al.}(2017){Bah{\'e}}, {Schaye}, {Crain}, {McCarthy},
  {Bower}, {Theuns}, {McGee}, \& {Trayford}}]{Bahe17b}
{Bah{\'e}} Y.~M., {Schaye} J., {Crain} R.~A., {McCarthy} I.~G., {Bower} R.~G.,
  {Theuns} T., {McGee} S.~L., {Trayford} J.~W., 2017, \mnras, 464, 508

\bibitem[{{Barsanti} {et~al.}(2020){Barsanti}, {Owers}, {McDermid}, {Bekki},
  {Bland-Hawthorn}, {Brough}, {Bryant}, {Cortese}, {Croom}, {Foster},
  {Lawrence}, {L{\'o}pez-S{\'a}nchez}, {Oh}, {Robotham}, {Scott}, {Sweet}, \&
  {van de Sande}}]{Barsanti20}
{Barsanti} S., {Owers} M.~S., {McDermid} R.~M., {Bekki} K., {Bland-Hawthorn}
  J., {Brough} S., {Bryant} J.~J., {Cortese} L. {et~al}, 2020, arXiv e-prints,
  arXiv:2011.04873

\bibitem[{{Bernardi} {et~al.}(2019){Bernardi}, {Dom{\'\i}nguez S{\'a}nchez},
  {Brownstein}, {Drory}, \& {Sheth}}]{Bernardi19}
{Bernardi} M., {Dom{\'\i}nguez S{\'a}nchez} H., {Brownstein} J.~R., {Drory} N.,
  {Sheth} R.~K., 2019, \mnras, 489, 5633

\bibitem[{{B{\'e}thermin} {et~al.}(2015){B{\'e}thermin}, {Daddi}, {Magdis},
  {Lagos}, {Sargent}, {Albrecht}, {Aussel}, {Bertoldi}, {Buat}, {Galametz},
  {Heinis}, {Ilbert}, {Karim}, {Koekemoer}, {Lacey}, {Le Floc'h}, {Navarrete},
  {Pannella}, {Schreiber}, {Smol{\v c}i{\'c}}, {Symeonidis}, \&
  {Viero}}]{Bethermin14}
{B{\'e}thermin} M., {Daddi} E., {Magdis} G., {Lagos} C., {Sargent} M.,
  {Albrecht} M., {Aussel} H., {Bertoldi} F. {et~al}, 2015, \aap, 573, A113

\bibitem[{{Bois} {et~al.}(2011){Bois}, {Emsellem}, {Bournaud}, {Alatalo},
  {Blitz}, {Bureau}, {Cappellari}, {Davies}, {Davis}, {de Zeeuw}, {Duc},
  {Khochfar}, {Krajnovi{\'c}}, {Kuntschner}, {Lablanche}, {McDermid},
  {Morganti}, {Naab}, {Oosterloo}, {Sarzi}, {Scott}, {Serra}, {Weijmans}, \&
  {Young}}]{Bois11}
{Bois} M., {Emsellem} E., {Bournaud} F., {Alatalo} K., {Blitz} L., {Bureau} M.,
  {Cappellari} M., {Davies} R.~L. {et~al}, 2011, \mnras, 416, 1654

\bibitem[{{Bower} {et~al.}(2017){Bower}, {Schaye}, {Frenk}, {Theuns},
  {Schaller}, {Crain}, \& {McAlpine}}]{Bower17}
{Bower} R.~G., {Schaye} J., {Frenk} C.~S., {Theuns} T., {Schaller} M., {Crain}
  R.~A., {McAlpine} S., 2017, \mnras, 465, 32

\bibitem[{{Brough} {et~al.}(2017){Brough}, {van de Sande}, {Owers},
  {d'Eugenio}, {Sharp}, {Cortese}, {Scott}, {Croom}, {Bassett}, {Bekki},
  {Bryant}, {Davies}, {Drinkwater}, {Driver}, {Foster}, {Goldstein},
  {Lopez-Sanchez}, {Medling}, {Sweet}, {Taranu}, {Tonini}, {Yi}, {Goodwin},
  {Lawrence}, \& {Richards}}]{Brough17}
{Brough} S., {van de Sande} J., {Owers} M.~S., {d'Eugenio} F., {Sharp} R.,
  {Cortese} L., {Scott} N., {Croom} S.~M. {et~al}, 2017, ArXiv e-prints

\bibitem[{{Bruzual} \& {Charlot}(2003)}]{Bruzual03}
{Bruzual} G., {Charlot} S., 2003, \mnras, 344, 1000

\bibitem[{{Bryant} {et~al.}(2016){Bryant}, {Bland-Hawthorn}, {Lawrence},
  {Croom}, {Brown}, {Venkatesan}, {Gillingham}, {Zhelem}, {Content},
  {Saunders}, {Staszak}, {van de Sande}, {Couch}, {Leon-Saval}, {Tims},
  {McDermid}, \& {Schaefer}}]{Bryant16}
{Bryant} J.~J., {Bland-Hawthorn} J., {Lawrence} J., {Croom} S., {Brown} D.,
  {Venkatesan} S., {Gillingham} P.~R., {Zhelem} R., {Content} R., {Saunders}
  W., {Staszak} N.~F., {van de Sande} J., {Couch} W., {Leon-Saval} S., {Tims}
  J., {McDermid} R., {Schaefer} A., 2016, in Society of Photo-Optical
  Instrumentation Engineers (SPIE) Conference Series, Vol. 9908, Ground-based
  and Airborne Instrumentation for Astronomy VI, {Evans} C.~J., {Simard} L.,
  {Takami} H., eds., p. 99081F

\bibitem[{{Bryant} {et~al.}(2015){Bryant}, {Owers}, {Robotham}, {Croom},
  {Driver}, {Drinkwater}, {Lorente}, {Cortese}, {Scott}, {Colless}, {Schaefer},
  {Taylor}, {Konstantopoulos}, {Allen}, {Baldry}, {Barnes}, {Bauer},
  {Bland-Hawthorn}, {Bloom}, {Brooks}, {Brough}, {Cecil}, {Couch}, {Croton},
  {Davies}, {Ellis}, {Fogarty}, {Foster}, {Glazebrook}, {Goodwin}, {Green},
  {Gunawardhana}, {Hampton}, {Ho}, {Hopkins}, {Kewley}, {Lawrence},
  {Leon-Saval}, {Leslie}, {McElroy}, {Lewis}, {Liske}, {L{\'o}pez-S{\'a}nchez},
  {Mahajan}, {Medling}, {Metcalfe}, {Meyer}, {Mould}, {Obreschkow}, {O'Toole},
  {Pracy}, {Richards}, {Shanks}, {Sharp}, {Sweet}, {Thomas}, {Tonini}, \&
  {Walcher}}]{Bryant15}
{Bryant} J.~J., {Owers} M.~S., {Robotham} A.~S.~G., {Croom} S.~M., {Driver}
  S.~P., {Drinkwater} M.~J., {Lorente} N.~P.~F., {Cortese} L. {et~al}, 2015,
  \mnras, 447, 2857

\bibitem[{{Bundy} {et~al.}(2015){Bundy}, {Bershady}, {Law}, {Yan}, {Drory},
  {MacDonald}, {Wake}, {Cherinka}, {S{\'a}nchez-Gallego}, {Weijmans}, {Thomas},
  {Tremonti}, {Masters}, {Coccato}, {Diamond-Stanic}, {Arag{\'o}n-Salamanca},
  {Avila-Reese}, {Badenes}, {Falc{\'o}n-Barroso}, {Belfiore}, {Bizyaev},
  {Blanc}, {Bland-Hawthorn}, {Blanton}, {Brownstein}, {Byler}, {Cappellari},
  {Conroy}, {Dutton}, {Emsellem}, {Etherington}, {Frinchaboy}, {Fu}, {Gunn},
  {Harding}, {Johnston}, {Kauffmann}, {Kinemuchi}, {Klaene}, {Knapen},
  {Leauthaud}, {Li}, {Lin}, {Maiolino}, {Malanushenko}, {Malanushenko}, {Mao},
  {Maraston}, {McDermid}, {Merrifield}, {Nichol}, {Oravetz}, {Pan}, {Parejko},
  {Sanchez}, {Schlegel}, {Simmons}, {Steele}, {Steinmetz}, {Thanjavur},
  {Thompson}, {Tinker}, {van den Bosch}, {Westfall}, {Wilkinson}, {Wright},
  {Xiao}, \& {Zhang}}]{Bundy15}
{Bundy} K., {Bershady} M.~A., {Law} D.~R., {Yan} R., {Drory} N., {MacDonald}
  N., {Wake} D.~A., {Cherinka} B. {et~al}, 2015, \apj, 798, 7

\bibitem[{{Ca{\~n}as} {et~al.}(2019){Ca{\~n}as}, {Elahi}, {Welker}, {del P
  Lagos}, {Power}, {Dubois}, \& {Pichon}}]{Canas19}
{Ca{\~n}as} R., {Elahi} P.~J., {Welker} C., {del P Lagos} C., {Power} C.,
  {Dubois} Y., {Pichon} C., 2019, \mnras, 482, 2039

\bibitem[{{Cappellari}(2016)}]{Cappellari16}
{Cappellari} M., 2016, \araa, 54, 597

\bibitem[{{Cappellari} {et~al.}(2007){Cappellari}, {Emsellem}, {Bacon},
  {Bureau}, {Davies}, {de Zeeuw}, {Falc{\'o}n-Barroso}, {Krajnovi{\'c}},
  {Kuntschner}, {McDermid}, {Peletier}, {Sarzi}, {van den Bosch}, \& {van de
  Ven}}]{Cappellari07}
{Cappellari} M., {Emsellem} E., {Bacon} R., {Bureau} M., {Davies} R.~L., {de
  Zeeuw} P.~T., {Falc{\'o}n-Barroso} J., {Krajnovi{\'c}} D. {et~al}, 2007,
  \mnras, 379, 418

\bibitem[{{Cappellari} {et~al.}(2011){Cappellari}, {Emsellem}, {Krajnovi{\'c}},
  {McDermid}, {Scott}, {Verdoes Kleijn}, {Young}, {Alatalo}, {Bacon}, {Blitz},
  {Bois}, {Bournaud}, {Bureau}, {Davies}, {Davis}, {de Zeeuw}, {Duc},
  {Khochfar}, {Kuntschner}, {Lablanche}, {Morganti}, {Naab}, {Oosterloo},
  {Sarzi}, {Serra}, \& {Weijmans}}]{Cappellari11}
{Cappellari} M., {Emsellem} E., {Krajnovi{\'c}} D., {McDermid} R.~M., {Scott}
  N., {Verdoes Kleijn} G.~A., {Young} L.~M., {Alatalo} K. {et~al}, 2011,
  \mnras, 413, 813

\bibitem[{{Choi} \& {Yi}(2017{\natexlab{a}})}]{Choi17}
{Choi} H., {Yi} S.~K., 2017{\natexlab{a}}, \apj, 837, 68

\bibitem[{{Choi} \& {Yi}(2017{\natexlab{b}})}]{Choi17b}
---, 2017{\natexlab{b}}, \apj, 837, 68

\bibitem[{{Cole} {et~al.}(2020){Cole}, {Bezanson}, {van der Wel}, {Bell},
  {D'Eugenio}, {Franx}, {Gallazzi}, {van Houdt}, {Muzzin}, {Pacifici}, {van de
  Sande}, {Sobral}, {Straatman}, \& {Wu}}]{Cole20}
{Cole} J., {Bezanson} R., {van der Wel} A., {Bell} E., {D'Eugenio} F., {Franx}
  M., {Gallazzi} A., {van Houdt} J. {et~al}, 2020, \apjl, 890, L25

\bibitem[{{Collacchioni} {et~al.}(2020){Collacchioni}, {Lagos}, {Mitchell},
  {Schaye}, {Wisnioski}, {Cora}, \& {Correa}}]{Collacchioni20}
{Collacchioni} F., {Lagos} C. D.~P., {Mitchell} P.~D., {Schaye} J., {Wisnioski}
  E., {Cora} S.~A., {Correa} C.~A., 2020, \mnras, 495, 2827

\bibitem[{{Correa} \& {Schaye}(2020)}]{Correa20}
{Correa} C.~A., {Schaye} J., 2020, \mnras, 499, 3578

\bibitem[{{Correa} {et~al.}(2019){Correa}, {Schaye}, \& {Trayford}}]{Correa19}
{Correa} C.~A., {Schaye} J., {Trayford} J.~W., 2019, \mnras, 484, 4401

\bibitem[{{Cortese} {et~al.}(2019){Cortese}, {van de Sande}, {Lagos},
  {Catinella}, {Davies}, {Croom}, {Brough}, {Bryant}, {Lawrence}, {Owers},
  {Richards}, {Sweet}, \& {Bland -Hawthorn}}]{Cortese19}
{Cortese} L., {van de Sande} J., {Lagos} C.~P., {Catinella} B., {Davies}
  L.~J.~M., {Croom} S.~M., {Brough} S., {Bryant} J.~J. {et~al}, 2019, \mnras,
  485, 2656

\bibitem[{{Crain} {et~al.}(2015){Crain}, {Schaye}, {Bower}, {Furlong},
  {Schaller}, {Theuns}, {Dalla Vecchia}, {Frenk}, {McCarthy}, {Helly},
  {Jenkins}, {Rosas-Guevara}, {White}, \& {Trayford}}]{Crain15}
{Crain} R.~A., {Schaye} J., {Bower} R.~G., {Furlong} M., {Schaller} M.,
  {Theuns} T., {Dalla Vecchia} C., {Frenk} C.~S. {et~al}, 2015, \mnras, 450,
  1937

\bibitem[{{Croom} {et~al.}(2012){Croom}, {Lawrence}, {Bland-Hawthorn},
  {Bryant}, {Fogarty}, {Richards}, {Goodwin}, {Farrell}, {Miziarski}, {Heald},
  {Jones}, {Lee}, {Colless}, {Brough}, {Hopkins}, {Bauer}, {Birchall}, {Ellis},
  {Horton}, {Leon-Saval}, {Lewis}, {L{\'o}pez-S{\'a}nchez}, {Min}, {Trinh}, \&
  {Trowland}}]{Croom12}
{Croom} S.~M., {Lawrence} J.~S., {Bland-Hawthorn} J., {Bryant} J.~J., {Fogarty}
  L., {Richards} S., {Goodwin} M., {Farrell} T. {et~al}, 2012, \mnras, 421, 872

\bibitem[{{Dalla Vecchia} \& {Schaye}(2012)}]{DallaVecchia12}
{Dalla Vecchia} C., {Schaye} J., 2012, \mnras, 426, 140

\bibitem[{{Davies} {et~al.}(2018){Davies}, {Robotham}, {Driver}, {Lagos},
  {Cortese}, {Mannering}, {Foster}, {Lidman}, {Hashemizadeh}, {Koushan},
  {O'Toole}, {Baldry}, {Bilicki}, {Bland -Hawthorn}, {Bremer}, {Brown},
  {Bryant}, {Catinella}, {Croom}, {Grootes}, {Holwerda}, {Jarvis}, {Maddox},
  {Meyer}, {Moffett}, {Phillipps}, {Taylor}, {Windhorst}, \& {Wolf}}]{Davies18}
{Davies} L.~J.~M., {Robotham} A.~S.~G., {Driver} S.~P., {Lagos} C.~P.,
  {Cortese} L., {Mannering} E., {Foster} C., {Lidman} C. {et~al}, 2018, \mnras,
  480, 768

\bibitem[{{Davies} {et~al.}(2019){Davies}, {Robotham}, {Lagos}, {Driver},
  {Stevens}, {Bah{\'e}}, {Alpaslan}, {Bremer}, {Brown}, {Brough},
  {Bland-Hawthorn}, {Cortese}, {Elahi}, {Grootes}, {Holwerda}, {Ludlow},
  {McGee}, {Owers}, \& {Phillipps}}]{Davies19}
{Davies} L.~J.~M., {Robotham} A.~S.~G., {Lagos} C. d.~P., {Driver} S.~P.,
  {Stevens} A.~R.~H., {Bah{\'e}} Y.~M., {Alpaslan} M., {Bremer} M.~N. {et~al},
  2019, \mnras, 483, 5444

\bibitem[{{Davison} {et~al.}(2020){Davison}, {Norris}, {Pfeffer}, {Davies}, \&
  {Crain}}]{Davison20}
{Davison} T.~A., {Norris} M.~A., {Pfeffer} J.~L., {Davies} J.~J., {Crain}
  R.~A., 2020, \mnras, 497, 81

\bibitem[{{De Rossi} {et~al.}(2017){De Rossi}, {Bower}, {Font}, {Schaye}, \&
  {Theuns}}]{derossi17}
{De Rossi} M.~E., {Bower} R.~G., {Font} A.~S., {Schaye} J., {Theuns} T., 2017,
  \mnras, 472, 3354

\bibitem[{{Deeley} {et~al.}(2017){Deeley}, {Drinkwater}, {Cunnama},
  {Bland-Hawthorn}, {Brough}, {Cluver}, {Colless}, {Davies}, {Driver},
  {Foster}, {Grootes}, {Hopkins}, {Kafle}, {Lara-Lopez}, {Liske}, {Mahajan},
  {Phillipps}, {Power}, \& {Robotham}}]{Deeley17}
{Deeley} S., {Drinkwater} M.~J., {Cunnama} D., {Bland-Hawthorn} J., {Brough}
  S., {Cluver} M., {Colless} M., {Davies} L. J.~M. {et~al}, 2017, \mnras, 467,
  3934

\bibitem[{{Di Matteo} {et~al.}(2009){Di Matteo}, {Jog}, {Lehnert}, {Combes}, \&
  {Semelin}}]{DiMatteo09}
{Di Matteo} P., {Jog} C.~J., {Lehnert} M.~D., {Combes} F., {Semelin} B., 2009,
  \aap, 501, L9

\bibitem[{{Dolag} {et~al.}(2009){Dolag}, {Borgani}, {Murante}, \&
  {Springel}}]{Dolag09}
{Dolag} K., {Borgani} S., {Murante} G., {Springel} V., 2009, \mnras, 399, 497

\bibitem[{{Dressler}(1980)}]{Dressler80}
{Dressler} A., 1980, \apj, 236, 351

\bibitem[{{Ebrov{\'a}} {et~al.}(2020){Ebrov{\'a}}, {{\L}okas}, \&
  {Eli{\'a}{\v{s}}ek}}]{Ebrova20}
{Ebrov{\'a}} I., {{\L}okas} E.~L., {Eli{\'a}{\v{s}}ek} J., 2020, arXiv
  e-prints, arXiv:2010.02222

\bibitem[{{Emsellem} {et~al.}(2011){Emsellem}, {Cappellari}, {Krajnovi{\'c}},
  {Alatalo}, {Blitz}, {Bois}, {Bournaud}, {Bureau}, {Davies}, {Davis}, {de
  Zeeuw}, {Khochfar}, {Kuntschner}, {Lablanche}, {McDermid}, {Morganti},
  {Naab}, {Oosterloo}, {Sarzi}, {Scott}, {Serra}, {van de Ven}, {Weijmans}, \&
  {Young}}]{Emsellem11}
{Emsellem} E., {Cappellari} M., {Krajnovi{\'c}} D., {Alatalo} K., {Blitz} L.,
  {Bois} M., {Bournaud} F., {Bureau} M. {et~al}, 2011, \mnras, 414, 888

\bibitem[{{Emsellem} {et~al.}(2007){Emsellem}, {Cappellari}, {Krajnovi{\'c}},
  {van de Ven}, {Bacon}, {Bureau}, {Davies}, {de Zeeuw}, {Falc{\'o}n-Barroso},
  {Kuntschner}, {McDermid}, {Peletier}, \& {Sarzi}}]{Emsellem07}
{Emsellem} E., {Cappellari} M., {Krajnovi{\'c}} D., {van de Ven} G., {Bacon}
  R., {Bureau} M., {Davies} R.~L., {de Zeeuw} P.~T. {et~al}, 2007, \mnras, 379,
  401

\bibitem[{{Falc{\'o}n-Barroso} {et~al.}(2017){Falc{\'o}n-Barroso}, {Lyubenova},
  {van de Ven}, {Mendez-Abreu}, {Aguerri}, {Garc{\'\i}a-Lorenzo},
  {Bekerait{\'e}}, {S{\'a}nchez}, {Husemann}, {Garc{\'\i}a-Benito}, {Mast},
  {Walcher}, {Zibetti}, {Barrera-Ballesteros}, {Galbany},
  {S{\'a}nchez-Bl{\'a}zquez}, {Singh}, {van den Bosch}, {Wild}, {Zhu},
  {Bland-Hawthorn}, {Cid Fernandes}, {de Lorenzo-C{\'a}ceres}, {Gallazzi},
  {Gonz{\'a}lez Delgado}, {Marino}, {M{\'a}rquez}, {P{\'e}rez}, {P{\'e}rez},
  {Roth}, {Rosales-Ortega}, {Ruiz-Lara}, {Wisotzki}, {Ziegler}, \& {Califa
  Collaboration}}]{Falcon-Barroso17}
{Falc{\'o}n-Barroso} J., {Lyubenova} M., {van de Ven} G., {Mendez-Abreu} J.,
  {Aguerri} J.~A.~L., {Garc{\'\i}a-Lorenzo} B., {Bekerait{\'e}} S.,
  {S{\'a}nchez} S.~F. {et~al}, 2017, \aap, 597, A48

\bibitem[{{Foster} {et~al.}(2020){Foster}, {Mendel}, {Lagos}, {Wisnioski},
  {Yuan}, {D'Eugenio}, {Barone}, {Harborne}, {Vaughan}, {Schulze}, {Remus},
  {Gupta}, {Collacchioni}, {Khim}, {Taylor}, {Bassett}, {Croom}, {McDermid},
  {Poci}, {Battisti}, {Bland-Hawthorn}, {Bellstedt}, {Colless}, {Davies},
  {Driver}, {Ferr{\'e}-Mateu}, {Fisher}, {Gjergo}, {Johnston}, {Khalid},
  {Kobayashi}, {Oh}, {Peng}, {Robotham}, {Sweet}, {Taylor}, {Tran}, {Trayford},
  {van de Sande}, {Yi}, \& {Zanisi}}]{Foster20}
{Foster} C., {Mendel} J.~T., {Lagos} C.~D.~P., {Wisnioski} E., {Yuan} T.,
  {D'Eugenio} F., {Barone} T.~M., {Harborne} K.~E. {et~al}, 2020, arXiv
  e-prints, arXiv:2011.13567

\bibitem[{{Foster} {et~al.}(2017){Foster}, {van de Sande}, {D'Eugenio},
  {Cortese}, {McDermid}, {Bland-Hawthorn}, {Brough}, {Bryant}, {Croom},
  {Goodwin}, {Konstantopoulos}, {Lawrence}, {L{\'o}pez-S{\'a}nchez}, {Medling},
  {Owers}, {Richards}, {Scott}, {Taranu}, {Tonini}, \& {Zafar}}]{Foster17}
{Foster} C., {van de Sande} J., {D'Eugenio} F., {Cortese} L., {McDermid} R.~M.,
  {Bland-Hawthorn} J., {Brough} S., {Bryant} J. {et~al}, 2017, \mnras, 472, 966

\bibitem[{{Furlong} {et~al.}(2017){Furlong}, {Bower}, {Crain}, {Schaye},
  {Theuns}, {Trayford}, {Qu}, {Schaller}, {Berthet}, \& {Helly}}]{Furlong15}
{Furlong} M., {Bower} R.~G., {Crain} R.~A., {Schaye} J., {Theuns} T.,
  {Trayford} J.~W., {Qu} Y., {Schaller} M. {et~al}, 2017, \mnras, 465, 722

\bibitem[{{Furlong} {et~al.}(2015){Furlong}, {Bower}, {Theuns}, {Schaye},
  {Crain}, {Schaller}, {Dalla Vecchia}, {Frenk}, {McCarthy}, {Helly},
  {Jenkins}, \& {Rosas-Guevara}}]{Furlong14}
{Furlong} M., {Bower} R.~G., {Theuns} T., {Schaye} J., {Crain} R.~A.,
  {Schaller} M., {Dalla Vecchia} C., {Frenk} C.~S. {et~al}, 2015, \mnras, 450,
  4486

\bibitem[{{Gonz{\'a}lez Delgado} {et~al.}(2015){Gonz{\'a}lez Delgado},
  {Garc{\'\i}a-Benito}, {P{\'e}rez}, {Cid Fernandes}, {de Amorim},
  {Cortijo-Ferrero}, {Lacerda}, {L{\'o}pez Fern{\'a}ndez}, {Vale-Asari},
  {S{\'a}nchez}, {Moll{\'a}}, {Ruiz-Lara}, {S{\'a}nchez-Bl{\'a}zquez},
  {Walcher}, {Alves}, {Aguerri}, {Bekerait{\'e}}, {Bland-Hawthorn}, {Galbany},
  {Gallazzi}, {Husemann}, {Iglesias-P{\'a}ramo}, {Kalinova},
  {L{\'o}pez-S{\'a}nchez}, {Marino}, {M{\'a}rquez}, {Masegosa}, {Mast},
  {M{\'e}ndez-Abreu}, {Mendoza}, {del Olmo}, {P{\'e}rez}, {Quirrenbach}, \&
  {Zibetti}}]{Gonzalez-Delgado15}
{Gonz{\'a}lez Delgado} R.~M., {Garc{\'\i}a-Benito} R., {P{\'e}rez} E., {Cid
  Fernandes} R., {de Amorim} A.~L., {Cortijo-Ferrero} C., {Lacerda} E.~A.~D.,
  {L{\'o}pez Fern{\'a}ndez} R. {et~al}, 2015, \aap, 581, A103

\bibitem[{{Graham} {et~al.}(2019){Graham}, {Cappellari}, {Bershady}, \&
  {Drory}}]{Graham19}
{Graham} M.~T., {Cappellari} M., {Bershady} M.~A., {Drory} N., 2019, arXiv
  e-prints, arXiv:1910.05139

\bibitem[{{Graham} {et~al.}(2018){Graham}, {Cappellari}, {Li}, {Mao},
  {Bershady}, {Bizyaev}, {Brinkmann}, {Brownstein}, {Bundy}, {Drory}, {Law},
  {Pan}, {Thomas}, {Wake}, {Weijmans}, {Westfall}, \& {Yan}}]{Graham18}
{Graham} M.~T., {Cappellari} M., {Li} H., {Mao} S., {Bershady} M.~A., {Bizyaev}
  D., {Brinkmann} J., {Brownstein} J.~R. {et~al}, 2018, \mnras, 477, 4711

\bibitem[{{Green} {et~al.}(2018){Green}, {Croom}, {Scott}, {Cortese},
  {Medling}, {D'Eugenio}, {Bryant}, {Bland-Hawthorn}, {Allen}, {Sharp}, {Ho},
  {Groves}, {Drinkwater}, {Mannering}, {Harischand ra}, {van de Sande},
  {Thomas}, {O'Toole}, {McDermid}, {Vuong}, {Sealey}, {Bauer}, {Brough},
  {Catinella}, {Cecil}, {Colless}, {Couch}, {Driver}, {Federrath}, {Foster},
  {Goodwin}, {Hampton}, {Hopkins}, {Jones}, {Konstantopoulos}, {Lawrence},
  {Leon-Saval}, {Liske}, {L{\'o}pez-S{\'a}nchez}, {Lorente}, {Mould},
  {Obreschkow}, {Owers}, {Richards}, {Robotham}, {Schaefer}, {Sweet}, {Taranu},
  {Tescari}, {Tonini}, \& {Zafar}}]{Green18}
{Green} A.~W., {Croom} S.~M., {Scott} N., {Cortese} L., {Medling} A.~M.,
  {D'Eugenio} F., {Bryant} J.~J., {Bland-Hawthorn} J. {et~al}, 2018, \mnras,
  475, 716

\bibitem[{{Greene} {et~al.}(2015){Greene}, {Janish}, {Ma}, {McConnell},
  {Blakeslee}, {Thomas}, \& {Murphy}}]{Greene15}
{Greene} J.~E., {Janish} R., {Ma} C.-P., {McConnell} N.~J., {Blakeslee} J.~P.,
  {Thomas} J., {Murphy} J.~D., 2015, \apj, 807, 11

\bibitem[{{Greene} {et~al.}(2017){Greene}, {Leauthaud}, {Emsellem}, {Ge},
  {Arag'on-Salamanca}, {Greco}, {Lin}, {Mao}, {Masters}, {Merrifield}, {More},
  {Okabe}, {Schneider}, {Thomas}, {Wake}, {Pan}, {Bizyaev}, {Oravetz},
  {Simmons}, \& {Yan}}]{Greene17}
{Greene} J.~E., {Leauthaud} A., {Emsellem} E., {Ge} J., {Arag'on-Salamanca} A.,
  {Greco} J.~P., {Lin} Y.-T., {Mao} S. {et~al}, 2017, ArXiv e-prints

\bibitem[{{Harborne} {et~al.}(2020{\natexlab{a}}){Harborne}, {Power}, \&
  {Robotham}}]{Harborne20a}
{Harborne} K.~E., {Power} C., {Robotham} A. S.~G., 2020{\natexlab{a}}, \pasa,
  37, e016

\bibitem[{{Harborne} {et~al.}(2020{\natexlab{b}}){Harborne}, {van de Sande},
  {Cortese}, {Power}, {Robotham}, {Lagos}, \& {Croom}}]{Harborne20b}
{Harborne} K.~E., {van de Sande} J., {Cortese} L., {Power} C., {Robotham}
  A.~S.~G., {Lagos} C.~D.~P., {Croom} S., 2020{\natexlab{b}}, \mnras, 497, 2018

\bibitem[{{Jesseit} {et~al.}(2009){Jesseit}, {Cappellari}, {Naab}, {Emsellem},
  \& {Burkert}}]{Jesseit09}
{Jesseit} R., {Cappellari} M., {Naab} T., {Emsellem} E., {Burkert} A., 2009,
  \mnras, 397, 1202

\bibitem[{{Jiang} {et~al.}(2014){Jiang}, {Helly}, {Cole}, \& {Frenk}}]{Jiang14}
{Jiang} L., {Helly} J.~C., {Cole} S., {Frenk} C.~S., 2014, \mnras, 440, 2115

\bibitem[{{Karademir} {et~al.}(2019){Karademir}, {Remus}, {Burkert}, {Dolag},
  {Hoffmann}, {Moster}, {Steinwandel}, \& {Zhang}}]{Karademir19}
{Karademir} G.~S., {Remus} R.-S., {Burkert} A., {Dolag} K., {Hoffmann} T.~L.,
  {Moster} B.~P., {Steinwandel} U.~P., {Zhang} J., 2019, \mnras, 487, 318

\bibitem[{{Katsianis} {et~al.}(2017){Katsianis}, {Blanc}, {Lagos}, {Tejos},
  {Bower}, {Alavi}, {Gonzalez}, {Theuns}, {Schaller}, \& {Lopez}}]{Katsianis17}
{Katsianis} A., {Blanc} G., {Lagos} C.~P., {Tejos} N., {Bower} R.~G., {Alavi}
  A., {Gonzalez} V., {Theuns} T. {et~al}, 2017, \mnras, 472, 919

\bibitem[{{Krajnovi{\'c}} {et~al.}(2018){Krajnovi{\'c}}, {Emsellem}, {den
  Brok}, {Marino}, {Schmidt}, {Steinmetz}, \& {Weilbacher}}]{Krajnovic18}
{Krajnovi{\'c}} D., {Emsellem} E., {den Brok} M., {Marino} R.~A., {Schmidt}
  K.~B., {Steinmetz} M., {Weilbacher} P.~M., 2018, \mnras, 477, 5327

\bibitem[{{Krajnovi{\'c}} {et~al.}(2020){Krajnovi{\'c}}, {Ural}, {Kuntschner},
  {Goudfrooij}, {Wolfe}, {Cappellari}, {Davies}, {de Zeeuw}, {Duc}, {Emsellem},
  {Karick}, {McDermid}, {Mei}, \& {Naab}}]{Krajnovic20}
{Krajnovi{\'c}} D., {Ural} U., {Kuntschner} H., {Goudfrooij} P., {Wolfe} M.,
  {Cappellari} M., {Davies} R., {de Zeeuw} T.~P. {et~al}, 2020, \aap, 635, A129

\bibitem[{{Kuntschner} {et~al.}(2010){Kuntschner}, {Emsellem}, {Bacon},
  {Cappellari}, {Davies}, {de Zeeuw}, {Falc{\'o}n-Barroso}, {Krajnovi{\'c}},
  {McDermid}, {Peletier}, {Sarzi}, {Shapiro}, {van den Bosch}, \& {van de
  Ven}}]{Kuntschner10}
{Kuntschner} H., {Emsellem} E., {Bacon} R., {Cappellari} M., {Davies} R.~L.,
  {de Zeeuw} P.~T., {Falc{\'o}n-Barroso} J., {Krajnovi{\'c}} D. {et~al}, 2010,
  \mnras, 408, 97

\bibitem[{{Lagos} {et~al.}(2015){Lagos}, {Crain}, {Schaye}, {Furlong}, {Frenk},
  {Bower}, {Schaller}, {Theuns}, {Trayford}, {Bah{\'e}}, \& {Dalla
  Vecchia}}]{Lagos15}
{Lagos} C.~d.~P., {Crain} R.~A., {Schaye} J., {Furlong} M., {Frenk} C.~S.,
  {Bower} R.~G., {Schaller} M., {Theuns} T. {et~al}, 2015, \mnras, 452, 3815

\bibitem[{{Lagos} {et~al.}(2018{\natexlab{a}}){Lagos}, {Schaye}, {Bah{\'e}},
  {Van de Sande}, {Kay}, {Barnes}, {Davis}, \& {Dalla Vecchia}}]{Lagos18b}
{Lagos} C. d.~P., {Schaye} J., {Bah{\'e}} Y., {Van de Sande} J., {Kay} S.~T.,
  {Barnes} D., {Davis} T.~A., {Dalla Vecchia} C., 2018{\natexlab{a}}, \mnras,
  476, 4327

\bibitem[{{Lagos} {et~al.}(2018{\natexlab{b}}){Lagos}, {Stevens}, {Bower},
  {Davis}, {Contreras}, {Padilla}, {Obreschkow}, {Croton}, {Trayford},
  {Welker}, \& {Theuns}}]{Lagos18a}
{Lagos} C. d.~P., {Stevens} A. R.~H., {Bower} R.~G., {Davis} T.~A., {Contreras}
  S., {Padilla} N.~D., {Obreschkow} D., {Croton} D. {et~al},
  2018{\natexlab{b}}, \mnras, 473, 4956

\bibitem[{{Lagos} {et~al.}(2017){Lagos}, {Theuns}, {Stevens}, {Cortese},
  {Padilla}, {Davis}, {Contreras}, \& {Croton}}]{Lagos16b}
{Lagos} C.~d.~P., {Theuns} T., {Stevens} A.~R.~H., {Cortese} L., {Padilla}
  N.~D., {Davis} T.~A., {Contreras} S., {Croton} D., 2017, \mnras, 464, 3850

\bibitem[{{Lange} {et~al.}(2016){Lange}, {Moffett}, {Driver}, {Robotham},
  {Lagos}, {Kelvin}, {Conselice}, {Margalef-Bentabol}, {Alpaslan}, {Baldry},
  {Bland-Hawthorn}, {Bremer}, {Brough}, {Cluver}, {Colless}, {Davies},
  {H{\"a}u{\ss}ler}, {Holwerda}, {Hopkins}, {Kafle}, {Kennedy}, {Liske},
  {Phillipps}, {Popescu}, {Taylor}, {Tuffs}, {van Kampen}, \&
  {Wright}}]{Lange16}
{Lange} R., {Moffett} A.~J., {Driver} S.~P., {Robotham} A. S.~G., {Lagos} C.
  d.~P., {Kelvin} L.~S., {Conselice} C., {Margalef-Bentabol} B. {et~al}, 2016,
  \mnras, 462, 1470

\bibitem[{{Li} {et~al.}(2018{\natexlab{a}}){Li}, {Mao}, {Cappellari}, {Graham},
  {Emsellem}, \& {Long}}]{Li18c}
{Li} H., {Mao} S., {Cappellari} M., {Graham} M.~T., {Emsellem} E., {Long}
  R.~J., 2018{\natexlab{a}}, \apjl, 863, L19

\bibitem[{{Li} {et~al.}(2018{\natexlab{b}}){Li}, {Mao}, {Emsellem}, {Xu},
  {Springel}, \& {Krajnovi{\'c}}}]{Li18}
{Li} H., {Mao} S., {Emsellem} E., {Xu} D., {Springel} V., {Krajnovi{\'c}} D.,
  2018{\natexlab{b}}, \mnras, 473, 1489

\bibitem[{{Ma} {et~al.}(2014){Ma}, {Greene}, {McConnell}, {Janish},
  {Blakeslee}, {Thomas}, \& {Murphy}}]{Ma14}
{Ma} C.-P., {Greene} J.~E., {McConnell} N., {Janish} R., {Blakeslee} J.~P.,
  {Thomas} J., {Murphy} J.~D., 2014, \apj, 795, 158

\bibitem[{{Marasco} {et~al.}(2016){Marasco}, {Crain}, {Schaye}, {Bah{\'e}},
  {van der Hulst}, {Theuns}, \& {Bower}}]{Marasco16}
{Marasco} A., {Crain} R.~A., {Schaye} J., {Bah{\'e}} Y.~M., {van der Hulst} T.,
  {Theuns} T., {Bower} R.~G., 2016, \mnras, 461, 2630

\bibitem[{{McAlpine} {et~al.}(2015){McAlpine}, {Helly}, {Schaller}, {Trayford},
  {Qu}, {Furlong}, {Bower}, {Crain}, {Schaye}, {Theuns}, {Dalla Vecchia},
  {Frenk}, {McCarthy}, {Jenkins}, {Rosas-Guevara}, {White}, {Baes}, {Camps}, \&
  {Lemson}}]{McAlpine15}
{McAlpine} S., {Helly} J.~C., {Schaller} M., {Trayford} J.~W., {Qu} Y.,
  {Furlong} M., {Bower} R.~G., {Crain} R.~A. {et~al}, 2015, ArXiv:1510.01320

\bibitem[{{Naab} {et~al.}(2014){Naab}, {Oser}, {Emsellem}, {Cappellari},
  {Krajnovi{\'c}}, {McDermid}, {Alatalo}, {Bayet}, {Blitz}, {Bois}, {Bournaud},
  {Bureau}, {Crocker}, {Davies}, {Davis}, {de Zeeuw}, {Duc}, {Hirschmann},
  {Johansson}, {Khochfar}, {Kuntschner}, {Morganti}, {Oosterloo}, {Sarzi},
  {Scott}, {Serra}, {Ven}, {Weijmans}, \& {Young}}]{Naab14}
{Naab} T., {Oser} L., {Emsellem} E., {Cappellari} M., {Krajnovi{\'c}} D.,
  {McDermid} R.~M., {Alatalo} K., {Bayet} E. {et~al}, 2014, \mnras, 444, 3357

\bibitem[{{Peng} {et~al.}(2010){Peng}, {Lilly}, {Kova{\v c}}, {Bolzonella},
  {Pozzetti}, {Renzini}, {Zamorani}, {Ilbert}, {Knobel}, {Iovino}, {Maier},
  {Cucciati}, {Tasca}, {Carollo}, {Silverman}, {Kampczyk}, {de Ravel},
  {Sanders}, {Scoville}, {Contini}, {Mainieri}, {Scodeggio}, {Kneib}, {Le
  F{\`e}vre}, {Bardelli}, {Bongiorno}, {Caputi}, {Coppa}, {de la Torre},
  {Franzetti}, {Garilli}, {Lamareille}, {Le Borgne}, {Le Brun}, {Mignoli},
  {Perez Montero}, {Pello}, {Ricciardelli}, {Tanaka}, {Tresse}, {Vergani},
  {Welikala}, {Zucca}, {Oesch}, {Abbas}, {Barnes}, {Bordoloi}, {Bottini},
  {Cappi}, {Cassata}, {Cimatti}, {Fumana}, {Hasinger}, {Koekemoer},
  {Leauthaud}, {Maccagni}, {Marinoni}, {McCracken}, {Memeo}, {Meneux}, {Nair},
  {Porciani}, {Presotto}, \& {Scaramella}}]{Peng10}
{Peng} Y.-j., {Lilly} S.~J., {Kova{\v c}} K., {Bolzonella} M., {Pozzetti} L.,
  {Renzini} A., {Zamorani} G., {Ilbert} O. {et~al}, 2010, \apj, 721, 193

\bibitem[{{Penoyre} {et~al.}(2017){Penoyre}, {Moster}, {Sijacki}, \&
  {Genel}}]{Penoyre17}
{Penoyre} Z., {Moster} B.~P., {Sijacki} D., {Genel} S., 2017, \mnras, 468, 3883

\bibitem[{{Pillepich} {et~al.}(2018){Pillepich}, {Springel}, {Nelson}, {Genel},
  {Naiman}, {Pakmor}, {Hernquist}, {Torrey}, {Vogelsberger}, {Weinberger}, \&
  {Marinacci}}]{Pillepich17}
{Pillepich} A., {Springel} V., {Nelson} D., {Genel} S., {Naiman} J., {Pakmor}
  R., {Hernquist} L., {Torrey} P. {et~al}, 2018, \mnras, 473, 4077

\bibitem[{{Planck Collaboration}(2014)}]{Planck14}
{Planck Collaboration}, 2014, \aap, 571, A16

\bibitem[{{Pulsoni} {et~al.}(2020){Pulsoni}, {Gerhard}, {Arnaboldi},
  {Pillepich}, {Nelson}, {Hernquist}, \& {Springel}}]{Pulsoni20}
{Pulsoni} C., {Gerhard} O., {Arnaboldi} M., {Pillepich} A., {Nelson} D.,
  {Hernquist} L., {Springel} V., 2020, \aap, 641, A60

\bibitem[{{Qu} {et~al.}(2017){Qu}, {Helly}, {Bower}, {Theuns}, {Crain},
  {Frenk}, {Furlong}, {McAlpine}, {Schaller}, {Schaye}, \& {White}}]{Qu17}
{Qu} Y., {Helly} J.~C., {Bower} R.~G., {Theuns} T., {Crain} R.~A., {Frenk}
  C.~S., {Furlong} M., {McAlpine} S. {et~al}, 2017, \mnras, 464, 1659

\bibitem[{{Rahmati} {et~al.}(2013){Rahmati}, {Pawlik}, {Raicevic}, \&
  {Schaye}}]{Rahmati13}
{Rahmati} A., {Pawlik} A.~H., {Raicevic} M., {Schaye} J., 2013, \mnras, 430,
  2427

\bibitem[{{Robotham} {et~al.}(2020){Robotham}, {Bellstedt}, {Lagos}, {Thorne},
  {Davies}, {Driver}, \& {Bravo}}]{Robotham20}
{Robotham} A.~S.~G., {Bellstedt} S., {Lagos} C. d.~P., {Thorne} J.~E., {Davies}
  L.~J., {Driver} S.~P., {Bravo} M., 2020, \mnras, 495, 905

\bibitem[{{Rosas-Guevara} {et~al.}(2015){Rosas-Guevara}, {Bower}, {Schaye},
  {Furlong}, {Frenk}, {Booth}, {Crain}, {Dalla Vecchia}, {Schaller}, \&
  {Theuns}}]{Rosas-Guevara13}
{Rosas-Guevara} Y.~M., {Bower} R.~G., {Schaye} J., {Furlong} M., {Frenk} C.~S.,
  {Booth} C.~M., {Crain} R.~A., {Dalla Vecchia} C. {et~al}, 2015, \mnras, 454,
  1038

\bibitem[{{Rosito} {et~al.}(2019){Rosito}, {Tissera}, {Pedrosa}, \&
  {Lagos}}]{Rosito19}
{Rosito} M.~S., {Tissera} P.~B., {Pedrosa} S.~E., {Lagos} C.~D.~P., 2019, \aap,
  629, L3

\bibitem[{{S{\'a}nchez} {et~al.}(2012){S{\'a}nchez}, {Kennicutt}, {Gil de Paz},
  {van de Ven}, {V{\'{\i}}lchez}, {Wisotzki}, {Walcher}, {Mast}, {Aguerri},
  {Albiol-P{\'e}rez}, {Alonso-Herrero}, {Alves}, {Bakos}, {Bart{\'a}kov{\'a}},
  {Bland-Hawthorn}, {Boselli}, {Bomans}, {Castillo-Morales}, {Cortijo-Ferrero},
  {de Lorenzo-C{\'a}ceres}, {Del Olmo}, {Dettmar}, {D{\'{\i}}az}, {Ellis},
  {Falc{\'o}n-Barroso}, {Flores}, {Gallazzi}, {Garc{\'{\i}}a-Lorenzo},
  {Gonz{\'a}lez Delgado}, {Gruel}, {Haines}, {Hao}, {Husemann},
  {Igl{\'e}sias-P{\'a}ramo}, {Jahnke}, {Johnson}, {Jungwiert}, {Kalinova},
  {Kehrig}, {Kupko}, {L{\'o}pez-S{\'a}nchez}, {Lyubenova}, {Marino},
  {M{\'a}rmol-Queralt{\'o}}, {M{\'a}rquez}, {Masegosa}, {Meidt},
  {Mendez-Abreu}, {Monreal-Ibero}, {Montijo}, {Mour{\~a}o}, {Palacios-Navarro},
  {Papaderos}, {Pasquali}, {Peletier}, {P{\'e}rez}, {P{\'e}rez}, {Quirrenbach},
  {Rela{\~n}o}, {Rosales-Ortega}, {Roth}, {Ruiz-Lara},
  {S{\'a}nchez-Bl{\'a}zquez}, {Sengupta}, {Singh}, {Stanishev}, {Trager},
  {Vazdekis}, {Viironen}, {Wild}, {Zibetti}, \& {Ziegler}}]{Sanchez12}
{S{\'a}nchez} S.~F., {Kennicutt} R.~C., {Gil de Paz} A., {van de Ven} G.,
  {V{\'{\i}}lchez} J.~M., {Wisotzki} L., {Walcher} C.~J., {Mast} D. {et~al},
  2012, \aap, 538, A8

\bibitem[{{Santucci} {et~al.}(2020){Santucci}, {Brough}, {Scott}, {Montes},
  {Owers}, {van Sande}, {Bland-Hawthorn}, {Bryant}, {Croom}, {Ferreras},
  {Lawrence}, {L{\'o}pez-S{\'a}nchez}, \& {Richards}}]{Santucci20}
{Santucci} G., {Brough} S., {Scott} N., {Montes} M., {Owers} M.~S., {van Sande}
  J., {Bland-Hawthorn} J., {Bryant} J.~J. {et~al}, 2020, \apj, 896, 75

\bibitem[{{Schaye} {et~al.}(2015){Schaye}, {Crain}, {Bower}, {Furlong},
  {Schaller}, {Theuns}, {Dalla Vecchia}, {Frenk}, {McCarthy}, {Helly},
  {Jenkins}, {Rosas-Guevara}, {White}, {Baes}, {Booth}, {Camps}, {Navarro},
  {Qu}, {Rahmati}, {Sawala}, {Thomas}, \& {Trayford}}]{Schaye14}
{Schaye} J., {Crain} R.~A., {Bower} R.~G., {Furlong} M., {Schaller} M.,
  {Theuns} T., {Dalla Vecchia} C., {Frenk} C.~S. {et~al}, 2015, \mnras, 446,
  521

\bibitem[{{Schaye} \& {Dalla Vecchia}(2008)}]{Schaye08}
{Schaye} J., {Dalla Vecchia} C., 2008, \mnras, 383, 1210

\bibitem[{{Schulze} {et~al.}(2020){Schulze}, {Remus}, {Dolag}, {Bellstedt},
  {Burkert}, \& {Forbes}}]{Schulze20}
{Schulze} F., {Remus} R.-S., {Dolag} K., {Bellstedt} S., {Burkert} A., {Forbes}
  D.~A., 2020, \mnras, 493, 3778

\bibitem[{{Schulze} {et~al.}(2018){Schulze}, {Remus}, {Dolag}, {Burkert},
  {Emsellem}, \& {van de Ven}}]{Schulze18}
{Schulze} F., {Remus} R.-S., {Dolag} K., {Burkert} A., {Emsellem} E., {van de
  Ven} G., 2018, \mnras, 480, 4636

\bibitem[{{Scott} {et~al.}(2018){Scott}, {van de Sande}, {Croom}, {Groves},
  {Owers}, {Poetrodjojo}, {D'Eugenio}, {Medling}, {Barat}, {Barone},
  {Bland-Hawthorn}, {Brough}, {Bryant}, {Cortese}, {Foster}, {Green}, {Oh},
  {Colless}, {Drinkwater}, {Driver}, {Goodwin}, {Gunawardhana}, {Federrath},
  {Harischand ra}, {Jin}, {Lawrence}, {Lorente}, {Mannering}, {O'Toole},
  {Richards}, {Sanchez}, {Schaefer}, {Sealey}, {Sharp}, {Sweet}, {Taranu}, \&
  {Varidel}}]{Scott18}
{Scott} N., {van de Sande} J., {Croom} S.~M., {Groves} B., {Owers} M.~S.,
  {Poetrodjojo} H., {D'Eugenio} F., {Medling} A.~M. {et~al}, 2018, \mnras, 481,
  2299

\bibitem[{{Segers} {et~al.}(2016){Segers}, {Schaye}, {Bower}, {Crain},
  {Schaller}, \& {Theuns}}]{Segers16}
{Segers} M.~C., {Schaye} J., {Bower} R.~G., {Crain} R.~A., {Schaller} M.,
  {Theuns} T., 2016, \mnras, 461, L102

\bibitem[{{Smethurst} {et~al.}(2018){Smethurst}, {Masters}, {Lintott},
  {Weijmans}, {Merrifield}, {Penny}, {Arag{\'o}n-Salamanca}, {Brownstein},
  {Bundy}, {Drory}, {Law}, \& {Nichol}}]{Smethurst18}
{Smethurst} R.~J., {Masters} K.~L., {Lintott} C.~J., {Weijmans} A.,
  {Merrifield} M., {Penny} S.~J., {Arag{\'o}n-Salamanca} A., {Brownstein} J.
  {et~al}, 2018, \mnras, 473, 2679

\bibitem[{{Sparre} \& {Springel}(2016)}]{Sparre16}
{Sparre} M., {Springel} V., 2016, \mnras, 462, 2418

\bibitem[{{Springel} {et~al.}(2001){Springel}, {White}, {Tormen}, \&
  {Kauffmann}}]{Springel01}
{Springel} V., {White} S.~D.~M., {Tormen} G., {Kauffmann} G., 2001, \mnras,
  328, 726

\bibitem[{{Tacchella} {et~al.}(2019){Tacchella}, {Diemer}, {Hernquist},
  {Genel}, {Marinacci}, {Nelson}, {Pillepich}, {Rodriguez-Gomez}, {Sales},
  {Springel}, \& {Vogelsberger}}]{Tacchella19}
{Tacchella} S., {Diemer} B., {Hernquist} L., {Genel} S., {Marinacci} F.,
  {Nelson} D., {Pillepich} A., {Rodriguez-Gomez} V. {et~al}, 2019, \mnras, 487,
  5416

\bibitem[{{The EAGLE team}(2017)}]{EAGLE17}
{The EAGLE team}, 2017, arXiv e-prints, arXiv:1706.09899

\bibitem[{{Thob} {et~al.}(2019){Thob}, {Crain}, {McCarthy}, {Schaller},
  {Lagos}, {Schaye}, {Talens}, {James}, {Theuns}, \& {Bower}}]{Thob19}
{Thob} A. C.~R., {Crain} R.~A., {McCarthy} I.~G., {Schaller} M., {Lagos} C.
  D.~P., {Schaye} J., {Talens} G. J.~J., {James} P.~A. {et~al}, 2019, \mnras,
  485, 972

\bibitem[{{Tissera} {et~al.}(2019){Tissera}, {Rosas-Guevara}, {Bower}, {Crain},
  {del P Lagos}, {Schaller}, {Schaye}, \& {Theuns}}]{Tissera19}
{Tissera} P.~B., {Rosas-Guevara} Y., {Bower} R.~G., {Crain} R.~A., {del P
  Lagos} C., {Schaller} M., {Schaye} J., {Theuns} T., 2019, \mnras, 482, 2208

\bibitem[{{Trayford} {et~al.}(2016){Trayford}, {Theuns}, {Bower}, {Crain},
  {Lagos}, {Schaller}, \& {Schaye}}]{Trayford16}
{Trayford} J.~W., {Theuns} T., {Bower} R.~G., {Crain} R.~A., {Lagos} C.~d.~P.,
  {Schaller} M., {Schaye} J., 2016, \mnras, 460, 3925

\bibitem[{{Trayford} {et~al.}(2015){Trayford}, {Theuns}, {Bower}, {Schaye},
  {Furlong}, {Schaller}, {Frenk}, {Crain}, {Vecchia}, \&
  {McCarthy}}]{Trayford15}
{Trayford} J.~W., {Theuns} T., {Bower} R.~G., {Schaye} J., {Furlong} M.,
  {Schaller} M., {Frenk} C.~S., {Crain} R.~A. {et~al}, 2015, \mnras, 452, 2879

\bibitem[{{van de Sande} {et~al.}(2017{\natexlab{a}}){van de Sande},
  {Bland-Hawthorn}, {Brough}, {Croom}, {Cortese}, {Foster}, {Scott}, {Bryant},
  {d'Eugenio}, {Tonini}, {Goodwin}, {Konstantopoulos}, {Lawrence}, {Medling},
  {Owers}, {Richards}, {Schaefer}, \& {Yi}}]{VandeSande17b}
{van de Sande} J., {Bland-Hawthorn} J., {Brough} S., {Croom} S.~M., {Cortese}
  L., {Foster} C., {Scott} N., {Bryant} J.~J. {et~al}, 2017{\natexlab{a}},
  \mnras, 472, 1272

\bibitem[{{van de Sande} {et~al.}(2017{\natexlab{b}}){van de Sande},
  {Bland-Hawthorn}, {Fogarty}, {Cortese}, {d'Eugenio}, {Croom}, {Scott},
  {Allen}, {Brough}, {Bryant}, {Cecil}, {Colless}, {Couch}, {Davies}, {Elahi},
  {Foster}, {Goldstein}, {Goodwin}, {Groves}, {Ho}, {Jeong}, {Jones},
  {Konstantopoulos}, {Lawrence}, {Leslie}, {L{\'o}pez-S{\'a}nchez}, {McDermid},
  {McElroy}, {Medling}, {Oh}, {Owers}, {Richards}, {Schaefer}, {Sharp},
  {Sweet}, {Taranu}, {Tonini}, {Walcher}, \& {Yi}}]{VandeSande17}
{van de Sande} J., {Bland-Hawthorn} J., {Fogarty} L.~M.~R., {Cortese} L.,
  {d'Eugenio} F., {Croom} S.~M., {Scott} N., {Allen} J.~T. {et~al},
  2017{\natexlab{b}}, \apj, 835, 104

\bibitem[{{van de Sande} {et~al.}(2019){van de Sande}, {Lagos}, {Welker},
  {Bland-Hawthorn}, {Schulze}, {Remus}, {Bah{\'e}}, {Brough}, {Bryant},
  {Cortese}, {Croom}, {Devriendt}, {Dubois}, {Goodwin}, {Konstantopoulos},
  {Lawrence}, {Medling}, {Pichon}, {Richards}, {Sanchez}, {Scott}, \&
  {Sweet}}]{vandeSande19}
{van de Sande} J., {Lagos} C. D.~P., {Welker} C., {Bland-Hawthorn} J.,
  {Schulze} F., {Remus} R.-S., {Bah{\'e}} Y., {Brough} S. {et~al}, 2019,
  \mnras, 484, 869

\bibitem[{{van de Sande} {et~al.}(2021){van de Sande}, {Vaughan}, {Cortese},
  {Scott}, {Bland-Hawthorn}, {Croom}, {Lagos}, {Brough}, {Bryant}, {Devriendt},
  {Dubois}, {D'Eugenio}, {Foster}, {Fraser-McKelvie}, {Harborne}, {Lawrence},
  {Oh}, {Owers}, {Poci}, {Remus}, {Richards}, {Schulze}, {Sweet}, {Varidel}, \&
  {Welker}}]{vandeSande20}
{van de Sande} J., {Vaughan} S.~P., {Cortese} L., {Scott} N., {Bland-Hawthorn}
  J., {Croom} S.~M., {Lagos} C. D.~P., {Brough} S. {et~al}, 2021, \mnras, 505,
  3078

\bibitem[{{Veale} {et~al.}(2017){Veale}, {Ma}, {Thomas}, {Greene}, {McConnell},
  {Walsh}, {Ito}, {Blakeslee}, \& {Janish}}]{Veale17}
{Veale} M., {Ma} C.-P., {Thomas} J., {Greene} J.~E., {McConnell} N.~J., {Walsh}
  J., {Ito} J., {Blakeslee} J.~P. {et~al}, 2017, \mnras, 464, 356

\bibitem[{{Walo-Mart{\'\i}n} {et~al.}(2020){Walo-Mart{\'\i}n},
  {Falc{\'o}n-Barroso}, {Dalla Vecchia}, {P{\'e}rez}, \&
  {Negri}}]{Walo-Martin20}
{Walo-Mart{\'\i}n} D., {Falc{\'o}n-Barroso} J., {Dalla Vecchia} C., {P{\'e}rez}
  I., {Negri} A., 2020, \mnras, 494, 5652

\bibitem[{{Wang} {et~al.}(2020){Wang}, {Cappellari}, {Peng}, \&
  {Graham}}]{Wang20}
{Wang} B., {Cappellari} M., {Peng} Y., {Graham} M., 2020, \mnras, 495, 1958

\bibitem[{{Wang} {et~al.}(2018){Wang}, {Wang}, {Mo}, {Lim}, {van den Bosch},
  {Kong}, {Wang}, {Yang}, \& {Chen}}]{Wang18}
{Wang} E., {Wang} H., {Mo} H., {Lim} S.~H., {van den Bosch} F.~C., {Kong} X.,
  {Wang} L., {Yang} X. {et~al}, 2018, \apj, 860, 102

\bibitem[{{Weijmans} {et~al.}(2014){Weijmans}, {de Zeeuw}, {Emsellem},
  {Krajnovi{\'c}}, {Lablanche}, {Alatalo}, {Blitz}, {Bois}, {Bournaud},
  {Bureau}, {Cappellari}, {Crocker}, {Davies}, {Davis}, {Duc}, {Khochfar},
  {Kuntschner}, {McDermid}, {Morganti}, {Naab}, {Oosterloo}, {Sarzi}, {Scott},
  {Serra}, {Verdoes Kleijn}, \& {Young}}]{Weijmans14}
{Weijmans} A.-M., {de Zeeuw} P.~T., {Emsellem} E., {Krajnovi{\'c}} D.,
  {Lablanche} P.-Y., {Alatalo} K., {Blitz} L., {Bois} M. {et~al}, 2014, \mnras,
  444, 3340

\bibitem[{{Wiersma} {et~al.}(2009{\natexlab{a}}){Wiersma}, {Schaye}, \&
  {Smith}}]{Wiersma09b}
{Wiersma} R.~P.~C., {Schaye} J., {Smith} B.~D., 2009{\natexlab{a}}, \mnras,
  393, 99

\bibitem[{{Wiersma} {et~al.}(2009{\natexlab{b}}){Wiersma}, {Schaye}, {Theuns},
  {Dalla Vecchia}, \& {Tornatore}}]{Wiersma09}
{Wiersma} R.~P.~C., {Schaye} J., {Theuns} T., {Dalla Vecchia} C., {Tornatore}
  L., 2009{\natexlab{b}}, \mnras, 399, 574

\bibitem[{{Wright} {et~al.}(2019){Wright}, {Lagos}, {Davies}, {Power},
  {Trayford}, \& {Wong}}]{Wright19}
{Wright} R.~J., {Lagos} C. d.~P., {Davies} L. J.~M., {Power} C., {Trayford}
  J.~W., {Wong} O.~I., 2019, \mnras, 487, 3740

\end{thebibliography}








\bsp	
\label{lastpage}
\end{document}